\documentclass[reprint, superscriptaddress, amsmath,amssymb, aps, prx, longbibliography,]{revtex4-1}
\usepackage{braket}
\usepackage{appendix}
\usepackage{graphicx}
\usepackage{amsmath}
\usepackage{amssymb}
\usepackage{algorithm}
\usepackage{babel}
\usepackage{algpseudocode}
\usepackage{bbm}
\usepackage{times}
\usepackage[colorlinks=true,urlcolor=blue,linkcolor=blue,citecolor=blue]{hyperref}
\usepackage[margin=0.75in]{geometry}

\newcommand{\tr}{\operatorname{tr}}

\DeclareMathOperator\arctanh{arctanh}

\begin{document}

\preprint{APS/123-QI}

\title{Stronger speed limit for observables: 
Tighter bound for the capacity of entanglement,\\
the modular Hamiltonian and 
the charging of a quantum battery }


\author{Divyansh Shrimali}
\email{divyanshshrimali@hri.res.in}
\affiliation{Harish Chandra Research Institute,\\ A CI of Homi Bhabha National Institute, Chhatnag Road, Jhunsi, Prayagraj 211019, India
}

%\collaboration{%}%\noaffiliation

\author{Biswaranjan Panda}
 
\email{biswaranjanpanda2002@gmail.com}
 
\affiliation{Indian Institute of Science Education and Research (IISER),\\
Berhampur, Odisha, India}%
\affiliation{Center for Quantum Engineering, Research and  Education (CQuERE),  \\
TCG CREST, Kolkata, India}
%\affiliation{
%}%

\author{Arun Kumar Pati}

\email{arun.pati@tcgcrest.org}

\affiliation{Center for Quantum Engineering, Research and  Education (CQuERE),  \\
TCG CREST, Kolkata, India}
%\affiliation{}

%\collaboration{%}%\noaffiliation

\date{\today}% It is always \today, today,

\begin{abstract}
How fast an observable can evolve in time is answered by so-called ``observable speed limit". Here, we prove a stronger version of the observable speed limit and show that the previously obtained bound is a special case of the new bound. The stronger quantum speed limit for the state also follows from the stronger quantum speed limit for observables (SQSLO).  We apply this to prove a stronger bound for the entanglement rate using the notion of capacity of entanglement (the quantum information theoretic counterpart of the heat capacity), and show that it outperforms previous bounds. 
Furthermore, we apply the SQSLO for the rate of modular Hamiltonian and in the context of interacting qubits in a quantum battery. These illustrative examples reveal that the speed limit for the modular energy and the time required to charge the battery can be exactly predicted using the new bound. This shows that for estimating the charging time of quantum battery, SQSLO is actually tight, i.e. it saturates. Our findings can have important applications in quantum thermodynamics, the complexity of operator growth, predicting the time rate of quantum correlation growth, and quantum technology in general. 

\end{abstract}

\maketitle
%\tableofcontents

\section{Introduction} \label{into}

Since the inception of scientific explorations, time has remained a paramount and fundamental notion in the study of physical systems. However, understanding time presents a considerable challenge, as it is not an operator but rather a parameter. New insights on the nature of time emerged after the formulation of the geometric uncertainty relation between energy fluctuation and time, imposing limitations on the rate at which a quantum system evolves. This concept was later formalized as the Quantum Speed Limit (QSL), which delineates the minimal time required for the evolution of a quantum system. The Mandelstam and Tamm derived a time-energy uncertainty relation that bounds the speed of evolution in terms of the energy dispersion~\cite{Mandelstam-45}. And some years later, another speed limit was identified for quantum state evolution, which incorporates the average energy in the ground state of the Hamiltonian~ \cite{Pati_geometry_phases-91,Margolus-98}. There exist few protocols involving quantum controls also that have been utilized to provide optimal value and controls to reach the target state within minimum time for entanglement production~\cite{Bao-18} and charging of Quantum Batteries~\cite{Rodriguez-24,Evangelakos-23,Mazzoncini-23}. Our work in this article though strictly relates with QSL which depends on the shortest path connecting the initial and final states of a given quantum system which depends on the fluctuation in the Hamiltonian and thus provides crucial insights into the dynamics of quantum processes. 

During the nascent stages of research, the bounds of the QSL were primarily established for the unitary dynamics of pure states for quantum systems~\cite{Mandelstam-45,Aharonov-90,Pati_geometry_phases-91,Margolus-98, Levitin-09,Gislason-85,Eberly-73,Bauer-78,Bhattacharyya-83,Leubner-85,Vaidman-92,Uhlmann-92,Uffink-93,Pfeifer-95,Horesh-98,Pati-99,Jonas-99,Andrecut-04,Gray-05,Magdalena-06,Andrews-07,Yurtsever-10,Shuang-10,Poggi-13}. Subsequently, researchers delved into investigating QSL within the framework of unitary dynamics for mixed states~\cite{Kupferman-08,Jones-10,Chau-10,Deffner-13,Fung-14,Andersson-14,Mondal-16,Mondal-16PLA,Deffner-17PA, Campaioli-18PRL}. The significance of QSL extends beyond theoretical explorations as it plays a pivotal role in the advancement of quantum technologies and devices, among other applications. Indeed, QSL finds diverse applications, including but not limited to, quantum computing~\cite{Ashhab-12}, quantum thermodynamics~\cite{Chiranjib-18,Funo-19}, quantum control theory~\cite{Caneva-09,Campbell-17}, quantum metrology~\cite{Campbell-18}, and beyond. 

Later, following the discovery of the stronger uncertainty relation Ref.~\cite{Pati-14}, a more robust QSL was unveiled~\cite{dimpi-22a}, which presented a tighter bound than the previously established Mandelstam and Tamm (MT) and Margolus-Levitin (ML) bounds. These advancements were made within the Schrödinger picture, where the state vector evolves over time. Subsequently, the exploration of QSL within the Heisenberg picture where observables evolve in time rather than the state, gathered interest.

Henceforth, leveraging the Robertson-Schrödinger uncertainty for observables utilizing Mandelstam and Tamm (MT) bound, a novel QSL bound was established, which is termed as the Quantum Speed Limit for observables (QSLO)~\cite{Brij-22}. This development prompted a natural inquiry: could we derive another bound using the stronger uncertainty relation? This question arises because the SQSL is already tighter than the MT bound, and while the QSLO is approximately equally as tight as the MT bound, there remains a need for a tighter bound for observables in the Heisenberg picture. Indeed, not only have we derived this new stronger bound, but have shown that it gives a significant improvement over QSLO while examining few prominent examples provided in this paper.
% but we have also demonstrated its significantly better performance through the examination of two prominent examples provided in this paper.

Entanglement is considered a very useful resource in information-processing tasks. Hence over the years, how to create and quantify entanglement has been a subject of major exploration~\cite{Horodecki-09,Sreetama-17}. The creation of quantum entanglement between two particles depends upon the choice of the initial state and suitable non-local interaction between them, but the designing of suitable interacting Hamiltonian is not always easy, which renders the production of entanglement a non-trivial task. Thus, for a given non-local Hamiltonian, what can be the best way to utilize this Hamiltonian to create entanglement. One way to answer this query, is by making use of the capacity of entanglement that was originally proposed to characterize topologically ordered states in the context of Kitaev model~\cite{capacity_def-19}. For a given pure bipartite entangled state $\rho_{AB}$, the capacity of entanglement is defined as the second cumulant of the entanglement spectrum, i.e., associated with the reduced density matrix, with $\{\lambda_{i}\}'$s, the eigenvalues of the reduced density matrix of any one of the subsystem, the capacity of entanglement $C_{E}$ is defined as the second cumulant of this entanglement spectrum, i.e; \(C_{E}=\sum_{i}\lambda_{i}{\rm log}^{2}\lambda_{i} - S_{EE}^{2}\), where $S_{EE}=-\sum_{i}\lambda_{i}{\rm log}\lambda_{i}$ is the well-known entanglement entropy. $C_{E}$ is similar in form to the heat capacity of thermal systems and can be thought of as the variance of the distribution of $-\log\lambda_{i}$ with probability $\lambda_{i}$ and hence contains information about the width of the eigenvalue distribution of reduced density matrix. It was shown in Ref.~\cite{capacity-22} that the quantum speed limit for creating the entanglement depends inversely on the fluctuation in the non-local Hamiltonian as well as on the average of the square root of the capacity of entanglement. It was, thus, inferred that the more the capacity of entanglement, the shorter the time duration system may take to produce the desired amount of entanglement.

Our first illustration involves readdressing the entanglement rate which was bounded by fluctuation in the non-local Hamiltonian and the capacity of entanglement as defined in Ref.~\cite{capacity_def}.  It is to be seen whether we can achieve a tighter bound for entanglement generation or degradation with the stronger uncertainty relation. If so, what can be the physical implication for the new expression, and under what choice of parameters we can achieve a tighter bound for a greater duration? Furthermore, a similar object was studied under the Heisenberg picture and the subsequent bound was interpreted in the form of the generation of modular energy, defined in the present context as a mean of composite modular Hamiltonian. Needless to say, the notion of capacity of entanglement has applications in diverse areas of physics ranging from condensed matter systems~\cite{Nicolas-16} to conformal field theories~\cite{capacity_def, Yao-10}, and alike. 

For the final example case, we analyse the ergotropy and bound on the charging process of quantum batteries. The various traditional batteries we make use of such as Lithium-ion, alkaline, and lead-acid batteries operate based on electrochemical reactions involving the movement of electrons between two electrodes through electrolytes. The performance of these batteries depends on factors like electrolyte composition, electrode materials, and overall design. The Quantum batteries (QBs) represent a new frontier, grounded in quantum mechanical principles such as tunneling effects, entanglement, qubit-based technologies, and more~\cite{Alicki-13}. Theoretical models propose that these batteries can leverage quantum superposition and entanglement to store and recover energy, offering enhanced efficiency compared to conventional batteries~\cite{Binder-15,Campaioli-17,Rossini-20,Gyhm-22,Gyhm-24, Gyhm-24Min}. However, despite their potential advantages, QBs are still in the early stages of development due to technological limitations. Numerous challenges, including issues related to stability, scalability, and practical implementation, need to be addressed for their widespread usage~\cite{Lewenstein-20,Campaioli-23}.

The QB model comprises two essential components: a battery charger and a battery holder. However, energy loss is also accounted for in the subsequent stages, achievable by isolating the quantum system from the environment, treated as a dissipation-less subsystem. The effective coupling of the battery holder with the battery charger is crucial for energy acquisition. The focus of recent theoretical research has been on exploring basic bipartite state models and other related models in the realm of quantum batteries~\cite{Andolina-18,Thao-18,Zhang-19,Barra-19,Santos-19,Andolina-19,Alba-20,Santos-20,Santos-21,Dou-22,Barra-22,Carrasco-22,Shaghaghi-22,Carla-23,Santos-23,Kamin-23,Downing-23,Hadipour-23a, Kamin-20A}. Theoretical evidence already supports the notion that in a collective charging scheme, QBs can demonstrate accelerated charging leveraging quantum correlations~\cite{Binder-15,Campaioli-17,Iran-20}. Presently, diverse models of QBs have been proposed, including quantum cavities, spin chains, the Sachdev-Ye-Kitaev model and quantum oscillators~\cite{Pirmoradian-19,Zhang-23,Crescente-20,Joshi-22,Lewenstein-20,BrijM-21,Niedenzu-18,Andolina-19,Ferraro-18,Zhao-21,Rossini-19,Zakavati-21,Zhao-22}. However, experimental investigations are limited, with fewer models explored, such as the cavity-assisted charging of an organic quantum battery~\cite{Quach-22}. In this article, we take the example of entanglement based QBs under different charging regimes. It is examined whether stronger quantum speed limit for observables (SQSLO) gives any significant improvement over existing QSLO~\cite{Brij-22, Carabba-22} for charging time.

This paper is organised as follows. In Section~\ref{s2}, we discuss all the basic concepts utilized in this paper. Subsequently, in Section~\ref{s3}, we derive the SQSLO bound by employing a stronger uncertainty relation and compare it with other previously established bounds. In the next Section~\ref{sca}, we have given the SQSL for states and demonstrated a better bound for entanglement generation with capacity of entanglement. Following this, in Section~\ref{s4}, we present two applications of the QSLO bound for the modular energy and charging time of the quantum battery. Finally, in Section~\ref{s5}, we conclude our paper.

\section{Definitions and Relations}\label{s2}

\textbf{Stronger Uncertainty Relation:}  
Unlike classical system, where all observables can be measured with arbitrary accuracy, the same is not true for quantum systems. For a given quantum state there are restrictions on the results of the measurements of non-commuting observables. The uncertainty relation captures such a restriction for two incompatible observables.

The Heisenberg-Robertson uncertainty relation provides a lower bound by merely yielding the product of two variances of observables based on their commutator. This proves that it is impossible to prepare a quantum state for  which variances of two non-commuting observables can be arbitrarily reduced simultaneously. In contrast, a stronger uncertainty relation offers a more comprehensive approach by considering the sum of variances. This approach ensures that the lower bound remains non-trivial, especially when dealing with two observables that are incompatible within the state of the system. Thus, it provides a more nuanced understanding of uncertainty, particularly in cases where traditional relations fall short. However, we will not be using the sum form of the stronger uncertainty relation. One of the stronger uncertainty relation in the product form as given in~\cite{Pati-14} has the form
\begin{equation}\label{MPU}
    \Delta A\Delta B \ge \pm \frac{i}{2}\frac{\langle [A, B]\rangle}{\left( 1- \frac{1}{2}\left|\langle\Psi^{\perp}|\frac{A}{\Delta A} \mp i\frac{B}{\Delta B}|\Psi\rangle\right|^2\right)}, 
\end{equation}
where $A$ and $B$ are two incompatible observables with \(\Delta A=\sqrt{\langle A^{2}\rangle-\langle A\rangle^{2}},\hspace{0.1cm}\Delta B=\sqrt{\langle B^{2}\rangle-\langle B\rangle^{2}}\), and the averages are defined in the state $|\Psi\rangle$ for the given quantum system. This Eqn.~(\ref{MPU}) can be reduced to the Heisenberg-Robertson uncertainty relation when it minimizes the lower bound over $|\Psi^{\perp}\rangle$ and becomes an equality when maximizes it.
The above relation is stronger than the standard Heisenberg-Robertson uncertainty relation. We will be using this to prove our stronger quantum speed limit for observable.

\vspace{0.5cm}
\textbf{Capacity of entanglement: }
Let us consider a composite system $AB$ with pure state $|\Psi\rangle_{AB}$. The amount of entanglement between subsystems $A$ and $B$ can be quantified via the entanglement entropy which is defined as the von Neumann entropy of the reduced density operator $\rho_{A}= \sum_{n}\lambda_{n}\ket{\psi_{n}}_{A}\bra{\psi_{n}}  $ (or $\rho_{B}$), i.e., 
\begin{equation}
 S_{EE} = S(\rho_{A})=  - \tr(\rho_{A} \log\rho_{A}) = -\sum_{n}\lambda_{n}\log\lambda_{n}
\end{equation}
which is invariant under local unitary transformations on $\rho_{A}$. %Density operator $\rho$ can always be digonalized  by unitary transformation i.e., it can be written in the form  $\rho= \sum_{n}\lambda_{n}\ket{\psi_n}\bra{\psi_n}$. Then von Neumann entropy can be re-expressed as 
The von Neumann entropy vanishes when density operator $\rho_{A}$ is a pure state. For a completely mixed density operator, the von Neumann entropy attains its maximum value of $\rm \log d_{A}$, where $\rm d_{A} = dim(\cal{H}_{A})$.

For any density operator $ \rho_{A}$ associated with quantum system $A$, we can define a formal “Hamiltonian” $K_{A}$, called the modular Hamiltonian, with respect to which the density operator $\rho_{A}$ is a Gibbs like state (with $\beta=1$)
$$\rho_{A}=\frac{e^{-K_{A}}}{Z},$$ 
where $Z=\tr(e^{-K_{A}}).$ Note that any density matrix can be written in this form for some choice of Hermitian operator $K_{A}$. With slight adjustments in the above equation, the modular Hamiltonian $K_{A}$ can be written as $K_{A}=-\log\rho_{A}$.  In this case, the entanglement entropy of the system is equivalent to the thermodynamic entropy of a system described by Hamiltonian $ K_{A} $ (with $\beta=1$). 
Writing in terms of modular Hamiltonian $K_A =-\log\rho_{A}$, the entanglement entropy
becomes the expectation value of the modular Hamiltonian
\begin{equation}
    S_{EE} = -\tr(\rho_{A}\log\rho_{A}) = \tr(\rho_{A} K_{A})= \langle K_{A}\rangle \,.
\end{equation}

The capacity of entanglement is another information-theoretic quantity that has gained some interest in recent time~\cite{Pawel-22}. It is defined as the variance of the modular Hamiltonian $K_{A}$~\cite{capacity_def} in the state $\ket{\Psi}_{AB}$ and can be expressed as
\begin{align}
 C_{E}(\rho_{A}) & =\bra{\Psi}(K_{A}\otimes \mathcal{I}_{B})^{2} \ket{\Psi}-\bra{\Psi}(K_{A}\otimes \mathcal{I}_{B})\ket{\Psi}^{2} \nonumber \\
 & = \tr[\rho_{A}(-\log\rho_{A})^2]-[\tr(-\rho_{A}\,\log\rho_{A})]^{2} \\
 & = \tr[\rho_{A}K_{A}^2]-[\tr(\rho_{A}K_{A})]^{2}\nonumber \\      \label{equ:Entanglement_Capacity}
 &=\langle K_{A}^{2}\rangle-\langle K_{A}\rangle^{2} = \Delta K_{A}^2.
\end{align}
The capacity of entanglement has also been defined in terms of variance of the relative surprisal between two density matrices $V(\rho||\sigma)$:
\begin{equation}
    V(\rho||\sigma)=\tr\left(\rho(\log(\rho)-\log(\sigma))^{2}\right) - D(\rho||\sigma)^{2}.
\end{equation}
Here, if one of the density matrices becomes maximally mixed (i.e., either $\rho$ or $\sigma$ becomes $\mathcal{I}/d$), then the variance of the relative surprisal becomes the capacity of entanglement. For further details and properties of capacity of entanglement, readers are advised to go through Ref.~\cite{Divyansh-22}.

\vspace{0.5cm}
\textbf{Extractable work from quantum batteries:}
Let the quantum system representing the battery be of dimension $d$ with the corresponding Hilbert space $\mathcal{H}$. We further pick a standard basis for describing the system Hamiltonian
\begin{equation}
    H=\sum_{j=1}^{d} h_{j} |j\rangle\langle j|\hspace{0.25cm}{\rm with}\hspace{0.25cm}h_{j+1}>h_j ,
\end{equation}
where the assumption is that the energy levels are non-degenerate. 

To extract the energy from the battery, the time-dependent fields that are used can be described as $V(t)=V^{\dagger}(t)$ where such fields are switched on for time interval $0 \leq t\leq \tau$. The initial state of the battery is described by a density matrix $\rho$ which is time evolved from the Liouville-von Neumann equation
\begin{equation}
    \frac{d}{dt}\rho(t)=-i[H+V(t),\rho(t)],\hspace{0.3cm}\rho(0)=\rho.
\end{equation}
The work extraction by this procedure is then
\begin{equation}
    W={\rm tr}(\rho H)-{\rm tr}(\rho(\tau)H),
\end{equation}
where time evolved state is given as \(\rho(\tau)=U(\tau)\hspace{0.1cm}\rho\hspace{0.1cm} U^{\dagger}(\tau)\).

Further, through a proper choice of $V$, which is termed as controlling term and gives rise to action of local unitaries, any unitary $U$ can be obtained for $U(\tau)$. Therefore the maximal amount of extractable work, called \textit{ergotropy}~\cite{Alicki-13}, can be defined as
\begin{equation}
    W_{max}:={\rm tr}(\rho H) - \min{\rm tr}(U\rho U^{\dagger}H),
\end{equation}
where the minimum is taken over all unitary transformations of $\mathcal{H}$.

\section{Stronger QSL for Observables (SQSLO)}\label{s3}

\subsection{Derivation of Stronger Quantum Speed Limit for Observables}\label{se1}

Let us consider a quantum system with a state vector $|\Psi \rangle \in \mathcal{H}^N$. 
In the Heisenberg picture, we can imagine that the operators representing the observables evolve in time, while
the vectors in the Hilbert space (quantum states) remain independent of
time. This is opposite to the Schr\"{o}dinger picture, where the
observables are independent of time and the states evolve in time.
In the Heisenberg picture, each self-adjoint operator
evolves in time according to the operator-valued differential equation.

As we are dealing with the Heisenberg picture, the observable $O(t)$undergoes an unitary evolution as given by the Heisenberg equation of motion
\begin{equation}
    i\hbar\frac{d O(t)}{dt} = [ O(t), H],
\end{equation}
where $H$ is the Hamiltonian operator of the system, and where $[O, H]$ is the
commutator. If $O(t)$ commutes with the Hamiltonian, then it remains constant
in time. 
%As established in the Heisenberg picture, observables undergo changes. Utilizing the Robertson-Heisenberg uncertainty relation, we have already derived the Quantum Speed Limit for observables ($T_{QSLO}$). 
In this section, we aim to derive a more stringent QSL bound for observables, surpassing the previously obtained limit. This bound stems from  the stronger uncertainty relation, applicable to any two incompatible observables $A(t)$ and $B(t)$ in the Heisenberg picture. This is given by

\begin{equation}
    \Delta A(t)\hspace{0.1cm}\Delta B(t)\hspace{0.1cm}(1 - R(t) )  \ge \pm \frac{i}{2}\langle\Psi |[A(t), B(t)]|\Psi \rangle,
    \label{eqn:stronger_uncertainty}
\end{equation}
\noindent where 
\begin{equation}
    R(t) = \frac{1}{2}\left|\langle\Psi^{\perp}|\frac{A(t)}{\Delta A(t)} \mp i\frac{B(t)}{\Delta B(t)}|\Psi \rangle\right|^2 ,
    \label{eqn:R_t}
\end{equation}
 $\Delta A(t) $ = $\sqrt{\langle A(t)^2\rangle - \langle A(t) \rangle^2}$, $\Delta B(t) $ = $\sqrt{\langle B(t)^2\rangle - \langle B(t) \rangle^2}$, $|\Psi \rangle$ is the state of the system  in which averages are calculated and $|\Psi^{\perp}\rangle$ is the orthogonal
 state to  $|\Psi \rangle$.
 %Based on the sign of the expression $\pm\frac{i}{2}\frac{\langle\Psi(t)|[A, B]|\Psi(t)\rangle}{1-R(t)}$, the pair of signs $(\pm, \mp)$ can be chosen accordingly. 
 %Stronger uncertainty relation in the product form looks similar to the Robertson uncertainty relation having an extra term $R(t)$. 
 We will prove that the bound obtained from the above equation is tighter than the existing bound $T_{QSLO}$ which was derived by using Robertson uncertainty relation. 

Now, consider the desired observable, denoted as $A = O(t)$, and an another operator, $B = H$. 
Using the stronger uncertainty relation, we can obtain

\begin{equation}
    \Delta O(t)\hspace{0.1cm}\Delta H \hspace{0.1cm}(1 - R(t)) \ge \frac{\hbar}{2}\left|\frac{d\langle O(t)\rangle}{dt}\right|.
    \label{eqn:S_C_R}
\end{equation}

From the above expression, we obtain the stronger quantum speed limit for observable (SQSLO) as given by 

\begin{equation}\label{osqsl}
    T \ge T_{SQSL} ^O = \frac{\hbar}{2\Delta H} \int_0 ^T \frac{|d\langle O(t)\rangle|}{\Delta O(t)\hspace{0.1cm}\eta (t)},
\end{equation}
where $\eta (t)$ = $(1 - R(t))$, $\Delta O(t)$ = $\sqrt{\langle O(t)^2\rangle - \langle O(t)\rangle^2}$ and $\Delta H$ = $\sqrt{\langle H^2\rangle - \langle H\rangle^2}$. Here, the time $T$ denotes the time we consider for the evolution of quantum system. This SQSLO can be written as
\begin{equation}
    T_{SQSL} ^O = \frac{\hbar}{2\Delta H} \frac{|\langle O(T)\rangle - \langle O(0)\rangle|}{\langle\langle \Delta O(t)\hspace{0.1cm}\eta (t)\rangle\rangle_T},
\end{equation}
where $\langle\langle \Delta O(t)\hspace{0.1cm}\eta (t)\rangle\rangle_T$ = $\frac{1}{T}\int_0 ^T  \Delta O(T)\hspace{0.1cm}\eta(t)dt$, is the time average of the quantity $\Delta O(t)\hspace{0.1cm}\eta(t)$~\cite{pati-23}. 

Alternatively, through Eqn.(\eqref{eqn:S_C_R}), we can rewrite SQSLO as
% $$
%  T \Delta H \left(\frac{1}{T}\int_0 ^T(1 - R(t)) dt \right) \ge \frac{\hbar}{2} \int_0 ^T \frac{|d\langle O(t)\rangle |}{\Delta O(t)} .
% $$
% Then again by some simple interchange of steps, we get

% $$
% T \Delta H (1 - \overline{R}(t)) \ge \frac{\hbar}{2} \int_0 ^T \frac{|d\langle O(t)\rangle |}{\Delta O(t)} .
% $$
% Using Eqn.(\ref{eqn:S_C_R}) and following the previous step, in an alternative way, we can write SQSLO as 
\begin{equation}\label{asq}
    T \geq T_{SQSL} ^O = \frac{\hbar\hspace{0.1cm}\Lambda(T)}{2\Delta H}\int_{0}^{T}\frac{|d\langle O(t)\rangle|}{\Delta O(t)} ,
\end{equation}

\noindent where $\Lambda(T)=\frac{1}{1-\overline{R}(t)}$, with $\overline{R}(t)=\frac{1}{T}\int_{0}^{T}R(t)\hspace{0.1cm}dt$.

Now, we can show that the previously derived bound of the QSLO~\cite{Brij-22} follows from the stronger QSLO. This will ensure that SQSLO is an improvement over QSLO. %Subsequently, in the application section, we visualize these speed limits through the plotted representation. The plots serve as a demonstrative tool, allowing us to %observe and conclude that SQSLO is consistently tighter than QSLO. 
As evident from Eqn.~(\ref{osqsl}), an additional factor of $\eta(t)=1 - R(t)$ is present in SQSLO, with $0\leq R(t) \leq 1\hspace{0.1cm}\forall\hspace{0.1cm}t$,  we have
$\eta(t)\in [0,1]$. This results in the final expression

\begin{equation}
    T \ge \frac{\hbar}{2\Delta H} \int_{0}^{T} \frac{|d\langle O(t)\rangle|}{\Delta O(t)\eta(t)} \ge \frac{\hbar}{2\Delta H} \int_0 ^T \frac{|d\langle O(t)\rangle|}{\Delta O(t)}.
\end{equation}
Therefore, we have 
\begin{equation}
    T \ge  T_{SQSL} ^{O} \ge T_{QSL} ^{O}
\end{equation}
This shows that indeed SQSLO is tighter than QSLO.

\section{Stronger QSL for states and Entanglement Capacity}\label{sca}
In this section, we delve into the relationship between SQSL for observables and SQSL concerning states. Notably, SQSL for state emerges as a distinctive instance within the broader framework of SQSL for observable, when we consider the observable as the projector of the initial state. For realising that, let us consider a quantum system with an initial state $|\Psi \rangle = \sum_i c_i |i\rangle$. We continue with the observable taking the form of a projector, i.e., $O(0) = P$. Consequently, the probability of finding the system in state $|i\rangle$ at time $t=0$ becomes $|c_i|^2$, upon performing measurement with projector defined as $P=|i\rangle\langle i|$. Now we wish to study the bound on speed limit for the projector for the quantum system evolving a state $\ket{i}$ unitarily in time. Using Eqn.~\eqref{asq} in an alternative way, we can express the quantum speed limit for the projector as given by
\begin{equation}
    T \geq \frac{\hbar\hspace{0.1cm}\Lambda(T)}{2\Delta H}\int_{0}^{T}\frac{|d\langle P(t)\rangle|}{\Delta P(t)},
\end{equation}
where $P(t)$ = $U(t) P(0) U(t)^{\dagger}$ and $\langle P(t)\rangle$ = $p(t)$ is the probability of the quantum system in state $|i\rangle$ at some later time $t$. 

The above bound can be expressed as
\begin{equation}
    T\geq \frac{\hbar \Lambda(T)}{\Delta H}\left|\arcsin(\sqrt{p(T)})-\arcsin(\sqrt{p(0)})\right|,
\end{equation}
\noindent where $\Lambda(T)=\frac{1}{1-\overline{R}(t)}$, with $\overline{R}(t)=\frac{1}{T}\int_{0}^{T}R(t)\hspace{0.1cm}dt$. Now if we choose $p(0)=1$, i.e, $\ket{\Psi}=\ket{i}$, then the above inequality results in the following bound
\begin{equation}
    T\geq \frac{\hbar \Lambda(T)}{\Delta H} \arccos(\sqrt{p(T)}).
\end{equation}
This is equivalent to the stronger speed limit for the state obtained in Ref.~\cite{Brij-22}. As we know the Mandelstam and Tamm bound is a special case of the stronger speed limit for the state, we can say that the stronger speed limit for the state and MT bound, both are special cases of the stronger quantum speed limit for the observable. Thus, our result unifies the previous known bounds on the observable and state.

Next, we apply the SQSL for state to provide stronger bound for the entanglement rate using capacity of entanglement.

\subsection*{Improved Bounds on Rate of Entanglement through capacity of entanglement using SQSL for states}

The dynamics of entanglement under two-qubit nonlocal Hamiltonian has been adressed in Ref.~\cite{rate_ent}. Further, the inquiry on capacity of entanglement for two-qubit non-local Hamiltonian and its properties have been adressed in Ref.~\cite{capacity-22}. It was discovered that the defined capacity of entanglement indeed played a role in giving parameter free bound to quantum speed limit for creating entanglement.
In this section, we address the following question: Can we improve upon the quantum speed limit bound, using the stronger uncertainty relation. As we shall see, indeed one can get a tighter bound in such case and it again bolsters the point that the capacity of entanglement has a physical meaning in deciding how much time a bipartite state takes in order to produce a certain amount of entanglement.

Let us briefly discuss about the two-qubit system, for which the non-local Hamiltonian can be expressed as (except for trivial constants)
\begin{equation}
    H = \vec{\alpha}\cdot\vec{\sigma}^{A}\otimes\mathcal{I}_{B}+\mathcal{I}_{A}\otimes\vec{\beta}\cdot\vec{\sigma}^{B}+\sum_{i,j=1}^{3} \gamma_{ij}\sigma_{i}^{A}\otimes\sigma_{j}^{B}, \label{bipartite_Hamiltonian} 
\end{equation}
where $\vec{\alpha},\vec{\beta}$ are real vectors, $\gamma$ is a real matrix and, $\mathcal{I}_{A}$ and $\mathcal{I}_{B}$ are identity operator acting on $\mathcal{H}_{A}$ and $\mathcal{H}_{B }$. The above Hamiltonian can be rewritten in one of the two canonical forms under the action of local unitaries acting on each qubits~\cite{bennett-02,rate_ent}. This is given by
\begin{equation}
H^{\pm} = \mu_1\sigma_1^{A}\otimes\sigma_1^{B}\pm\mu_2\sigma_2^{A}\otimes\sigma_2^{B}+\mu_3\sigma_3^{A}\otimes\sigma_3^{B},
\label{eqn:Non_loc_hamiltonian}
\end{equation}
where \(\mu_1\geq\mu_2\geq\mu_3\geq 0\) are the singular values of matrix $\gamma$~\cite{rate_ent}. Using the Schmidt-decomposition, 
any two qubit pure state can be written as
\begin{equation}
    \ket{\Psi}_{AB} = \sqrt{p} \ket{\phi} \ket{\chi} + \sqrt{1-p} \ket{ \phi^{\perp}} \ket{\chi^{\perp}}.
    \label{2qubit_general}
\end{equation}
We can utilize the form of Hamiltonian in Eqn.~\eqref{eqn:Non_loc_hamiltonian} and choose $H^{+}$ (i.e. assuming $\det(\gamma)\geq 0$) to evolve the state in Eqn.~\eqref{2qubit_general} without loosing any generality \cite{rate_ent}. To further showcase a specific example, let us choose \(\ket{\phi}=\ket{0}\) and \(\ket{\chi}=\ket{0}\). Thus, the state at time $t=0$ takes the form
\begin{equation}
 \ket{\Psi(0)}_{AB}=\sqrt{p}\ket{0}\ket{0}+\sqrt{1-p}\ket{1}\ket{1}.  
 \label{equ:time zero state}
\end{equation}
Under the action of the non-local Hamiltonian, the joint state at time $t$ can be written as ($\hbar=1$)
\begin{align}
\ket{\Psi(t)}_{AB} = e^{-iHt}\ket{\Psi}_{AB} =\alpha(t)\ket{0}\ket{0}+\beta(t)\ket{1}\ket{1},
\label{equ:time evolved state}
\end{align}
where $\alpha(t) = e^{-i \mu _3 t} \left(\sqrt{p} \cos (\theta t) -i \sqrt{1-p} \sin( \theta t) \right)$ , $\beta(t) =e^{-i \mu _3 t} \left(\sqrt{1-p} \cos(\theta t) -i \sqrt{p} \sin(\theta t) \right)$ and $\theta = (\mu _1-\mu _2)$.
To evaluate the capacity of entanglement, we would require the reduced density matrix of the two qubit evolved state, $\rho_{A}(t)=\tr_{B}(\rho_{AB}(t))$, which is given by
\begin{align}
   \rho_{A}(t) &=  \lambda_{1}(t) \ket{0}\bra{0} + \lambda_{2}(t) \ket{1}\bra{1}, 
\end{align}
where $\lambda_1(t) = \left|\alpha(t)^2\right| $ and $\lambda_2(t) = \left|\beta(t)^2\right|$, thus read as     

\begin{align}
    \lambda_{1}(t) &= \frac{1}{2} \left[1 - (1-2 p) \cos \left(2\theta t\right) \right], \nonumber\\
    \lambda_{2}(t) &= \frac{1}{2} \left[1 + (1-2 p) \cos \left(2\theta t\right) \right]. \nonumber
\end{align}
The capacity of entanglement at a later time t can be calculated from the variance of modular Hamiltonian $ K_{A}$ defined as $K_{A}=-\log\rho_{A}$. This is given by
\begin{align}
  C_{E}(t)&=\tr(\rho_{A}(t)(-\log\rho_{A}(t))^2)-(\tr(-\rho_{A}(t)\log\rho_{A}(t)))^{2}\, , \nonumber\\
  &  = \sum_{i=1}^{2}\lambda_{i}(t) \log^{2}\lambda_{i}(t) - \left(-\sum_{i=1}^{2}\lambda_{i}(t) \log\lambda_{i}(t)\right)^{2} .
\end{align}

Further for Eqn.~\eqref{equ:time evolved state}, one can evaluate the entanglement entropy and capacity as:

\begin{align}
    C_{E}(t) =  -\frac{1}{2}&{\rm{Tanh}}^{-1}[(2p-1)\cos(2\theta t)]^{2} \times \nonumber\\
    & \left(-1+4p(p-1)+(1-2p)^{2}\cos(4\theta t)\right)\nonumber \\
    S_{EE}  = \frac{e^{-2i t\theta}}{4}\Bigg[&\left(-1-2e^{2it\theta}+2p+e^{4it\theta}(-1+2p)\right)\times\nonumber\\
    & \log\left[\frac{1+(1-2p)\cos(2t\theta)}{2}\right] -\nonumber\\
    & \left(-1+2e^{2it\theta}+2p+e^{4it\theta}(-1+2p)\right)\times\nonumber\\
    & \log\left[\frac{1-(1-2p)\cos(2t\theta)}{2}\right]\Bigg]\label{eqn:CE_S_expr_fin}
\end{align}
\noindent for chosen parameters $p\hspace{0.1cm}{\rm and}\hspace{0.1cm}\theta$. 
%  \begin{equation}
%     C_{E}(t) = -\frac{1}{2}\rm{Tanh}^{-1}[(2p-1)\cos(2\theta t)]^{2}\left(-1+4p(p-1)+(1-2p)^{2}\cos(4\theta t)\right),
% \end{equation}

\begin{figure}[h!]
    \centering
    \includegraphics[width=9cm]{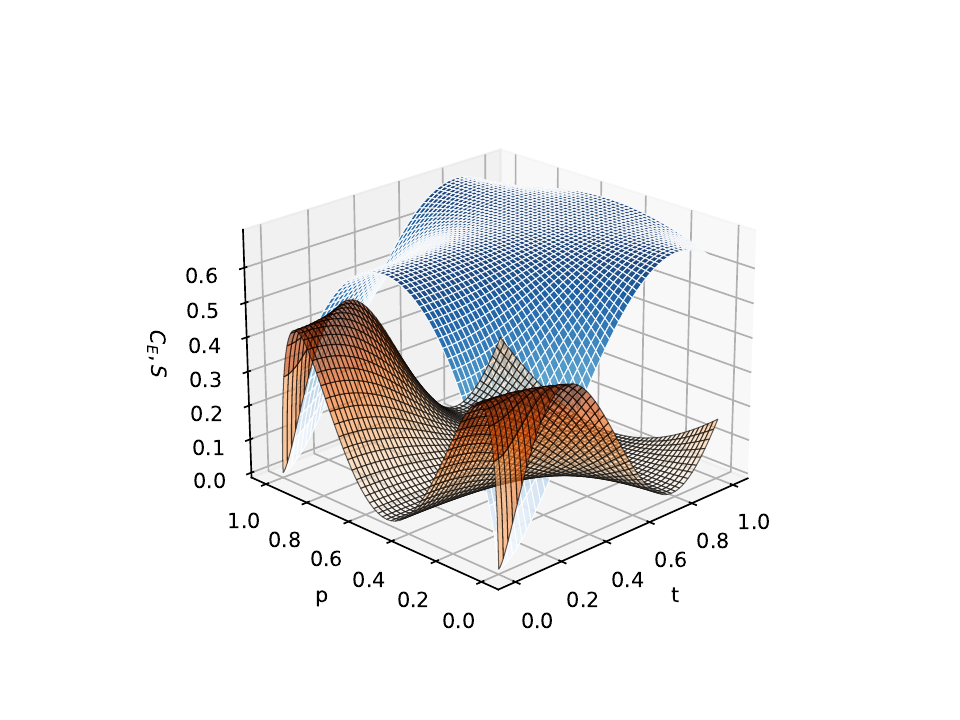}
    \caption{Surface-plot for capacity of entanglement $C_{E}(p,t)$ (`darker' orange surface plot) and entanglement entropy $S_{EE}(p,t)$ (`lighter' blue surface plot) vs p and t taking $\theta$ = 1}
    \label{fig:S+CE}
\end{figure}

The plot in Fig.\ref{fig:S+CE} shows how entanglement entropy and capacity of entanglement varies for some chosen value of $\theta=1$, where capacity reduces to zero for when the state is either separable or stationary~\cite{capacity-22, rate_ent}.

Now, for evaluating bound on the rate of entanglement, we use the stronger-uncertainty relation in the form of Eqn.~\eqref{eqn:stronger_uncertainty} for the two non-commuting operators $K_{AB}=K_{A}\otimes\mathcal{I}$ and $H_{AB}$. This leads to
\begin{equation}
    \frac{1}{2}|\langle\Psi(t)|[K_{A}\otimes\mathcal{I}, H_{AB}]| \Psi(t)\rangle|\leq \Delta K_{A}\hspace{0.1cm}\Delta H_{AB}\hspace{0.1cm}(1-R(t)).
    \label{equ:uncertainty_relation_for_modularhamiltonian_and_non-local_Hamiltonian}
\end{equation}
% Recall that the evolution of average of any self adjoint operator $O$ is given by 
% \begin{equation}
%     i \hbar\frac{{{\rm d} \langle O\rangle}}{{\rm d}t} = \langle[O,H]\rangle \, . \label{equ:equation_of_motion_of_K}
% \end{equation}
Using Eqn.~\eqref{eqn:S_C_R} (for $O = K_{A}$) in Eqn.~\eqref{equ:uncertainty_relation_for_modularhamiltonian_and_non-local_Hamiltonian}, we then obtain
\begin{equation}
    \frac{\hbar}{2}\left|\frac{{\rm d} \langle K_{A}\rangle}{{\rm d}t}\right|\leq \Delta K_{A}\hspace{0.1cm}\Delta H_{AB}\hspace{0.1cm}(1-R(t))\,.
\end{equation}

Let $\Gamma(t)$ denote the rate of entanglement. Recall that the average of the modular Hamiltonian is the entanglement entropy $S_{EE}$. In terms of the entanglement rate $\Gamma(t)$, the above equation can be written as
 \begin{equation}
    \left| \Gamma(t)\right| \leq \frac{2}{\hbar}\Delta K_{A}\hspace{0.1cm}\Delta H_{AB}\hspace{0.1cm}(1-R(t)) \, .
 \end{equation}
Since the square of the standard deviation of modular Hamiltonian is the capacity of entanglement, so in terms of the capacity of entanglement, we can write above bound  as
\begin{equation}
    \left|\Gamma(t)\right| \leq \frac{2}{\hbar}\sqrt{C_{E}(t)} \hspace{0.1cm}\Delta H_{AB}\hspace{0.1cm}(1-R(t))\, .     
    \label{equ:entanglement_speed}
\end{equation}
We can interpret the above formula by noting that one can define the speed of transportation of bipartite pure entangled state on projective Hilbert space of the given system by the expression $\frac{2}{\hbar}\Delta H_{AB}$. Further using the Fubini-Study metric for two nearby states, one can define the infinitesimal distance between two nearby states~\cite{Aharonov-90, Pati_geometry_phases-91, pati-95} as
\begin{equation}
    {\rm d}S^2 = 4\left( 1 -\left|\langle{\Psi(t)}\ket{\Psi(t+{\rm dt})}\right|^2 \right) = \frac{4}{\hbar^2} \Delta H_{AB}^{2}\hspace{0.1cm}{\rm  d}t^2.
\end{equation}
Therefore, the speed of transportation as measured by the Fubini-Study metric is given by $ V= \frac{{\rm d}S}{{\rm d}t} = \frac{2}{\hbar} \Delta H_{AB}$. Thus, the entanglement rate is upper bounded by the speed of quantum evolution \cite{Deffner-17} and the square root of the capacity of entanglement and correction factor due to stronger uncertainty, i.e., $|\Gamma(t)|\leq\sqrt{C_{E}(t)}\hspace{0.1cm}V\hspace{0.1cm}(1-R(t))$.

We know from Ref.~\cite{Bravyi-07} that for an ancilla unassisted case, the entanglement rate is upper bounded by $c\|H\|\log d$, where $d={\rm min(dim}{\cal{H}}_{A},{\rm dim}{\cal{H}}_{B})$, $c$ being a constant between the value 1 and 2, and $\|H\|$ is operator norm of Hamiltonian which corresponds to $p=\infty$ of the Schatten p-norm of $H$ which is defined as $\|H \|_p = [\tr \left(\sqrt{H^\dag H}\right)^p ]^\frac{1}{p}$. Now, using the fact that the maximum value of capacity of entanglement  is proportional to $S_{\rm max}(\rho_{A})^{2}$~\cite{capacity_def}, where $S_{\rm{max}}(\rho_{A})$ is maximum value of von Neumann entropy of subsystem which is upper bounded by $\log{d}_{A}$, where $d_{A}$ is the dimension of Hilbert space of subsystem $A$, and $\Delta H \leq \|H\|$, a similar bound on the entanglement rate can be obtained from Eqn.~\eqref{equ:entanglement_speed}. Further the factor $(1-R(t))$ varies between 0 and 1 for given standard choice of Eqn.~\eqref{equ:time evolved state}.  Thus, the bound on the entanglement rate given in Eqn.~\eqref{equ:entanglement_speed} is significantly stronger than the previously known bound and will be shown subsequently to be a improvement upon the bound found in Ref.~\cite{capacity-22}.

This bound on entanglement rate can be used to provide QSL which decides how fast a quantum state evolves in time from an initial state to a final state~\cite{Pfeifer-93}. Even though it was discovered by Mandelstam and Tamm, over last one decade, there have been active explorations on generalising the notion of quantum speed limit for mixed states~\cite{Wu-18,Mondal-16} and on resources that a quantum system might posses~\cite{Modi-20}. The notion of generalized quantum speed limit has been explored in Ref.~\cite{dimpi-22JPA}. Further, the quantum speed limit for observables has been defined and it was shown that the QSL for state evolution is a special case of the QSL for observable~\cite{Brij-22}. For a quantum system evolving under a given dynamics, there exists a fundamental limitation on the speed for entropy \(S(\rho)\), maximal information \(I(\rho)\), and quantum coherence \(C(\rho)\)~\cite{mohan-22} as well as on other quantum correlations like entanglement, quantum mutual information and Bell-CHSH correlation~\cite{Divyansh-22}. 

Now, we are in the position to give a stronger uncertainty based expression for QSL bound, 
% and highlight that rate at which entanglement can be generated or degraded, depends on capacity of entanglement along with correction factor based on $R(t)$,
\begin{equation}
  \int_{0}^{T} \left | \frac{{\rm d} S_{EE}(t)}{{\rm d}t}\right|{{\rm d}t} \leq \int_{0}^{T}\frac{2}{\hbar} \sqrt{C_{E}(t)}\hspace{0.1cm}\Delta H \hspace{0.1cm}(1-R(t))\hspace{0.1cm}{\rm d}t. \label{equ:integral_over_t}\\
\end{equation}
For the time independent Hamiltonian, we obtain the following bound for the stronger quantum speed limit for entanglement
\begin{align}
T \ge T_{\rm SQSLO}^{E} := \frac{\hbar\left|S_{EE}(T)-S_{EE}(0)\right|}{2\Delta H\frac{1}{T} \int_{0}^{T} \sqrt{C_{E}(t)}\hspace{0.1cm}(1-R(t)) {\rm d}t} .
\label{speeed_limit_entanglement}
\end{align}
It is thus clear that evolution speed for entanglement generation (or degradation) is a function of capacity of entanglement $C_{E}$ and a correction factor due to stronger uncertainty relation. Thus, we can say that $C_{E}$ with 
$(1-R(t))$ together controls how much time a system may take to produce a certain amount of entanglement. Extending the analysis of the bound, we have calculated $R(t)$ by using the above-prescribed expression. We note that under certain choice of $|\psi^{\perp}\rangle$ this corresponds to the system evolving along the geodesic path~\cite{dimpi-22a}. The corresponding choice is as
\begin{equation}
    |\psi^{\perp}(t)\rangle = \frac{O(t)-\langle O(t)\rangle}{\Delta O(t)}|\psi(t)\rangle ,
   \label{eqn:perp_prescription}
\end{equation}
and with this, the speed limit bound is the most optimized.

\begin{figure}[h!]
    \centering
    \includegraphics[width=9cm]{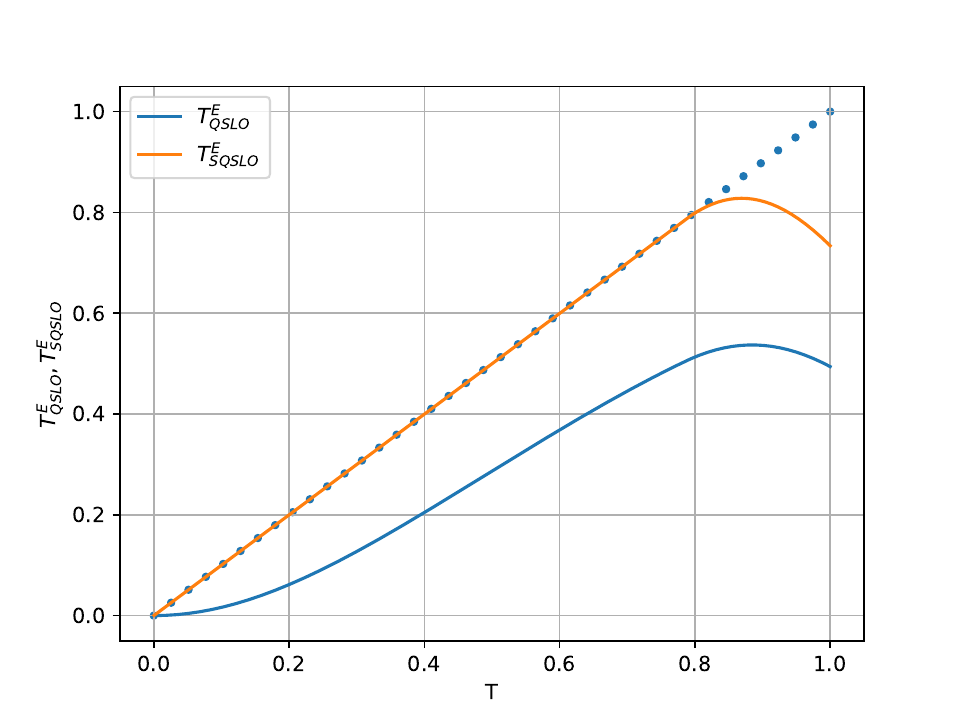}
    \caption{Here we depict $T_{QSLO}^{E}$ (upper line) ,$T_{SQSLO}^{E}$ (lower curve) vs T with p = 0.1 for $\theta$ = 1.0}
    \label{fig:speedlim+strongerQSL}
\end{figure}

To examine the tightness of the given QSL bound for generation of entanglement by taking an example of the state as given in Eqn.~\eqref{equ:time evolved state} for which we have estimated both capacity of entanglement $C_{E}$ and entanglement entropy $S_{EE}$ in Eqn.~\eqref{eqn:CE_S_expr_fin}. Further with  $\Delta H = \hspace{0.05cm}\left|\theta\hspace{0.1cm}(1-2p)\right|$ and evaluating $R(t)$ as defined in Eqn.~\eqref{eqn:R_t} making use of $\ket{\Psi^{\perp}(t)}$ through Eqn.~\eqref{eqn:perp_prescription} we plot for $T^{E}_{SQSLO}$ and $T^{E}_{QSLO}$ vs $T\in[0,1]$ in Fig.~\ref{fig:speedlim+strongerQSL} where $T^{E}_{QSLO}$ is given as ~\cite{capacity-22}
\begin{align}
T \ge T_{\rm QSLO}^{E} := \frac{\hbar\left|S_{EE}(T)-S_{EE}(0)\right|}{2\Delta H\frac{1}{T} \int_{0}^{T} \sqrt{C_{E}(t)} {\rm d}t} .
\label{speeed_limit_entanglement_old}
\end{align}

\begin{figure}[ht!]
    \centering
    \includegraphics[width=9cm]{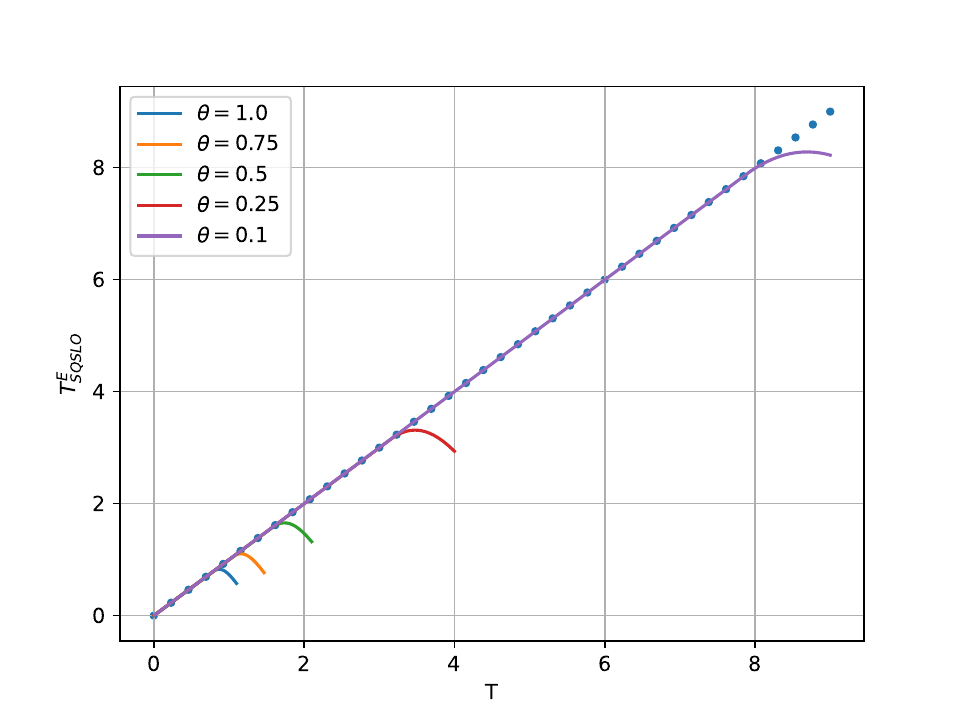}
    \caption{Speed limit plot for entanglement generation for varying $\theta$ keeping $p=0.1$ fixed where tight bound for longest time could be achieved for lowest $\theta = 0.1$ and shortest for largest $\theta=1.0$.}
    \label{fig:speed_lim_pfix}
\end{figure}

We clearly see that the operator stronger quantum speed limit $T^{E}_{SQSLO}$ which is with the correction factor gives a significantly tighter bound than the other $T^{E}_{QSLO}$. With the choice of state and Hamiltonian with $p=0.1$ and $\theta=1.0$, we indeed show that the bound is tighter and achievable. Further, in Fig.~\ref{fig:speed_lim_pfix}, we plot $T^E_{SQSLO}$ vs $T$ for a fixed $p$ but several $\theta$ values, which shows that we get better bounds for lower values of $\theta$ and each of this case, $T^{E}_{SQSLO}$ gives a tighter bound.

In the subsequent section, we demonstrate that one obtains better bounds through SQSLO compared to QSLO by illustrating through two important examples.

\section{Illustrations and Examples}\label{s4}

\subsection{Improved Bounds on Modular energy using SQSLO}
In the previous section, we have explored the study of entanglement generation using SQSL for state. Here, we would like to 
investigate how the modular Hamiltonian itself changes under unitary transformation in the Heisenberg picture. Consider a two-qubit system with a similar generalised canonical forms of non-local Hamiltonian $H^{\pm}$ as in Eqn.~\eqref{eqn:Non_loc_hamiltonian}. Further our state in this case retains the form $\ket{\Psi}_{AB}=\ket{\Psi(0)}_{AB}$ as in Eqn.~\eqref{equ:time zero state}. As such the state operator of the composite system also retains its form as
\begin{equation}
    \rho_{AB}=\rho_{AB}(0)=\ket{\Psi(0)}_{AB}\bra{\Psi(0)} .
\end{equation}
We again choose $H^{+}$ form of canonical Hamiltonian to time evolve the following composite modular Hamiltonian
\begin{equation}
    K_{AB}\equiv K_{AB}(0)=K_{A}\otimes\mathcal{I}_{B},
\end{equation} 
\noindent where $K_{A}\equiv K_{A}(0)=-\log{\rho_{A}}$ is the modular Hamiltonian and $\rho_{A}(0)=\tr_{B}(\rho_{AB}(0))$. In the Heisenberg picture, the operator of the composite $AB$ system evolves with unitary operator $U(t)=e^{-i H^{+} t}$
\begin{equation}
    K_{AB}(t)=U^{\dagger}(t)K_{AB}(0)U(t) .
\end{equation}

We interpret the quantity $\langle K_{AB}(t)\rangle = \tr(\rho_{AB}\hspace{0.1cm}K_{AB}(t))$ as the modular energy~$\mathcal{E_{M}}$ in the Heisenberg picture. Note that even though $\langle K_{AB}(0) \rangle = \tr(\rho_{A}(0)K_{A}(0))$ represents entanglement at $t=0$, $K_{AB}(t)$ does not represent entanglement at time $t$ in the Heisenberg picture.
 
Now, the variance of composite modular Hamiltonian~$\mathcal{C_{M}}$ can be written as
\begin{equation}
    \mathcal{C_{M}}(t)= \Delta K_{AB}(t)^{2} =\langle K_{AB}^{2}(t)\rangle - \langle K_{AB}(t)\rangle^{2} .
    \label{eqn:C_MAB}
\end{equation}
% The expression of entanglement entropy and capacity of entanglement can be re-expressed as,
% \begin{equation}
%     S(t) = \tr(\rho.K_{AB}(t))=\langle K_{AB}(t)\rangle
% \end{equation}
% \begin{equation}
%     C_{E}(t)=\langle K_{AB}^{2}(t)\rangle - \langle K_{AB}(t)\rangle^{2}
%     \label{eqn:CE_H}
% \end{equation}
The generalised expression for the $\mathcal{E_{M}}$ and $\mathcal{C_{M}}$ can be evaluated and expressed as
\begin{align}
    \mathcal{C_{M}}(t) = \frac{1}{4}&\log(-1+\frac{1}{p})^{2}\hspace{0.1cm}\left(4p(1-p)\hspace{0.1cm}\cos(2\theta)t)^{2} + \sin(2\theta t)^{2}\right)\nonumber \\
    \mathcal{E_{M}}(t) = & -(1-2p)\hspace{0.1cm}\arctanh(1-2p)\hspace{0.1cm}\cos(2\theta t)\nonumber \\
    &-\frac{1}{2}\log(p(1-p))
    \label{eqn:CE_S_expr_H} .
\end{align}
 
We begin by making use of stronger-uncertainty relation for the case of in general non-commuting Hamiltonians $K_{AB}(t)$ and $H_{AB}$ and derive SQSLO bound for modular energy. Denoting $\ket{\Psi}_{AB}=\ket{\Psi}$ for brevity, we have
\begin{equation}
    \frac{1}{2}|\langle\Psi|[K_{AB}(t), H_{AB}]|\Psi\rangle|\leq \Delta K_{AB} \Delta H_{AB} (1-R(t))
    \label{equ:capH_bound_expr1}
\end{equation}
which leads to
\begin{equation}
    \frac{\hbar}{2}\left|\frac{{\rm d} \langle K_{AB}(t)\rangle}{{\rm d}t}\right| = \frac{\hbar}{2}\left|\frac{{\rm d} \mathcal{E}_{M}}{{\rm d}t}\right| \leq \Delta K_{AB}(t)\Delta H_{AB} (1-R(t))\,.
\end{equation}
As $\Delta K_{AB}(t) = \sqrt{\mathcal{C}_{M}(t)}$, from Eqn.~\eqref{eqn:C_MAB}, we get

\begin{equation}
    \left|\frac{{\rm d} \mathcal{E}_{M}}{{\rm d}t}\right| \leq \frac{2}{\hbar}\sqrt{\mathcal{C}_{M}(t)}\Delta H_{AB} (1-R(t)) .
\end{equation}

This is a bound on  the rate of the modular energy in the Heisenberg picture. This is clearly distinct from the 
earlier case in the Schrödinger picture (as these two quantities are different).

\begin{figure}[h!]
    \centering
    \includegraphics[width=9cm]{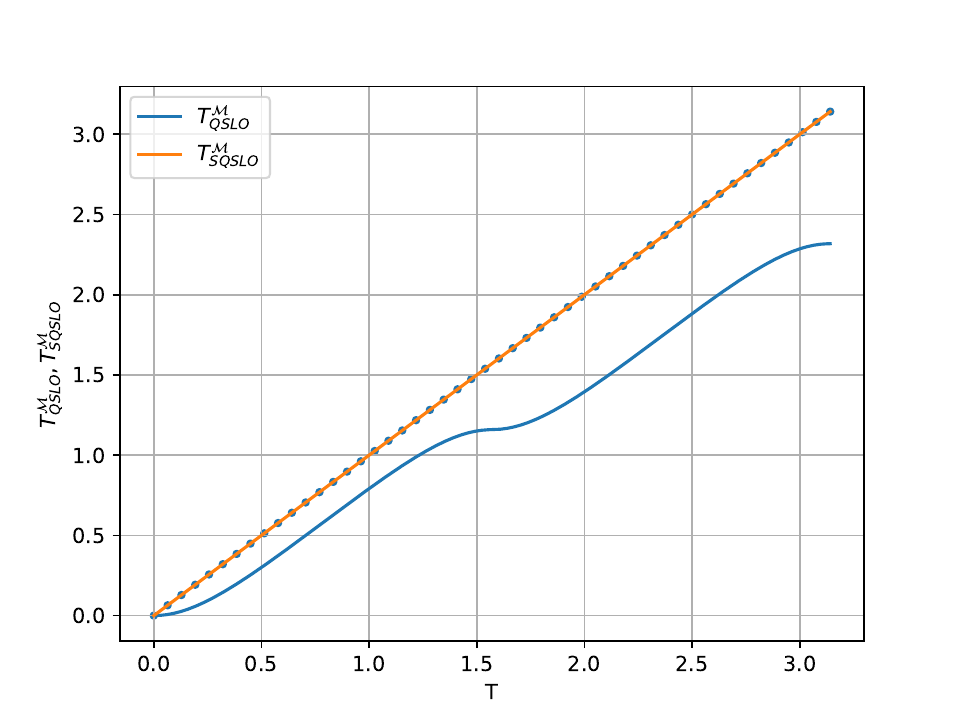}
    \caption{Here we depict $T_{QSLO}^{\mathcal{M}}$ (lower curve), $T_{SQSLO}^{\mathcal{M}}$ (upper saturated line) vs T with $p = 0.1$ for $\theta$ = 1.0}
    \label{fig:SQSLO_cap01}
\end{figure}

For the purpose of evaluating $R(t)$, we will need the optimized $\ket{\Psi^{\perp}}$ as prescribed in Eqn.~\eqref{eqn:perp_prescription}. It is important to mention that though the state vector of the system $\ket{\Psi}$ remains time independent in the considered picture, yet $\ket{\Psi^{\perp}}$ carries an explicit time dependence due to the prescription used involving operators that are time evolving themselves. So at each instant of time of the operator evaluation, it picks up a different $\ket{\Psi^{\perp}}$ while only maintaining that it be perpendicular to the taken choice of $\ket{\Psi}$. It goes without saying that such $\ket{\Psi^{\perp}}$ is not physically relevant to the system, as it plays no role in the description of it at any point in time. The general expression for this choice evaluates out as given in Appendix~\ref{ec}~\eqref{eqn:psi_perp_H_cap}. Now this leads us to the following SQSLO bound as given by
\begin{equation}
    T\geq T_{SQSLO}^{\mathcal{M}}:=\frac{\hbar}{2\hspace{0.1cm}\Delta H_{AB}}\int_{0}^{T}\frac{\left|{\rm d}\mathcal{E}_{M}\right|}{\sqrt{\mathcal{C}_{M}(t)}\hspace{0.1cm}\left(1-R(t)\right)} .
    \label{eqn:SQSLO_capH}
\end{equation}
We get the equivalent $T^{\mathcal{M}}_{QSLO}$ case from the Robertson-Schrödinger uncertainty relation which is akin to dropping the correction factor from Eqn.~\eqref{eqn:SQSLO_capH}
\begin{equation}
    T\geq T_{QSLO}^{\mathcal{M}}:=\frac{\hbar}{2\hspace{0.1cm}\Delta H_{AB}}\int_{0}^{T}\frac{\left|{\rm d}\mathcal{E}_{M}\right|}{\sqrt{\mathcal{C}_{M}(t)}\hspace{0.1cm}}.
    \label{eqn:QSLO_capH}
\end{equation}
Now, with $\Delta H_{AB}=\left|(1-2p)\hspace{0.1cm}\theta\right|$, we plot SQSLO and QSLO bounds in Fig.~\ref{fig:SQSLO_cap01} for the case of $p = 0.1$ and $\theta = 1.0$ for which $R(t)$ is given in Eqn.~\eqref{eqn:R01_H}.  
\noindent We observe that in the case of Heisenberg picture, the QSLO bound, i.e., $T_{QSLO}^{\mathcal{M}}$ turns out to be a bit loose whereas the SQSLO bound, i.e., $T_{SQSLO}^{\mathcal{M}}$, which is with the correction factor turns out to be saturated.

\begin{figure}[h!]
    \centering
    \includegraphics[width=9cm]{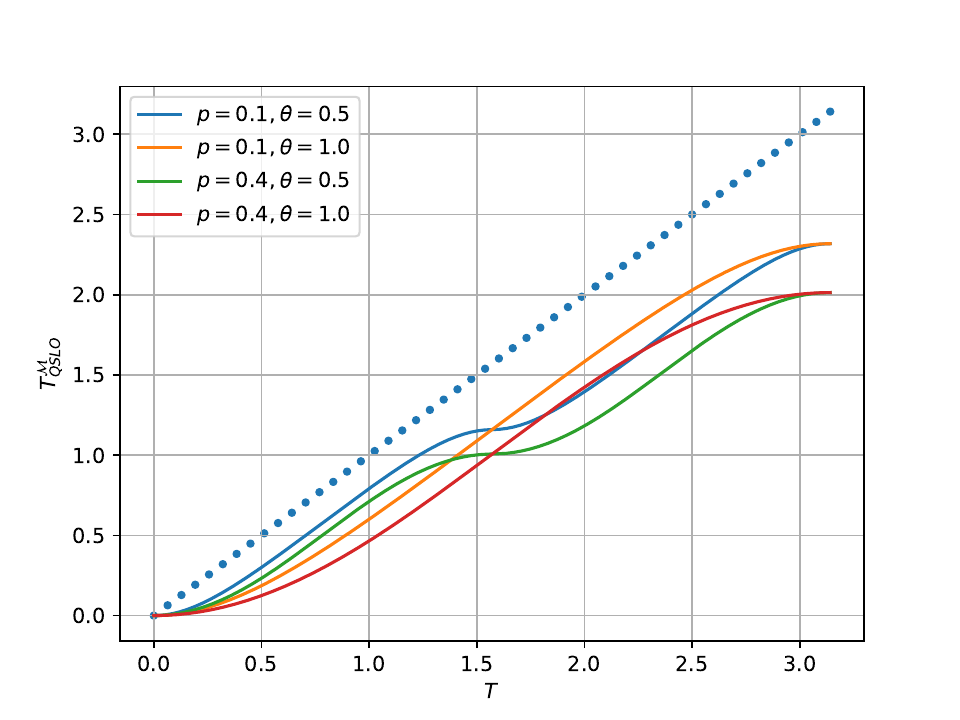}
    \caption{Depiction of $T_{QSLO}^{\mathcal{M}}$ for chosen values of $p$ and $\theta$. The dotted line is the reference ideal (saturated) case.}
    \label{fig:QSLO_ps_ths}
\end{figure}

We further plot QSLO bound for four cases of $p=\{0.1, 0.4\}$ and $\theta=\{0.5, 1.0\}$ cases using expressions for respective $R(t)$ from Eqn(s).\eqref{eqn:R01_H} to \eqref{eqn:R045_H} which is shown in Fig. \ref{fig:QSLO_ps_ths}. We observe and can conclude from the figure upon plotting with several example cases, that the $T^{\mathcal{M}}_{QSLO}$ bound is not optimal.
 
\begin{figure}[h!]
    \centering
    \includegraphics[width=9cm]{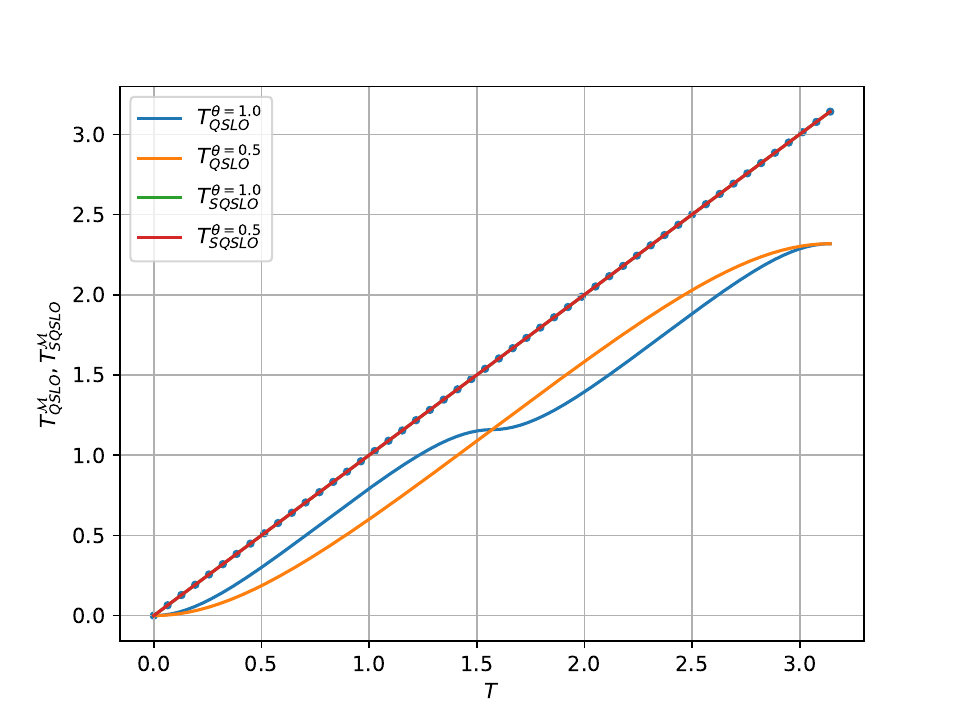}
    \caption{Depiction of $T_{QSLO}^{\mathcal{M}}$ (curves) and $T_{SQSLO}^{\mathcal{M}}$ (overlap with reference saturation case) for fixed $p = 0.1$ and varying $\theta$. We observe that $T_{QSLO}^{\mathcal{M}}$ for $\theta=0.5$ takes lead over $T_{QSLO}^{\mathcal{M}}$ for $\theta=1.0$ after $T=\pi/2$.}
    \label{fig:SQSLs_ths}
\end{figure}

Upon plotting for $T_{QSLO}^{\mathcal{M}}$ and $T_{SQSLO}^{\mathcal{M}}$ bounds for varying $\theta = \{0.5, 1.0\}$ and fixed $p = 0.1$ cases in Fig.\ref{fig:SQSLs_ths}. We observe that the SQSLO bound turns out to be saturated for any choice in parameters. This shows that the SQSLO for the rate of modular Hamiltonian is tight and saturated.

\subsection{Improved bounds on charging time of Quantum Batteries through SQSLO}

The models of quantum engines and refrigerators have been of great interest lately as they help in simulating theoretical efforts to formulate fundamental thermodynamical principles and bounds which are valid on micro or nano-scale. It has been found that these can differ from the standard ones and converge only in the limit of macroscopic systems~\cite{Alicki-79}. The amount of work that can be extracted from a small quantum mechanical system that is used to temporarily store energy and to transfer it from a production to a consumption center is the main content of a quantum battery. It is not coupled to external thermal baths in order to drive thermodynamical engines, but rather its dynamics is controlled by external time-dependent fields. 

The battery comes with its initial state $\rho$ and an internal Hamiltonian $H_{B}$. The process of energy extraction then follows when this system is reversibly evolved under some fields that are turned on during time interval $[0,T]$. The maximal amount of work that can be extracted by such a process has been explored in Ref.~\cite{Alicki-13}
Subsequently, numerous researchers have dedicated their efforts to furthering the understanding and exploitation of  non-classical features of quantum batteries such as in Ref.~\cite{Iran-20}. In many-body quantum systems characterized by multiple degrees of freedom, the presence of quantum batteries, capable of storing or releasing energy, is ubiquitous. In this section, we aim to determine the minimum achievable unitary charging time of the quantum battery utilizing the discussed SQSLO bound.\\

Consider a scenario where a quantum battery, with energy denoted by the Hamiltonian $H_B$, interacts with an external charging field represented by $H_C$. Consequently, the total energy of the system is determined by the combined Hamiltonian, expressed as follows:
\begin{equation}
    H_T = H_B + H_C.
\end{equation}

Now, the ergotropy is defined as the quantum system's capacity to extract energy via unitary operations from the quantum battery \cite{campaioli-18}, and is expressed as
\begin{equation}
    \mathcal{E} (t) = \langle\Psi(t)|H_B |\Psi(t)\rangle - \langle\Psi(0)|H_B |\Psi(0)\rangle,
\end{equation}
\noindent where $|\Psi(t)\rangle$ and  $|\Psi(0)\rangle$ are the final and initial state of the given quantum system.

While the aforementioned expression holds true in the Schrödinger picture, we would now like to switch over to its study in the Heisenberg picture where the expression for ergotropy takes the form
%study the problem of generation of ergotropy in Heisenberg picture. So under this new picture, the ergotropy can be defined as: it is imperative to note that our current analysis pertains to the Heisenberg picture. 
\begin{equation}
    \mathcal{E} (t) = \langle\Psi(0)|(H_B (t) - H_B (0))|\Psi(0)\rangle,
    \label{eqn:ergotropy}
\end{equation}

\noindent where $H_B (t)$ = $e^{\frac{iHt}{\hbar}} H_B (0) e^{-\frac{iHt}{\hbar}} $ and $H_B (0)$ = $H_B$.

The rate of change of ergotropy of quantum battery during the charging process  can be obtained as
\begin{equation}
    \frac{d \mathcal{E}(t)}{dt} = \frac{d}{dt} \langle\Psi(0)|H_B (t) |\Psi(0)\rangle.
\end{equation}

Using our bound we can write SQSLO for the ergotropy as
\begin{equation}
    T \ge \frac{\hbar}{2\Delta H_T} \int_0 ^T \frac{|d \mathcal{E} (t)|}{\Delta H_B(t) (1 - R(t))},
\end{equation}

\noindent where $R(t)$ = $\frac{1}{2}\left|\langle\Psi^{\perp}|\frac{H_B (t)}{\Delta H_B(t)} \mp i\frac{H_T}{\Delta H_T}|\Psi\rangle\right|^2$ and $T$ is the charging time period of the quantum battery.

This SQSLO for QBs can also be re-expressed as: 
\begin{equation}
    T_{SQSLO} ^{QB} = \frac{\hbar T}{2\Delta H_T} \left<\left<\frac{\left| \mathcal{E} (T) - \mathcal{E} (0) \right| }{ \Delta H_B(t) (1 -R(t))}\right>\right>_{T},
\end{equation}

\noindent where $\langle\langle A(t)\rangle\rangle_T$ = $\frac{1}{T}\int_0 ^T {\rm d}t\hspace{0.1cm}A(t)$, is the time average of the quantity $A(t)$.\\

Now that we have derived the Quantum speed limit (QSL) formula for Quantum Batteries (QB's) in a general case, let us take up a specific example where we apply our SQSLO bound on QB's. Our chosen example involves an entanglement-based QB consisting of two qubit cells and two coupled two-level systems. To charge the QB effectively, we must individually couple each cell with local fields. Consequently, our total Hamiltonian $H_T$ can be expressed as given in Ref.~\cite{Iran-20}
\begin{equation}
    H_T = H_B + H_C + H_{int},
\end{equation}
where \(H_B = \hbar\omega_0 \sum_{n=1} ^2 \sigma_n ^z \) being the battery Hamiltonian. Here, $\omega_0$ is the identical Larmor frequency for both the qubits. Let us label $\ket{\uparrow}$ and $\ket{\downarrow}$ as ground and excited states for a single qubit. With this one can define the fully charged state of the battery as $\ket{{\rm full}}=\ket{\uparrow\uparrow}$ with full energy $E_{{\rm full}}=2\hbar\omega$, and empty one as $\ket{{\rm emp}}=\ket{\downarrow\downarrow}$ with low energy $E_{{\rm emp}}=-2\hbar\omega$. Hence, the maximum energy that can be stored in the battery reads $\mathcal{E}_{{\rm max}}=4\hbar\omega$.

We consider the driving Hamiltonian to comprise of two parts, having charging part \(H_C = \hbar\Omega\sum_{n=1} ^2 \sigma_n ^x\), where $\Omega$ is some constant and nearest neighbour interaction part \(H_{int} = \hbar J(\sigma_1^x \sigma_2^x +\sigma_1^y \sigma_2^y +\sigma_1^z \sigma_2^z)\), where $J$ is the strength of two body interaction. The most general state of two qubits then reads as
\begin{equation}
\ket{\Psi(0)}=\mu\ket{\uparrow\uparrow}+\nu\ket{\uparrow\downarrow}+\eta\ket{\downarrow\uparrow}+\delta\ket{\downarrow\downarrow}.
\end{equation}
Let us consider the case for the most general two qubit non-entangled state with,
\begin{eqnarray}
    &\mu =  \sin(\theta_{1})\sin(\theta_{2})e^{i(\varphi_{1}+\varphi_{2})}\nonumber\\
    &\nu =  \sin(\theta_{1})\cos(\theta_{2})e^{i\varphi_{1}}\nonumber\\
    &\eta =  \cos(\theta_{1})\sin(\theta_{2})e^{i\varphi_{2}}\nonumber\\
    &\delta =  \cos(\theta_{1})\cos(\theta_{2})
    \label{eqn:psi0}
\end{eqnarray}
where $\theta_{1},\theta_{2}\in[0,\pi]$ and $\varphi_{1},\varphi_{2}\in[0,2\pi]$. For the purpose of illustration, let us assume that at the beginning of the charging process, the battery is assumed to be empty, i.e., $\rho(0)=\ket{{\rm emp}}\bra{{\rm emp}}$, which is achieved when we put $\theta_{1}=\theta_{2}=0$ in Eqn.~\eqref{eqn:psi0}.

The ergotropy Eqn.~\eqref{eqn:ergotropy} in this case under Heisenberg picture upon evaluation reads as
\begin{equation}
    \mathcal{E}(t) = \frac{4\omega\Omega^2}{\omega^2 + \Omega^2}\sin\left(\sqrt{\omega^2 +\Omega^2}\hspace{0.1cm}t\right)^2 .
\end{equation}

We would like to study the bounds on the charging time for mentioned scenario above. First we look at the QSLO bound studied in previous section, with a similar form to Eqn.~\eqref{eqn:QSLO_capH} for Quantum Batteries as 
\begin{equation}
    T_{QSLO} = \frac{\hbar}{2\Delta H_T} \int_0 ^T \frac{|d\mathcal{E} (t)|}{\Delta H_B (t)} ,
    \label{eqn:TOQSL} 
\end{equation}
where $\Delta H_{T}$ and $\Delta H_{B}(t)$ can be evaluated for chosen values of parameters $\omega$, $\Omega$ and $J$ in above bound.

\begin{figure}[h!]
    \centering
    \includegraphics[width=9cm]{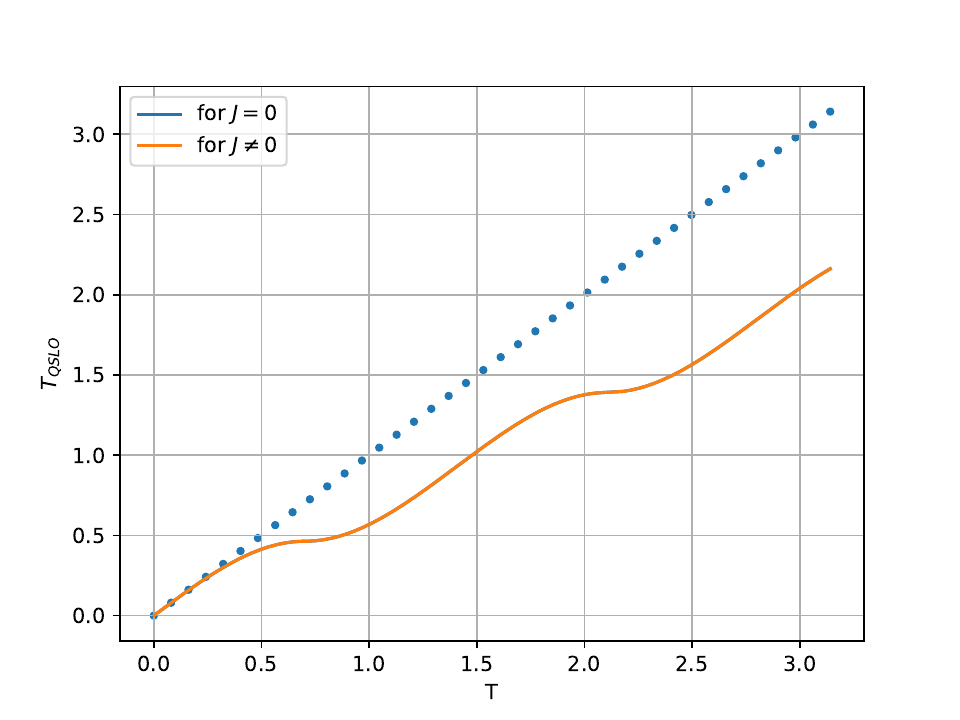}
    \caption{Figure shows that the QSLO bound plots (curves) for both parallel ($J=0$) and collective ($J\neq 0$) Quantum Battery scenarios overlap. The dotted curve is the reference ideal case.}
    \label{coqsl}
\end{figure}

With our general Hamiltonian for QB defined as above, let us take a case of \textit{parallel charging} when $J = 0$, rendering the interaction Hamiltonian inactive. Similarly, we can determine the QSLO for the case of \textit{collective charging} when $J \ne 0$ using Eqn.~\eqref{eqn:TOQSL}. Upon plotting these two-speed limit functions, as depicted in Fig. \ref{coqsl}, we observe that for the above two cases, the QSLO bound overlaps. Over that, there is a clear deviation from the reference ideal case, i.e., $T_{QSLO}=T$. It thus leaves the ground for improvement. This observation holds true for both scenarios of the QB Hamiltonian, namely the parallel and collective charging cases. Hence, we would like to compute the bounds by applying the SQSLO bound to both the cases.

Next, we study the scenarios involving \textit{coupled} and \textit{decoupled} cases. When $J=\Omega$, we say the system is coupled, whereas for $J \ne \Omega$, it represents the decoupled scenario. To compute the QSLO for both the coupled and decoupled Hamiltonians, we follow a similar procedure as we did for the parallel and collective QB cases. A novel aspect of our approach is the application of the SQSLO in both the coupled and decoupled Hamiltonians. The expression for the SQSLO bound is given as
\begin{equation}
    T_{SQSLO} = \frac{\hbar}{2\Delta H_T} \int_0 ^T \frac{\left|d\mathcal{E}(t)\right|}{\Delta H_B (t) (1-R(t))} ,
    \label{eqn:TSQSLO}
\end{equation}
where for the coupled case (with $J=\Omega=1$ and $\omega=2$) we obtain expression of $R(t)$ as
\begin{equation}
    R(t) = \frac{1}{2} \left( 
\begin{cases} 
2 - \frac{2\sqrt{10} \cos( \sqrt{5} t)}{\sqrt{9 + \cos(2 \sqrt{5} t)}} & \sin(\sqrt{5} t) < 0 \\ 
2 + \frac{2\sqrt{10} \cos(\sqrt{5} t)}{\sqrt{9 + \cos(2\sqrt{5} t)}} & \text{True}  ,
\end{cases}
\right) ,
\label{eqn:R_coup}
\end{equation}

\noindent where $|\Psi\rangle$ and $|\Psi^{\perp}\rangle$ are given in Appendix-\ref{qba}. For the purpose of evaluating $R(t)$, we need the optimized $\ket{\Psi^{\perp}}$ as prescribed in Eqn.~\eqref{eqn:perp_prescription}. 
% As discussed earlier, it is important to note that though the state vector of the system $\ket{\Psi}$ remains time independent in the considered picture, yet $\ket{\Psi^{\perp}}$ carries an explicit time dependence due to the prescription used involving operators that are time evolving themselves. So at each time stamp of the operator evaluation, it picks up a different $\ket{\Psi^{\perp}}$ while only maintaining that it be perpendicular to the given choice of $\ket{\Psi}$.
%It goes without saying that such $\ket{\Psi^{\perp}}$ is not physically relevant to the system, as it plays no role in the description of it at any point in time.

Again, for the decoupled case, i.e., $J \ne \Omega$ (taking $J=1,\hspace{0.1cm}\Omega=4,\hspace{0.1cm}\omega=2$) we evaluate the bound using the expression for bound in Eqn.~\eqref{eqn:TSQSLO}. The expression of $R(t)$ upon evaluation comes as

\begin{equation}
R(t) = \frac{1}{2} \left( 
\begin{cases} 
2 - \frac{4 \cos(2 \sqrt{2} t)}{\sqrt{3 + \cos(4 \sqrt{2} t)}} & \sin(2 \sqrt{2} t) < 0 \\ 
2 + \frac{4 \cos(2 \sqrt{2} t)}{\sqrt{3 + \cos(4 \sqrt{2} t)}} & \text{True}  ,
\end{cases}
\right) ,
\label{eqn:R_nocoup}
\end{equation} 

\noindent where $|\Psi\rangle$ and $|\Psi^{\perp}\rangle$ are given in Appendix-\ref{qba}.

\begin{figure}[h!]
    \centering
    \includegraphics[width=9cm]{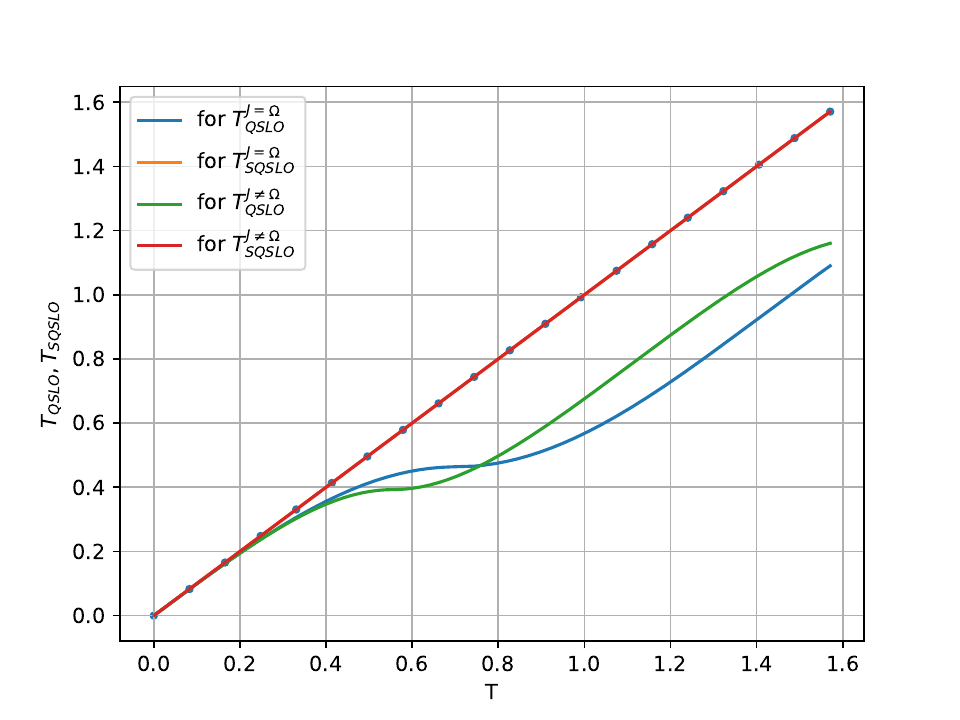}
    \caption{Figure depicts plots for both QSLO and SQSLO bounds for both coupling ($J=\Omega$) and decoupling ($J\neq \Omega$) Quantum Battery scenarios. The QSLO bounds overlap with reference ideal (saturated) case, whereas for QSLO curves, decoupling case takes the lead close to after $T=0.8$.}
    \label{cosqsl} 
\end{figure}

We have depicted both SQSLO curves for both the coupling and decoupling cases. Surprisingly, in both scenarios, the SQSLO plots exhibit remarkable accuracy as they overlap while showing saturation, as can be seen in Fig.~\ref{cosqsl}. As expected, QSLO does not yield optimally tight bounds for both coupling and decoupling cases as shown in the same figure. From these plots, we can conclude that SQSLO performs remarkably well, accurately representing the speed limit behavior in QB systems for both coupling and decoupling cases. This result is quite a significant improvement upon earlier bounds and is the optimal bound. Next, we will apply SQSLO in parallel and collective QB cases to further explore its behavior in those scenarios.

Having computed the QSLO for both parallel ($J=0$) and collective charging ($J\neq 0$) QB cases, we will now apply the SQSLO bound in both these cases. This involves evaluation of $R(t)$ for both scenarios. We have already given the expression for collective charging ($J\neq 0$) case earlier. For parallel charging ($J = 0$) case with $\Omega=1,\hspace{0.1cm}\omega=2$; this reads as
\begin{equation}
    R(t) = \frac{1}{2}\left(2+\frac{2\sqrt{10}\hspace{0.1cm}\left|\sin(\sqrt{5}t)\right|\cot(\sqrt{5}t)}{\sqrt{9+\cos(2\sqrt{5}t)}}\right) ,
    \label{R_J0} 
\end{equation}
% Upon calculating $R(t)$ for both scenarios, we have already given the expression of $R(t)$ for $J=0$ case which is $J=\Omega$. Hence  arrive at the expression for $R(t)$ in $J=0$ case, as shown below:
where the involved $|\Psi\rangle$ and $|\Psi^{\perp}\rangle$ in this case are given in Appendix-\ref{qba}.\\

% {\color{green}(SHOULD We give a plot for behavior of R's thus obtained in three cases !!!)}

From the previous expressions, we reiterate that it appears that $|\Psi^\perp\rangle$ exhibits time dependence. However, according to the Heisenberg picture, the state should not evolve, indicating that we should not observe time dependence in the orthogonal state. Fundamentally, we acknowledge the existence of multiple choices for the orthogonal state of a given state. Therefore, we have adopted the most widely accepted method to select the orthogonal state to optimize our parameter $R$. In Eqn.~\eqref{eqn:perp_prescription} , we notice that the observable $O$ is involved in the formula, and we employ its associated battery Hamiltonian $H_B(t)$, which evolves in the Heisenberg picture. Consequently, the time dependence on $|\Psi^\perp\rangle$ state arises. Through this selection, we achieve optimal value for the expression $R$.

%(\ref{os})

\begin{figure}[h!]
    \centering
    \includegraphics[width=9cm]{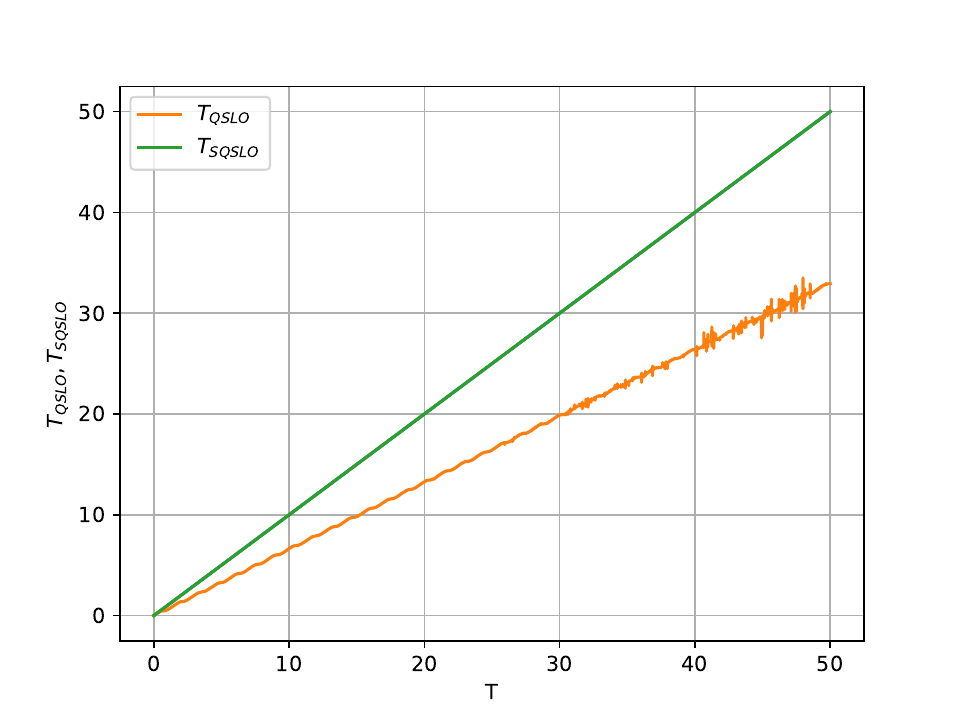}
    \caption{Figure shows the long time behaviour of  $T_{QSLO}$ (lower unsaturated curve), $T_{SQSLO}$ (tight and saturated overlap with reference ideal line).}
    \label{lot}
\end{figure}

Now, it is interesting to observe the behavior of SQSLO over a longer duration. We have plotted SQSLO for an extended period of time, and we observe optimal results, as illustrated in Fig.~\ref{lot}. Thus, one can affirm that for QB scenarios, SQSLO stands as the optimal choice—it represents the best bound for calculating the charging time of quantum batteries.

We have analysed various quantum battery scenarios, including parallel, collective, coupling and decoupling cases, and presented our findings for QSLO and SQSLO. It is evident that SQSLO consistently reveals the tightest bound for quantum speed limit. Consequently, we can assert that SQSLO outperforms in predicting the charging time.
% After having plotted all the curves for various QB scenarios, including parallel, collective, coupling, and decoupling cases, we have represented both QSLO and SQSLO in the plots. Analysing all these plots, it becomes evident that SQSLO consistently delivers the best of the results across all scenarios, closely aligning with the exact quantum speed limit. This alignment is clearly depicted in the shown plots. Consequently, we can thus confidently assert that the SQSLO is the optimal choice for charging Quantum Batteries.

%\vspace{5.0cm}
\section{Conclusion}\label{s5}
In conclusion, we have addressed the fundamental question of how fast an observable can evolve in time by invoking the concept of the observable speed limit. Our study presents a stronger version of this limit, demonstrating that previously derived bounds are special cases of our new bound. We have also shown that SQSLO can lead to stronger speed limit for states. By applying SQSLO, we have investigated its efficacy in evaluating the capacity of entanglement, akin to the heat capacity in quantum information theory. Notably, we have established a more robust bound for the entanglement rate, surpassing previous limitations. Moreover, our exploration extends to the realm of interacting qubits within quantum batteries. By leveraging SQSLO, we have accurately predicted the time required to charge the battery, showcasing the tightness of SQSLO in this context—it effectively saturates when estimating the charging time of quantum batteries. These findings hold significant implications across various domains, including quantum thermodynamics, the complexity of operator growth, prediction of quantum correlation growth rates, and the broader landscape of quantum technology. Thus, SQSLO emerges as a powerful tool with diverse applications, paving the way for advancements in quantum science and technology.

\vskip .5cm
\noindent
{\it Acknowledgments:}
DS acknowledges the support of the INFOSYS scholarship. BP acknowledges IIIT, Hyderabad and TCG CREST (CQuERE) for their support and hospitality  during the academic visit.

\bibliography{main.bib}

\begin{thebibliography}{111}%
\makeatletter
\providecommand \@ifxundefined [1]{%
 \@ifx{#1\undefined}
}%
\providecommand \@ifnum [1]{%
 \ifnum #1\expandafter \@firstoftwo
 \else \expandafter \@secondoftwo
 \fi
}%
\providecommand \@ifx [1]{%
 \ifx #1\expandafter \@firstoftwo
 \else \expandafter \@secondoftwo
 \fi
}%
\providecommand \natexlab [1]{#1}%
\providecommand \enquote  [1]{``#1''}%
\providecommand \bibnamefont  [1]{#1}%
\providecommand \bibfnamefont [1]{#1}%
\providecommand \citenamefont [1]{#1}%
\providecommand \href@noop [0]{\@secondoftwo}%
\providecommand \href [0]{\begingroup \@sanitize@url \@href}%
\providecommand \@href[1]{\@@startlink{#1}\@@href}%
\providecommand \@@href[1]{\endgroup#1\@@endlink}%
\providecommand \@sanitize@url [0]{\catcode `\\12\catcode `\$12\catcode `\&12\catcode `\#12\catcode `\^12\catcode `\_12\catcode `\%12\relax}%
\providecommand \@@startlink[1]{}%
\providecommand \@@endlink[0]{}%
\providecommand \url  [0]{\begingroup\@sanitize@url \@url }%
\providecommand \@url [1]{\endgroup\@href {#1}{\urlprefix }}%
\providecommand \urlprefix  [0]{URL }%
\providecommand \Eprint [0]{\href }%
\providecommand \doibase [0]{http://dx.doi.org/}%
\providecommand \selectlanguage [0]{\@gobble}%
\providecommand \bibinfo  [0]{\@secondoftwo}%
\providecommand \bibfield  [0]{\@secondoftwo}%
\providecommand \translation [1]{[#1]}%
\providecommand \BibitemOpen [0]{}%
\providecommand \bibitemStop [0]{}%
\providecommand \bibitemNoStop [0]{.\EOS\space}%
\providecommand \EOS [0]{\spacefactor3000\relax}%
\providecommand \BibitemShut  [1]{\csname bibitem#1\endcsname}%
\let\auto@bib@innerbib\@empty
%</preamble>
\bibitem [{\citenamefont {Mandelstam}\ and\ \citenamefont {Tamm}(1945)}]{Mandelstam-45}%
  \BibitemOpen
  \bibfield  {author} {\bibinfo {author} {\bibfnamefont {Leonid}\ \bibnamefont {Mandelstam}}\ and\ \bibinfo {author} {\bibfnamefont {IG}~\bibnamefont {Tamm}},\ }\bibfield  {title} {\enquote {\bibinfo {title} {The uncertainty relation between energy and time in non-relativistic quantum mechanics},}\ }\href {https://doi.org/10.1007/978-3-642-74626-0_8} {\bibfield  {journal} {\bibinfo  {journal} {J. Phys. (USSR)}\ }\textbf {\bibinfo {volume} {9}},\ \bibinfo {pages} {249} (\bibinfo {year} {1945})}\BibitemShut {NoStop}%
\bibitem [{\citenamefont {Pati}(1991)}]{Pati_geometry_phases-91}%
  \BibitemOpen
  \bibfield  {author} {\bibinfo {author} {\bibfnamefont {Arun~Kumar}\ \bibnamefont {Pati}},\ }\bibfield  {title} {\enquote {\bibinfo {title} {Relation between “phases” and “distance” in quantum evolution},}\ }\href {\doibase https://doi.org/10.1016/0375-9601(91)90255-7} {\bibfield  {journal} {\bibinfo  {journal} {Physics Letters A}\ }\textbf {\bibinfo {volume} {159}},\ \bibinfo {pages} {105--112} (\bibinfo {year} {1991})}\BibitemShut {NoStop}%
\bibitem [{\citenamefont {Margolus}\ and\ \citenamefont {Levitin}(1998)}]{Margolus-98}%
  \BibitemOpen
  \bibfield  {author} {\bibinfo {author} {\bibfnamefont {Norman}\ \bibnamefont {Margolus}}\ and\ \bibinfo {author} {\bibfnamefont {Lev~B.}\ \bibnamefont {Levitin}},\ }\bibfield  {title} {\enquote {\bibinfo {title} {The maximum speed of dynamical evolution},}\ }\href {\doibase https://doi.org/10.1016/S0167-2789(98)00054-2} {\bibfield  {journal} {\bibinfo  {journal} {Physica D: Nonlinear Phenomena}\ }\textbf {\bibinfo {volume} {120}},\ \bibinfo {pages} {188--195} (\bibinfo {year} {1998})}\BibitemShut {NoStop}%
\bibitem [{\citenamefont {Bao}\ \emph {et~al.}(2018)\citenamefont {Bao}, \citenamefont {Kleer}, \citenamefont {Wang},\ and\ \citenamefont {Rahmani}}]{Bao-18}%
  \BibitemOpen
  \bibfield  {author} {\bibinfo {author} {\bibfnamefont {Seraph}\ \bibnamefont {Bao}}, \bibinfo {author} {\bibfnamefont {Silken}\ \bibnamefont {Kleer}}, \bibinfo {author} {\bibfnamefont {Ruoyu}\ \bibnamefont {Wang}}, \ and\ \bibinfo {author} {\bibfnamefont {Armin}\ \bibnamefont {Rahmani}},\ }\bibfield  {title} {\enquote {\bibinfo {title} {Optimal control of superconducting gmon qubits using pontryagin's minimum principle: Preparing a maximally entangled state with singular bang-bang protocols},}\ }\href {\doibase 10.1103/PhysRevA.97.062343} {\bibfield  {journal} {\bibinfo  {journal} {Phys. Rev. A}\ }\textbf {\bibinfo {volume} {97}},\ \bibinfo {pages} {062343} (\bibinfo {year} {2018})}\BibitemShut {NoStop}%
\bibitem [{\citenamefont {Rodríguez}\ \emph {et~al.}(2024)\citenamefont {Rodríguez}, \citenamefont {Ahmadi}, \citenamefont {Suárez}, \citenamefont {Mazurek}, \citenamefont {Barzanjeh},\ and\ \citenamefont {Horodecki}}]{Rodriguez-24}%
  \BibitemOpen
  \bibfield  {author} {\bibinfo {author} {\bibfnamefont {R~R}\ \bibnamefont {Rodríguez}}, \bibinfo {author} {\bibfnamefont {B}~\bibnamefont {Ahmadi}}, \bibinfo {author} {\bibfnamefont {G}~\bibnamefont {Suárez}}, \bibinfo {author} {\bibfnamefont {P}~\bibnamefont {Mazurek}}, \bibinfo {author} {\bibfnamefont {S}~\bibnamefont {Barzanjeh}}, \ and\ \bibinfo {author} {\bibfnamefont {P}~\bibnamefont {Horodecki}},\ }\bibfield  {title} {\enquote {\bibinfo {title} {Optimal quantum control of charging quantum batteries},}\ }\href {\doibase 10.1088/1367-2630/ad3843} {\bibfield  {journal} {\bibinfo  {journal} {New Journal of Physics}\ }\textbf {\bibinfo {volume} {26}},\ \bibinfo {pages} {043004} (\bibinfo {year} {2024})}\BibitemShut {NoStop}%
\bibitem [{\citenamefont {Evangelakos}\ \emph {et~al.}(2023)\citenamefont {Evangelakos}, \citenamefont {Paspalakis},\ and\ \citenamefont {Stefanatos}}]{Evangelakos-23}%
  \BibitemOpen
  \bibfield  {author} {\bibinfo {author} {\bibfnamefont {Vasileios}\ \bibnamefont {Evangelakos}}, \bibinfo {author} {\bibfnamefont {Emmanuel}\ \bibnamefont {Paspalakis}}, \ and\ \bibinfo {author} {\bibfnamefont {Dionisis}\ \bibnamefont {Stefanatos}},\ }\bibfield  {title} {\enquote {\bibinfo {title} {Minimum-time generation of a uniform superposition in a qubit with only transverse field control},}\ }\href {\doibase 10.1103/PhysRevA.108.062425} {\bibfield  {journal} {\bibinfo  {journal} {Phys. Rev. A}\ }\textbf {\bibinfo {volume} {108}},\ \bibinfo {pages} {062425} (\bibinfo {year} {2023})}\BibitemShut {NoStop}%
\bibitem [{\citenamefont {Mazzoncini}\ \emph {et~al.}(2023)\citenamefont {Mazzoncini}, \citenamefont {Cavina}, \citenamefont {Andolina}, \citenamefont {Erdman},\ and\ \citenamefont {Giovannetti}}]{Mazzoncini-23}%
  \BibitemOpen
  \bibfield  {author} {\bibinfo {author} {\bibfnamefont {Francesco}\ \bibnamefont {Mazzoncini}}, \bibinfo {author} {\bibfnamefont {Vasco}\ \bibnamefont {Cavina}}, \bibinfo {author} {\bibfnamefont {Gian~Marcello}\ \bibnamefont {Andolina}}, \bibinfo {author} {\bibfnamefont {Paolo~Andrea}\ \bibnamefont {Erdman}}, \ and\ \bibinfo {author} {\bibfnamefont {Vittorio}\ \bibnamefont {Giovannetti}},\ }\bibfield  {title} {\enquote {\bibinfo {title} {Optimal control methods for quantum batteries},}\ }\href {\doibase 10.1103/PhysRevA.107.032218} {\bibfield  {journal} {\bibinfo  {journal} {Phys. Rev. A}\ }\textbf {\bibinfo {volume} {107}},\ \bibinfo {pages} {032218} (\bibinfo {year} {2023})}\BibitemShut {NoStop}%
\bibitem [{\citenamefont {Anandan}\ and\ \citenamefont {Aharonov}(1990)}]{Aharonov-90}%
  \BibitemOpen
  \bibfield  {author} {\bibinfo {author} {\bibfnamefont {Jeeva}\ \bibnamefont {Anandan}}\ and\ \bibinfo {author} {\bibfnamefont {Yakir}\ \bibnamefont {Aharonov}},\ }\bibfield  {title} {\enquote {\bibinfo {title} {Geometry of quantum evolution},}\ }\href {\doibase 10.1103/PhysRevLett.65.1697} {\bibfield  {journal} {\bibinfo  {journal} {Phys. Rev. Lett.}\ }\textbf {\bibinfo {volume} {65}},\ \bibinfo {pages} {1697--1700} (\bibinfo {year} {1990})}\BibitemShut {NoStop}%
\bibitem [{\citenamefont {Levitin}\ and\ \citenamefont {Toffoli}(2009)}]{Levitin-09}%
  \BibitemOpen
  \bibfield  {author} {\bibinfo {author} {\bibfnamefont {Lev~B.}\ \bibnamefont {Levitin}}\ and\ \bibinfo {author} {\bibfnamefont {Tommaso}\ \bibnamefont {Toffoli}},\ }\bibfield  {title} {\enquote {\bibinfo {title} {Fundamental limit on the rate of quantum dynamics: The unified bound is tight},}\ }\href {\doibase 10.1103/PhysRevLett.103.160502} {\bibfield  {journal} {\bibinfo  {journal} {Phys. Rev. Lett.}\ }\textbf {\bibinfo {volume} {103}},\ \bibinfo {pages} {160502} (\bibinfo {year} {2009})}\BibitemShut {NoStop}%
\bibitem [{\citenamefont {Gislason}\ \emph {et~al.}(1985)\citenamefont {Gislason}, \citenamefont {Sabelli},\ and\ \citenamefont {Wood}}]{Gislason-85}%
  \BibitemOpen
  \bibfield  {author} {\bibinfo {author} {\bibfnamefont {Eric~A.}\ \bibnamefont {Gislason}}, \bibinfo {author} {\bibfnamefont {Nora~H.}\ \bibnamefont {Sabelli}}, \ and\ \bibinfo {author} {\bibfnamefont {John~W.}\ \bibnamefont {Wood}},\ }\bibfield  {title} {\enquote {\bibinfo {title} {New form of the time-energy uncertainty relation},}\ }\href {\doibase 10.1103/PhysRevA.31.2078} {\bibfield  {journal} {\bibinfo  {journal} {Phys. Rev. A}\ }\textbf {\bibinfo {volume} {31}},\ \bibinfo {pages} {2078--2081} (\bibinfo {year} {1985})}\BibitemShut {NoStop}%
\bibitem [{\citenamefont {Eberly}\ and\ \citenamefont {Singh}(1973)}]{Eberly-73}%
  \BibitemOpen
  \bibfield  {author} {\bibinfo {author} {\bibfnamefont {Joseph~H.}\ \bibnamefont {Eberly}}\ and\ \bibinfo {author} {\bibfnamefont {L.~P.~S.}\ \bibnamefont {Singh}},\ }\bibfield  {title} {\enquote {\bibinfo {title} {Time operators, partial stationarity, and the energy-time uncertainty relation},}\ }\href {\doibase 10.1103/PhysRevD.7.359} {\bibfield  {journal} {\bibinfo  {journal} {Phys. Rev. D}\ }\textbf {\bibinfo {volume} {7}},\ \bibinfo {pages} {359--362} (\bibinfo {year} {1973})}\BibitemShut {NoStop}%
\bibitem [{\citenamefont {Bauer}\ and\ \citenamefont {Mello}(1978)}]{Bauer-78}%
  \BibitemOpen
  \bibfield  {author} {\bibinfo {author} {\bibfnamefont {M}~\bibnamefont {Bauer}}\ and\ \bibinfo {author} {\bibfnamefont {P.A}\ \bibnamefont {Mello}},\ }\bibfield  {title} {\enquote {\bibinfo {title} {The time-energy uncertainty relation},}\ }\href {\doibase https://doi.org/10.1016/0003-4916(78)90223-3} {\bibfield  {journal} {\bibinfo  {journal} {Annals of Physics}\ }\textbf {\bibinfo {volume} {111}},\ \bibinfo {pages} {38--60} (\bibinfo {year} {1978})}\BibitemShut {NoStop}%
\bibitem [{\citenamefont {Bhattacharyya}(1983)}]{Bhattacharyya-83}%
  \BibitemOpen
  \bibfield  {author} {\bibinfo {author} {\bibfnamefont {K}~\bibnamefont {Bhattacharyya}},\ }\bibfield  {title} {\enquote {\bibinfo {title} {Quantum decay and the mandelstam-tamm-energy inequality},}\ }\href {\doibase 10.1088/0305-4470/16/13/021} {\bibfield  {journal} {\bibinfo  {journal} {Journal of Physics A: Mathematical and General}\ }\textbf {\bibinfo {volume} {16}},\ \bibinfo {pages} {2993} (\bibinfo {year} {1983})}\BibitemShut {NoStop}%
\bibitem [{\citenamefont {Leubner}\ and\ \citenamefont {Kiener}(1985)}]{Leubner-85}%
  \BibitemOpen
  \bibfield  {author} {\bibinfo {author} {\bibfnamefont {C.}~\bibnamefont {Leubner}}\ and\ \bibinfo {author} {\bibfnamefont {C.}~\bibnamefont {Kiener}},\ }\bibfield  {title} {\enquote {\bibinfo {title} {Improvement of the eberly-singh time-energy inequality by combination with the mandelstam-tamm approach},}\ }\href {\doibase 10.1103/PhysRevA.31.483} {\bibfield  {journal} {\bibinfo  {journal} {Phys. Rev. A}\ }\textbf {\bibinfo {volume} {31}},\ \bibinfo {pages} {483--485} (\bibinfo {year} {1985})}\BibitemShut {NoStop}%
\bibitem [{\citenamefont {Vaidman}(1992)}]{Vaidman-92}%
  \BibitemOpen
  \bibfield  {author} {\bibinfo {author} {\bibfnamefont {Lev}\ \bibnamefont {Vaidman}},\ }\bibfield  {title} {\enquote {\bibinfo {title} {{Minimum time for the evolution to an orthogonal quantum state}},}\ }\href {\doibase 10.1119/1.16940} {\bibfield  {journal} {\bibinfo  {journal} {American Journal of Physics}\ }\textbf {\bibinfo {volume} {60}},\ \bibinfo {pages} {182--183} (\bibinfo {year} {1992})},\ \Eprint {http://arxiv.org/abs/https://pubs.aip.org/aapt/ajp/article-pdf/60/2/182/12124296/182\_1\_online.pdf} {https://pubs.aip.org/aapt/ajp/article-pdf/60/2/182/12124296/182\_1\_online.pdf} \BibitemShut {NoStop}%
\bibitem [{\citenamefont {Uhlmann}(1992)}]{Uhlmann-92}%
  \BibitemOpen
  \bibfield  {author} {\bibinfo {author} {\bibfnamefont {Armin}\ \bibnamefont {Uhlmann}},\ }\bibfield  {title} {\enquote {\bibinfo {title} {An energy dispersion estimate},}\ }\href {\doibase https://doi.org/10.1016/0375-9601(92)90555-Z} {\bibfield  {journal} {\bibinfo  {journal} {Physics Letters A}\ }\textbf {\bibinfo {volume} {161}},\ \bibinfo {pages} {329--331} (\bibinfo {year} {1992})}\BibitemShut {NoStop}%
\bibitem [{\citenamefont {Uffink}(1993)}]{Uffink-93}%
  \BibitemOpen
  \bibfield  {author} {\bibinfo {author} {\bibfnamefont {{Jozef B}}\ \bibnamefont {Uffink}},\ }\bibfield  {title} {{\selectlanguage {English}\enquote {\bibinfo {title} {The rate of evolution of a quantum state},}\ }}\href@noop {} {\bibfield  {journal} {\bibinfo  {journal} {American Journal of Physics}\ }\textbf {\bibinfo {volume} {61}},\ \bibinfo {pages} {935--936} (\bibinfo {year} {1993})}\BibitemShut {NoStop}%
\bibitem [{\citenamefont {Pfeifer}\ and\ \citenamefont {Fr\"ohlich}(1995)}]{Pfeifer-95}%
  \BibitemOpen
  \bibfield  {author} {\bibinfo {author} {\bibfnamefont {Peter}\ \bibnamefont {Pfeifer}}\ and\ \bibinfo {author} {\bibfnamefont {J\"urg}\ \bibnamefont {Fr\"ohlich}},\ }\bibfield  {title} {\enquote {\bibinfo {title} {Generalized time-energy uncertainty relations and bounds on lifetimes of resonances},}\ }\href {\doibase 10.1103/RevModPhys.67.759} {\bibfield  {journal} {\bibinfo  {journal} {Rev. Mod. Phys.}\ }\textbf {\bibinfo {volume} {67}},\ \bibinfo {pages} {759--779} (\bibinfo {year} {1995})}\BibitemShut {NoStop}%
\bibitem [{\citenamefont {Horesh}\ and\ \citenamefont {Mann}(1998)}]{Horesh-98}%
  \BibitemOpen
  \bibfield  {author} {\bibinfo {author} {\bibfnamefont {N}~\bibnamefont {Horesh}}\ and\ \bibinfo {author} {\bibfnamefont {A}~\bibnamefont {Mann}},\ }\bibfield  {title} {\enquote {\bibinfo {title} {Intelligent states for the anandan - aharonov parameter-based uncertainty relation},}\ }\href {\doibase 10.1088/0305-4470/31/36/003} {\bibfield  {journal} {\bibinfo  {journal} {Journal of Physics A: Mathematical and General}\ }\textbf {\bibinfo {volume} {31}},\ \bibinfo {pages} {L609} (\bibinfo {year} {1998})}\BibitemShut {NoStop}%
\bibitem [{\citenamefont {Pati}(1999)}]{Pati-99}%
  \BibitemOpen
  \bibfield  {author} {\bibinfo {author} {\bibfnamefont {Arun~Kumar}\ \bibnamefont {Pati}},\ }\bibfield  {title} {\enquote {\bibinfo {title} {Uncertainty relation of anandan–aharonov and intelligent states},}\ }\href {\doibase 10.1016/s0375-9601(99)00701-x} {\bibfield  {journal} {\bibinfo  {journal} {Physics Letters A}\ }\textbf {\bibinfo {volume} {262}},\ \bibinfo {pages} {296–301} (\bibinfo {year} {1999})}\BibitemShut {NoStop}%
\bibitem [{\citenamefont {S\"oderholm}\ \emph {et~al.}(1999)\citenamefont {S\"oderholm}, \citenamefont {Bj\"ork}, \citenamefont {Tsegaye},\ and\ \citenamefont {Trifonov}}]{Jonas-99}%
  \BibitemOpen
  \bibfield  {author} {\bibinfo {author} {\bibfnamefont {Jonas}\ \bibnamefont {S\"oderholm}}, \bibinfo {author} {\bibfnamefont {Gunnar}\ \bibnamefont {Bj\"ork}}, \bibinfo {author} {\bibfnamefont {Tedros}\ \bibnamefont {Tsegaye}}, \ and\ \bibinfo {author} {\bibfnamefont {Alexei}\ \bibnamefont {Trifonov}},\ }\bibfield  {title} {\enquote {\bibinfo {title} {States that minimize the evolution time to become an orthogonal state},}\ }\href {\doibase 10.1103/PhysRevA.59.1788} {\bibfield  {journal} {\bibinfo  {journal} {Phys. Rev. A}\ }\textbf {\bibinfo {volume} {59}},\ \bibinfo {pages} {1788--1790} (\bibinfo {year} {1999})}\BibitemShut {NoStop}%
\bibitem [{\citenamefont {Andrecut}\ and\ \citenamefont {Ali}(2004)}]{Andrecut-04}%
  \BibitemOpen
  \bibfield  {author} {\bibinfo {author} {\bibfnamefont {M}~\bibnamefont {Andrecut}}\ and\ \bibinfo {author} {\bibfnamefont {M~K}\ \bibnamefont {Ali}},\ }\bibfield  {title} {\enquote {\bibinfo {title} {The adiabatic analogue of the margolus–levitin theorem},}\ }\href {\doibase 10.1088/0305-4470/37/15/L01} {\bibfield  {journal} {\bibinfo  {journal} {Journal of Physics A: Mathematical and General}\ }\textbf {\bibinfo {volume} {37}},\ \bibinfo {pages} {L157} (\bibinfo {year} {2004})}\BibitemShut {NoStop}%
\bibitem [{\citenamefont {Gray}\ and\ \citenamefont {Vogt}(2005)}]{Gray-05}%
  \BibitemOpen
  \bibfield  {author} {\bibinfo {author} {\bibfnamefont {John}\ \bibnamefont {Gray}}\ and\ \bibinfo {author} {\bibfnamefont {Andrew}\ \bibnamefont {Vogt}},\ }\bibfield  {title} {\enquote {\bibinfo {title} {Mathematical analysis of the mandelstam-tamm time-energy uncertainty principle},}\ }\href {\doibase 10.1063/1.1897164} {\bibfield  {journal} {\bibinfo  {journal} {Journal of Mathematical Physics}\ }\textbf {\bibinfo {volume} {46}} (\bibinfo {year} {2005}),\ 10.1063/1.1897164}\BibitemShut {NoStop}%
\bibitem [{\citenamefont {Zieli\ifmmode~\acute{n}\else \'{n}\fi{}ski}\ and\ \citenamefont {Zych}(2006)}]{Magdalena-06}%
  \BibitemOpen
  \bibfield  {author} {\bibinfo {author} {\bibfnamefont {Bartosz}\ \bibnamefont {Zieli\ifmmode~\acute{n}\else \'{n}\fi{}ski}}\ and\ \bibinfo {author} {\bibfnamefont {Magdalena}\ \bibnamefont {Zych}},\ }\bibfield  {title} {\enquote {\bibinfo {title} {Generalization of the margolus-levitin bound},}\ }\href {\doibase 10.1103/PhysRevA.74.034301} {\bibfield  {journal} {\bibinfo  {journal} {Phys. Rev. A}\ }\textbf {\bibinfo {volume} {74}},\ \bibinfo {pages} {034301} (\bibinfo {year} {2006})}\BibitemShut {NoStop}%
\bibitem [{\citenamefont {Andrews}(2007)}]{Andrews-07}%
  \BibitemOpen
  \bibfield  {author} {\bibinfo {author} {\bibfnamefont {Mark}\ \bibnamefont {Andrews}},\ }\bibfield  {title} {\enquote {\bibinfo {title} {Bounds to unitary evolution},}\ }\href {\doibase 10.1103/PhysRevA.75.062112} {\bibfield  {journal} {\bibinfo  {journal} {Phys. Rev. A}\ }\textbf {\bibinfo {volume} {75}},\ \bibinfo {pages} {062112} (\bibinfo {year} {2007})}\BibitemShut {NoStop}%
\bibitem [{\citenamefont {Yurtsever}(2010)}]{Yurtsever-10}%
  \BibitemOpen
  \bibfield  {author} {\bibinfo {author} {\bibfnamefont {Ulvi}\ \bibnamefont {Yurtsever}},\ }\bibfield  {title} {\enquote {\bibinfo {title} {Fundamental limits on the speed of evolution of quantum states},}\ }\href {\doibase 10.1088/0031-8949/82/03/035008} {\bibfield  {journal} {\bibinfo  {journal} {Physica Scripta}\ }\textbf {\bibinfo {volume} {82}},\ \bibinfo {pages} {035008} (\bibinfo {year} {2010})}\BibitemShut {NoStop}%
\bibitem [{\citenamefont {Shuang-Shuang}\ \emph {et~al.}(2010)\citenamefont {Shuang-Shuang}, \citenamefont {Nan},\ and\ \citenamefont {Shun-Long}}]{Shuang-10}%
  \BibitemOpen
  \bibfield  {author} {\bibinfo {author} {\bibfnamefont {Fu}~\bibnamefont {Shuang-Shuang}}, \bibinfo {author} {\bibfnamefont {Li}~\bibnamefont {Nan}}, \ and\ \bibinfo {author} {\bibfnamefont {Luo}\ \bibnamefont {Shun-Long}},\ }\bibfield  {title} {\enquote {\bibinfo {title} {A note on fundamental limit of quantum dynamics rate},}\ }\href {\doibase 10.1088/0253-6102/54/4/15} {\bibfield  {journal} {\bibinfo  {journal} {Communications in Theoretical Physics}\ }\textbf {\bibinfo {volume} {54}},\ \bibinfo {pages} {661} (\bibinfo {year} {2010})}\BibitemShut {NoStop}%
\bibitem [{\citenamefont {Poggi}\ \emph {et~al.}(2013)\citenamefont {Poggi}, \citenamefont {Lombardo},\ and\ \citenamefont {Wisniacki}}]{Poggi-13}%
  \BibitemOpen
  \bibfield  {author} {\bibinfo {author} {\bibfnamefont {P.~M.}\ \bibnamefont {Poggi}}, \bibinfo {author} {\bibfnamefont {F.~C.}\ \bibnamefont {Lombardo}}, \ and\ \bibinfo {author} {\bibfnamefont {D.~A.}\ \bibnamefont {Wisniacki}},\ }\bibfield  {title} {\enquote {\bibinfo {title} {Quantum speed limit and optimal evolution time in a two-level system},}\ }\href {\doibase 10.1209/0295-5075/104/40005} {\bibfield  {journal} {\bibinfo  {journal} {Europhysics Letters}\ }\textbf {\bibinfo {volume} {104}},\ \bibinfo {pages} {40005} (\bibinfo {year} {2013})}\BibitemShut {NoStop}%
\bibitem [{\citenamefont {Kupferman}\ and\ \citenamefont {Reznik}(2008)}]{Kupferman-08}%
  \BibitemOpen
  \bibfield  {author} {\bibinfo {author} {\bibfnamefont {Judy}\ \bibnamefont {Kupferman}}\ and\ \bibinfo {author} {\bibfnamefont {Benni}\ \bibnamefont {Reznik}},\ }\bibfield  {title} {\enquote {\bibinfo {title} {Entanglement and the speed of evolution in mixed states},}\ }\href {\doibase 10.1103/PhysRevA.78.042305} {\bibfield  {journal} {\bibinfo  {journal} {Phys. Rev. A}\ }\textbf {\bibinfo {volume} {78}},\ \bibinfo {pages} {042305} (\bibinfo {year} {2008})}\BibitemShut {NoStop}%
\bibitem [{\citenamefont {Jones}\ and\ \citenamefont {Kok}(2010)}]{Jones-10}%
  \BibitemOpen
  \bibfield  {author} {\bibinfo {author} {\bibfnamefont {Philip~J.}\ \bibnamefont {Jones}}\ and\ \bibinfo {author} {\bibfnamefont {Pieter}\ \bibnamefont {Kok}},\ }\bibfield  {title} {\enquote {\bibinfo {title} {Geometric derivation of the quantum speed limit},}\ }\href {\doibase 10.1103/PhysRevA.82.022107} {\bibfield  {journal} {\bibinfo  {journal} {Phys. Rev. A}\ }\textbf {\bibinfo {volume} {82}},\ \bibinfo {pages} {022107} (\bibinfo {year} {2010})}\BibitemShut {NoStop}%
\bibitem [{\citenamefont {Chau}(2010)}]{Chau-10}%
  \BibitemOpen
  \bibfield  {author} {\bibinfo {author} {\bibfnamefont {H.~F.}\ \bibnamefont {Chau}},\ }\bibfield  {title} {\enquote {\bibinfo {title} {Tight upper bound of the maximum speed of evolution of a quantum state},}\ }\href {\doibase 10.1103/PhysRevA.81.062133} {\bibfield  {journal} {\bibinfo  {journal} {Phys. Rev. A}\ }\textbf {\bibinfo {volume} {81}},\ \bibinfo {pages} {062133} (\bibinfo {year} {2010})}\BibitemShut {NoStop}%
\bibitem [{\citenamefont {Deffner}\ and\ \citenamefont {Lutz}(2013)}]{Deffner-13}%
  \BibitemOpen
  \bibfield  {author} {\bibinfo {author} {\bibfnamefont {Sebastian}\ \bibnamefont {Deffner}}\ and\ \bibinfo {author} {\bibfnamefont {Eric}\ \bibnamefont {Lutz}},\ }\bibfield  {title} {\enquote {\bibinfo {title} {Energy–time uncertainty relation for driven quantum systems},}\ }\href {\doibase 10.1088/1751-8113/46/33/335302} {\bibfield  {journal} {\bibinfo  {journal} {Journal of Physics A: Mathematical and Theoretical}\ }\textbf {\bibinfo {volume} {46}},\ \bibinfo {pages} {335302} (\bibinfo {year} {2013})}\BibitemShut {NoStop}%
\bibitem [{\citenamefont {Fung}\ and\ \citenamefont {Chau}(2014)}]{Fung-14}%
  \BibitemOpen
  \bibfield  {author} {\bibinfo {author} {\bibfnamefont {Chi-Hang~Fred}\ \bibnamefont {Fung}}\ and\ \bibinfo {author} {\bibfnamefont {H.~F.}\ \bibnamefont {Chau}},\ }\bibfield  {title} {\enquote {\bibinfo {title} {Relation between physical time-energy cost of a quantum process and its information fidelity},}\ }\href {\doibase 10.1103/PhysRevA.90.022333} {\bibfield  {journal} {\bibinfo  {journal} {Phys. Rev. A}\ }\textbf {\bibinfo {volume} {90}},\ \bibinfo {pages} {022333} (\bibinfo {year} {2014})}\BibitemShut {NoStop}%
\bibitem [{\citenamefont {Andersson}\ and\ \citenamefont {Heydari}(2014)}]{Andersson-14}%
  \BibitemOpen
  \bibfield  {author} {\bibinfo {author} {\bibfnamefont {O}~\bibnamefont {Andersson}}\ and\ \bibinfo {author} {\bibfnamefont {H}~\bibnamefont {Heydari}},\ }\bibfield  {title} {\enquote {\bibinfo {title} {Quantum speed limits and optimal hamiltonians for driven systems in mixed states},}\ }\href {\doibase 10.1088/1751-8113/47/21/215301} {\bibfield  {journal} {\bibinfo  {journal} {Journal of Physics A: Mathematical and Theoretical}\ }\textbf {\bibinfo {volume} {47}},\ \bibinfo {pages} {215301} (\bibinfo {year} {2014})}\BibitemShut {NoStop}%
\bibitem [{\citenamefont {Mondal}\ \emph {et~al.}(2016)\citenamefont {Mondal}, \citenamefont {Datta},\ and\ \citenamefont {Sazim}}]{Mondal-16}%
  \BibitemOpen
  \bibfield  {author} {\bibinfo {author} {\bibfnamefont {Debasis}\ \bibnamefont {Mondal}}, \bibinfo {author} {\bibfnamefont {Chandan}\ \bibnamefont {Datta}}, \ and\ \bibinfo {author} {\bibfnamefont {Sk}~\bibnamefont {Sazim}},\ }\bibfield  {title} {\enquote {\bibinfo {title} {Quantum coherence sets the quantum speed limit for mixed states},}\ }\href {\doibase 10.1016/j.physleta.2015.12.015} {\bibfield  {journal} {\bibinfo  {journal} {Physics Letters A}\ }\textbf {\bibinfo {volume} {380}},\ \bibinfo {pages} {689–695} (\bibinfo {year} {2016})}\BibitemShut {NoStop}%
\bibitem [{\citenamefont {Mondal}\ and\ \citenamefont {Pati}(2016)}]{Mondal-16PLA}%
  \BibitemOpen
  \bibfield  {author} {\bibinfo {author} {\bibfnamefont {Debasis}\ \bibnamefont {Mondal}}\ and\ \bibinfo {author} {\bibfnamefont {Arun~Kumar}\ \bibnamefont {Pati}},\ }\bibfield  {title} {\enquote {\bibinfo {title} {Quantum speed limit for mixed states using an experimentally realizable metric},}\ }\href {\doibase https://doi.org/10.1016/j.physleta.2016.02.018} {\bibfield  {journal} {\bibinfo  {journal} {Physics Letters A}\ }\textbf {\bibinfo {volume} {380}},\ \bibinfo {pages} {1395--1400} (\bibinfo {year} {2016})}\BibitemShut {NoStop}%
\bibitem [{\citenamefont {Deffner}\ and\ \citenamefont {Campbell}(2017{\natexlab{a}})}]{Deffner-17PA}%
  \BibitemOpen
  \bibfield  {author} {\bibinfo {author} {\bibfnamefont {Sebastian}\ \bibnamefont {Deffner}}\ and\ \bibinfo {author} {\bibfnamefont {Steve}\ \bibnamefont {Campbell}},\ }\bibfield  {title} {\enquote {\bibinfo {title} {Quantum speed limits: from heisenberg’s uncertainty principle to optimal quantum control},}\ }\href {\doibase 10.1088/1751-8121/aa86c6} {\bibfield  {journal} {\bibinfo  {journal} {Journal of Physics A: Mathematical and Theoretical}\ }\textbf {\bibinfo {volume} {50}},\ \bibinfo {pages} {453001} (\bibinfo {year} {2017}{\natexlab{a}})}\BibitemShut {NoStop}%
\bibitem [{\citenamefont {Campaioli}\ \emph {et~al.}(2018{\natexlab{a}})\citenamefont {Campaioli}, \citenamefont {Pollock}, \citenamefont {Binder},\ and\ \citenamefont {Modi}}]{Campaioli-18PRL}%
  \BibitemOpen
  \bibfield  {author} {\bibinfo {author} {\bibfnamefont {Francesco}\ \bibnamefont {Campaioli}}, \bibinfo {author} {\bibfnamefont {Felix~A.}\ \bibnamefont {Pollock}}, \bibinfo {author} {\bibfnamefont {Felix~C.}\ \bibnamefont {Binder}}, \ and\ \bibinfo {author} {\bibfnamefont {Kavan}\ \bibnamefont {Modi}},\ }\bibfield  {title} {\enquote {\bibinfo {title} {Tightening quantum speed limits for almost all states},}\ }\href {\doibase 10.1103/PhysRevLett.120.060409} {\bibfield  {journal} {\bibinfo  {journal} {Phys. Rev. Lett.}\ }\textbf {\bibinfo {volume} {120}},\ \bibinfo {pages} {060409} (\bibinfo {year} {2018}{\natexlab{a}})}\BibitemShut {NoStop}%
\bibitem [{\citenamefont {Ashhab}\ \emph {et~al.}(2012)\citenamefont {Ashhab}, \citenamefont {de~Groot},\ and\ \citenamefont {Nori}}]{Ashhab-12}%
  \BibitemOpen
  \bibfield  {author} {\bibinfo {author} {\bibfnamefont {S.}~\bibnamefont {Ashhab}}, \bibinfo {author} {\bibfnamefont {P.~C.}\ \bibnamefont {de~Groot}}, \ and\ \bibinfo {author} {\bibfnamefont {Franco}\ \bibnamefont {Nori}},\ }\bibfield  {title} {\enquote {\bibinfo {title} {Speed limits for quantum gates in multiqubit systems},}\ }\href {\doibase 10.1103/PhysRevA.85.052327} {\bibfield  {journal} {\bibinfo  {journal} {Phys. Rev. A}\ }\textbf {\bibinfo {volume} {85}},\ \bibinfo {pages} {052327} (\bibinfo {year} {2012})}\BibitemShut {NoStop}%
\bibitem [{\citenamefont {Mukhopadhyay}\ \emph {et~al.}(2018)\citenamefont {Mukhopadhyay}, \citenamefont {Misra}, \citenamefont {Bhattacharya},\ and\ \citenamefont {Pati}}]{Chiranjib-18}%
  \BibitemOpen
  \bibfield  {author} {\bibinfo {author} {\bibfnamefont {Chiranjib}\ \bibnamefont {Mukhopadhyay}}, \bibinfo {author} {\bibfnamefont {Avijit}\ \bibnamefont {Misra}}, \bibinfo {author} {\bibfnamefont {Samyadeb}\ \bibnamefont {Bhattacharya}}, \ and\ \bibinfo {author} {\bibfnamefont {Arun~Kumar}\ \bibnamefont {Pati}},\ }\bibfield  {title} {\enquote {\bibinfo {title} {Quantum speed limit constraints on a nanoscale autonomous refrigerator},}\ }\href {\doibase 10.1103/PhysRevE.97.062116} {\bibfield  {journal} {\bibinfo  {journal} {Phys. Rev. E}\ }\textbf {\bibinfo {volume} {97}},\ \bibinfo {pages} {062116} (\bibinfo {year} {2018})}\BibitemShut {NoStop}%
\bibitem [{\citenamefont {Funo}\ \emph {et~al.}(2019)\citenamefont {Funo}, \citenamefont {Shiraishi},\ and\ \citenamefont {Saito}}]{Funo-19}%
  \BibitemOpen
  \bibfield  {author} {\bibinfo {author} {\bibfnamefont {Ken}\ \bibnamefont {Funo}}, \bibinfo {author} {\bibfnamefont {Naoto}\ \bibnamefont {Shiraishi}}, \ and\ \bibinfo {author} {\bibfnamefont {Keiji}\ \bibnamefont {Saito}},\ }\bibfield  {title} {\enquote {\bibinfo {title} {Speed limit for open quantum systems},}\ }\href {\doibase 10.1088/1367-2630/aaf9f5} {\bibfield  {journal} {\bibinfo  {journal} {New Journal of Physics}\ }\textbf {\bibinfo {volume} {21}},\ \bibinfo {pages} {013006} (\bibinfo {year} {2019})}\BibitemShut {NoStop}%
\bibitem [{\citenamefont {Caneva}\ \emph {et~al.}(2009)\citenamefont {Caneva}, \citenamefont {Murphy}, \citenamefont {Calarco}, \citenamefont {Fazio}, \citenamefont {Montangero}, \citenamefont {Giovannetti},\ and\ \citenamefont {Santoro}}]{Caneva-09}%
  \BibitemOpen
  \bibfield  {author} {\bibinfo {author} {\bibfnamefont {T.}~\bibnamefont {Caneva}}, \bibinfo {author} {\bibfnamefont {M.}~\bibnamefont {Murphy}}, \bibinfo {author} {\bibfnamefont {T.}~\bibnamefont {Calarco}}, \bibinfo {author} {\bibfnamefont {R.}~\bibnamefont {Fazio}}, \bibinfo {author} {\bibfnamefont {S.}~\bibnamefont {Montangero}}, \bibinfo {author} {\bibfnamefont {V.}~\bibnamefont {Giovannetti}}, \ and\ \bibinfo {author} {\bibfnamefont {G.~E.}\ \bibnamefont {Santoro}},\ }\bibfield  {title} {\enquote {\bibinfo {title} {Optimal control at the quantum speed limit},}\ }\href {\doibase 10.1103/physrevlett.103.240501} {\bibfield  {journal} {\bibinfo  {journal} {Physical Review Letters}\ }\textbf {\bibinfo {volume} {103}} (\bibinfo {year} {2009}),\ 10.1103/physrevlett.103.240501}\BibitemShut {NoStop}%
\bibitem [{\citenamefont {Campbell}\ and\ \citenamefont {Deffner}(2017)}]{Campbell-17}%
  \BibitemOpen
  \bibfield  {author} {\bibinfo {author} {\bibfnamefont {Steve}\ \bibnamefont {Campbell}}\ and\ \bibinfo {author} {\bibfnamefont {Sebastian}\ \bibnamefont {Deffner}},\ }\bibfield  {title} {\enquote {\bibinfo {title} {Trade-off between speed and cost in shortcuts to adiabaticity},}\ }\href {\doibase 10.1103/physrevlett.118.100601} {\bibfield  {journal} {\bibinfo  {journal} {Physical Review Letters}\ }\textbf {\bibinfo {volume} {118}} (\bibinfo {year} {2017}),\ 10.1103/physrevlett.118.100601}\BibitemShut {NoStop}%
\bibitem [{\citenamefont {Campbell}\ \emph {et~al.}(2018)\citenamefont {Campbell}, \citenamefont {Genoni},\ and\ \citenamefont {Deffner}}]{Campbell-18}%
  \BibitemOpen
  \bibfield  {author} {\bibinfo {author} {\bibfnamefont {Steve}\ \bibnamefont {Campbell}}, \bibinfo {author} {\bibfnamefont {Marco~G}\ \bibnamefont {Genoni}}, \ and\ \bibinfo {author} {\bibfnamefont {Sebastian}\ \bibnamefont {Deffner}},\ }\bibfield  {title} {\enquote {\bibinfo {title} {Precision thermometry and the quantum speed limit},}\ }\href {\doibase 10.1088/2058-9565/aaa641} {\bibfield  {journal} {\bibinfo  {journal} {Quantum Science and Technology}\ }\textbf {\bibinfo {volume} {3}},\ \bibinfo {pages} {025002} (\bibinfo {year} {2018})}\BibitemShut {NoStop}%
\bibitem [{\citenamefont {Maccone}\ and\ \citenamefont {Pati}(2014)}]{Pati-14}%
  \BibitemOpen
  \bibfield  {author} {\bibinfo {author} {\bibfnamefont {Lorenzo}\ \bibnamefont {Maccone}}\ and\ \bibinfo {author} {\bibfnamefont {Arun~K.}\ \bibnamefont {Pati}},\ }\bibfield  {title} {\enquote {\bibinfo {title} {Stronger uncertainty relations for all incompatible observables},}\ }\href {\doibase 10.1103/PhysRevLett.113.260401} {\bibfield  {journal} {\bibinfo  {journal} {Phys. Rev. Lett.}\ }\textbf {\bibinfo {volume} {113}},\ \bibinfo {pages} {260401} (\bibinfo {year} {2014})}\BibitemShut {NoStop}%
\bibitem [{\citenamefont {Thakuria}\ and\ \citenamefont {Pati}(2022)}]{dimpi-22a}%
  \BibitemOpen
  \bibfield  {author} {\bibinfo {author} {\bibfnamefont {Dimpi}\ \bibnamefont {Thakuria}}\ and\ \bibinfo {author} {\bibfnamefont {Arun~Kumar}\ \bibnamefont {Pati}},\ }\href@noop {} {\enquote {\bibinfo {title} {Stronger quantum speed limit},}\ } (\bibinfo {year} {2022}),\ \Eprint {http://arxiv.org/abs/2208.05469} {arXiv:2208.05469 [quant-ph]} \BibitemShut {NoStop}%
\bibitem [{\citenamefont {Mohan}\ and\ \citenamefont {Pati}(2022)}]{Brij-22}%
  \BibitemOpen
  \bibfield  {author} {\bibinfo {author} {\bibfnamefont {Brij}\ \bibnamefont {Mohan}}\ and\ \bibinfo {author} {\bibfnamefont {Arun~Kumar}\ \bibnamefont {Pati}},\ }\bibfield  {title} {\enquote {\bibinfo {title} {Quantum speed limits for observables},}\ }\href {\doibase 10.1103/PhysRevA.106.042436} {\bibfield  {journal} {\bibinfo  {journal} {Phys. Rev. A}\ }\textbf {\bibinfo {volume} {106}},\ \bibinfo {pages} {042436} (\bibinfo {year} {2022})}\BibitemShut {NoStop}%
\bibitem [{\citenamefont {Horodecki}\ \emph {et~al.}(2009)\citenamefont {Horodecki}, \citenamefont {Horodecki}, \citenamefont {Horodecki},\ and\ \citenamefont {Horodecki}}]{Horodecki-09}%
  \BibitemOpen
  \bibfield  {author} {\bibinfo {author} {\bibfnamefont {Ryszard}\ \bibnamefont {Horodecki}}, \bibinfo {author} {\bibfnamefont {Paweł}\ \bibnamefont {Horodecki}}, \bibinfo {author} {\bibfnamefont {Michał}\ \bibnamefont {Horodecki}}, \ and\ \bibinfo {author} {\bibfnamefont {Karol}\ \bibnamefont {Horodecki}},\ }\bibfield  {title} {\enquote {\bibinfo {title} {Quantum entanglement},}\ }\href {\doibase 10.1103/revmodphys.81.865} {\bibfield  {journal} {\bibinfo  {journal} {Reviews of Modern Physics}\ }\textbf {\bibinfo {volume} {81}},\ \bibinfo {pages} {865} (\bibinfo {year} {2009})}\BibitemShut {NoStop}%
\bibitem [{\citenamefont {Das}\ \emph {et~al.}(2016)\citenamefont {Das}, \citenamefont {Chanda}, \citenamefont {Lewenstein}, \citenamefont {Sanpera}, \citenamefont {Sen~De},\ and\ \citenamefont {Sen}}]{Sreetama-17}%
  \BibitemOpen
  \bibfield  {author} {\bibinfo {author} {\bibfnamefont {Sreetama}\ \bibnamefont {Das}}, \bibinfo {author} {\bibfnamefont {Titas}\ \bibnamefont {Chanda}}, \bibinfo {author} {\bibfnamefont {Maciej}\ \bibnamefont {Lewenstein}}, \bibinfo {author} {\bibfnamefont {Anna}\ \bibnamefont {Sanpera}}, \bibinfo {author} {\bibfnamefont {Aditi}\ \bibnamefont {Sen~De}}, \ and\ \bibinfo {author} {\bibfnamefont {Ujjwal}\ \bibnamefont {Sen}},\ }\bibfield  {title} {\enquote {\bibinfo {title} {The separability versus entanglement problem},}\ }\href {\doibase https://doi.org/10.1002/9783527805785.ch8} {\bibfield  {journal} {\bibinfo  {journal} {Quantum Information: From Foundations to Quantum Technology Applications}\ ,\ \bibinfo {pages} {127}} (\bibinfo {year} {2016})}\BibitemShut {NoStop}%
\bibitem [{\citenamefont {de~Boer}\ \emph {et~al.}(2019{\natexlab{a}})\citenamefont {de~Boer}, \citenamefont {J\"arvel\"a},\ and\ \citenamefont {Keski-Vakkuri}}]{capacity_def-19}%
  \BibitemOpen
  \bibfield  {author} {\bibinfo {author} {\bibfnamefont {Jan}\ \bibnamefont {de~Boer}}, \bibinfo {author} {\bibfnamefont {Jarkko}\ \bibnamefont {J\"arvel\"a}}, \ and\ \bibinfo {author} {\bibfnamefont {Esko}\ \bibnamefont {Keski-Vakkuri}},\ }\bibfield  {title} {\enquote {\bibinfo {title} {Aspects of capacity of entanglement},}\ }\href {\doibase 10.1103/PhysRevD.99.066012} {\bibfield  {journal} {\bibinfo  {journal} {Physical Review D}\ }\textbf {\bibinfo {volume} {99}},\ \bibinfo {pages} {066012} (\bibinfo {year} {2019}{\natexlab{a}})}\BibitemShut {NoStop}%
\bibitem [{\citenamefont {Shrimali}\ \emph {et~al.}(2022)\citenamefont {Shrimali}, \citenamefont {Bhowmick}, \citenamefont {Pandey},\ and\ \citenamefont {Pati}}]{capacity-22}%
  \BibitemOpen
  \bibfield  {author} {\bibinfo {author} {\bibfnamefont {Divyansh}\ \bibnamefont {Shrimali}}, \bibinfo {author} {\bibfnamefont {Swapnil}\ \bibnamefont {Bhowmick}}, \bibinfo {author} {\bibfnamefont {Vivek}\ \bibnamefont {Pandey}}, \ and\ \bibinfo {author} {\bibfnamefont {Arun~Kumar}\ \bibnamefont {Pati}},\ }\bibfield  {title} {\enquote {\bibinfo {title} {Capacity of entanglement for a nonlocal hamiltonian},}\ }\href {\doibase 10.1103/PhysRevA.106.042419} {\bibfield  {journal} {\bibinfo  {journal} {Phys. Rev. A}\ }\textbf {\bibinfo {volume} {106}},\ \bibinfo {pages} {042419} (\bibinfo {year} {2022})}\BibitemShut {NoStop}%
\bibitem [{\citenamefont {de~Boer}\ \emph {et~al.}(2019{\natexlab{b}})\citenamefont {de~Boer}, \citenamefont {J\"arvel\"a},\ and\ \citenamefont {Keski-Vakkuri}}]{capacity_def}%
  \BibitemOpen
  \bibfield  {author} {\bibinfo {author} {\bibfnamefont {Jan}\ \bibnamefont {de~Boer}}, \bibinfo {author} {\bibfnamefont {Jarkko}\ \bibnamefont {J\"arvel\"a}}, \ and\ \bibinfo {author} {\bibfnamefont {Esko}\ \bibnamefont {Keski-Vakkuri}},\ }\bibfield  {title} {\enquote {\bibinfo {title} {Aspects of capacity of entanglement},}\ }\href {\doibase 10.1103/PhysRevD.99.066012} {\bibfield  {journal} {\bibinfo  {journal} {Physical Review D}\ }\textbf {\bibinfo {volume} {99}},\ \bibinfo {pages} {066012} (\bibinfo {year} {2019}{\natexlab{b}})}\BibitemShut {NoStop}%
\bibitem [{\citenamefont {Laflorencie}(2016)}]{Nicolas-16}%
  \BibitemOpen
  \bibfield  {author} {\bibinfo {author} {\bibfnamefont {Nicolas}\ \bibnamefont {Laflorencie}},\ }\bibfield  {title} {\enquote {\bibinfo {title} {Quantum entanglement in condensed matter systems},}\ }\href {\doibase 10.1016/j.physrep.2016.06.008} {\bibfield  {journal} {\bibinfo  {journal} {Physics Reports}\ }\textbf {\bibinfo {volume} {646}},\ \bibinfo {pages} {1} (\bibinfo {year} {2016})}\BibitemShut {NoStop}%
\bibitem [{\citenamefont {Yao}\ and\ \citenamefont {Qi}(2010)}]{Yao-10}%
  \BibitemOpen
  \bibfield  {author} {\bibinfo {author} {\bibfnamefont {Hong}\ \bibnamefont {Yao}}\ and\ \bibinfo {author} {\bibfnamefont {Xiao-Liang}\ \bibnamefont {Qi}},\ }\bibfield  {title} {\enquote {\bibinfo {title} {Entanglement entropy and entanglement spectrum of the kitaev model},}\ }\href {\doibase 10.1103/PhysRevLett.105.080501} {\bibfield  {journal} {\bibinfo  {journal} {Physical Review Letters}\ }\textbf {\bibinfo {volume} {105}},\ \bibinfo {pages} {080501} (\bibinfo {year} {2010})}\BibitemShut {NoStop}%
\bibitem [{\citenamefont {Alicki}\ and\ \citenamefont {Fannes}(2013)}]{Alicki-13}%
  \BibitemOpen
  \bibfield  {author} {\bibinfo {author} {\bibfnamefont {Robert}\ \bibnamefont {Alicki}}\ and\ \bibinfo {author} {\bibfnamefont {Mark}\ \bibnamefont {Fannes}},\ }\bibfield  {title} {\enquote {\bibinfo {title} {Entanglement boost for extractable work from ensembles of quantum batteries},}\ }\href {\doibase 10.1103/PhysRevE.87.042123} {\bibfield  {journal} {\bibinfo  {journal} {Phys. Rev. E}\ }\textbf {\bibinfo {volume} {87}},\ \bibinfo {pages} {042123} (\bibinfo {year} {2013})}\BibitemShut {NoStop}%
\bibitem [{\citenamefont {Binder}\ \emph {et~al.}(2015)\citenamefont {Binder}, \citenamefont {Vinjanampathy}, \citenamefont {Modi},\ and\ \citenamefont {Goold}}]{Binder-15}%
  \BibitemOpen
  \bibfield  {author} {\bibinfo {author} {\bibfnamefont {Felix~C}\ \bibnamefont {Binder}}, \bibinfo {author} {\bibfnamefont {Sai}\ \bibnamefont {Vinjanampathy}}, \bibinfo {author} {\bibfnamefont {Kavan}\ \bibnamefont {Modi}}, \ and\ \bibinfo {author} {\bibfnamefont {John}\ \bibnamefont {Goold}},\ }\bibfield  {title} {\enquote {\bibinfo {title} {Quantacell: powerful charging of quantum batteries},}\ }\href {\doibase 10.1088/1367-2630/17/7/075015} {\bibfield  {journal} {\bibinfo  {journal} {New Journal of Physics}\ }\textbf {\bibinfo {volume} {17}},\ \bibinfo {pages} {075015} (\bibinfo {year} {2015})}\BibitemShut {NoStop}%
\bibitem [{\citenamefont {Campaioli}\ \emph {et~al.}(2017)\citenamefont {Campaioli}, \citenamefont {Pollock}, \citenamefont {Binder}, \citenamefont {C\'eleri}, \citenamefont {Goold}, \citenamefont {Vinjanampathy},\ and\ \citenamefont {Modi}}]{Campaioli-17}%
  \BibitemOpen
  \bibfield  {author} {\bibinfo {author} {\bibfnamefont {Francesco}\ \bibnamefont {Campaioli}}, \bibinfo {author} {\bibfnamefont {Felix~A.}\ \bibnamefont {Pollock}}, \bibinfo {author} {\bibfnamefont {Felix~C.}\ \bibnamefont {Binder}}, \bibinfo {author} {\bibfnamefont {Lucas}\ \bibnamefont {C\'eleri}}, \bibinfo {author} {\bibfnamefont {John}\ \bibnamefont {Goold}}, \bibinfo {author} {\bibfnamefont {Sai}\ \bibnamefont {Vinjanampathy}}, \ and\ \bibinfo {author} {\bibfnamefont {Kavan}\ \bibnamefont {Modi}},\ }\bibfield  {title} {\enquote {\bibinfo {title} {Enhancing the charging power of quantum batteries},}\ }\href {\doibase 10.1103/PhysRevLett.118.150601} {\bibfield  {journal} {\bibinfo  {journal} {Phys. Rev. Lett.}\ }\textbf {\bibinfo {volume} {118}},\ \bibinfo {pages} {150601} (\bibinfo {year} {2017})}\BibitemShut {NoStop}%
\bibitem [{\citenamefont {Rossini}\ \emph {et~al.}(2020)\citenamefont {Rossini}, \citenamefont {Andolina}, \citenamefont {Rosa}, \citenamefont {Carrega},\ and\ \citenamefont {Polini}}]{Rossini-20}%
  \BibitemOpen
  \bibfield  {author} {\bibinfo {author} {\bibfnamefont {Davide}\ \bibnamefont {Rossini}}, \bibinfo {author} {\bibfnamefont {Gian~Marcello}\ \bibnamefont {Andolina}}, \bibinfo {author} {\bibfnamefont {Dario}\ \bibnamefont {Rosa}}, \bibinfo {author} {\bibfnamefont {Matteo}\ \bibnamefont {Carrega}}, \ and\ \bibinfo {author} {\bibfnamefont {Marco}\ \bibnamefont {Polini}},\ }\bibfield  {title} {\enquote {\bibinfo {title} {Quantum advantage in the charging process of sachdev-ye-kitaev batteries},}\ }\href {\doibase 10.1103/PhysRevLett.125.236402} {\bibfield  {journal} {\bibinfo  {journal} {Phys. Rev. Lett.}\ }\textbf {\bibinfo {volume} {125}},\ \bibinfo {pages} {236402} (\bibinfo {year} {2020})}\BibitemShut {NoStop}%
\bibitem [{\citenamefont {Gyhm}\ \emph {et~al.}(2022)\citenamefont {Gyhm}, \citenamefont {\ifmmode~\check{S}\else \v{S}\fi{}afr\'anek},\ and\ \citenamefont {Rosa}}]{Gyhm-22}%
  \BibitemOpen
  \bibfield  {author} {\bibinfo {author} {\bibfnamefont {Ju-Yeon}\ \bibnamefont {Gyhm}}, \bibinfo {author} {\bibfnamefont {Dominik}\ \bibnamefont {\ifmmode~\check{S}\else \v{S}\fi{}afr\'anek}}, \ and\ \bibinfo {author} {\bibfnamefont {Dario}\ \bibnamefont {Rosa}},\ }\bibfield  {title} {\enquote {\bibinfo {title} {Quantum charging advantage cannot be extensive without global operations},}\ }\href {\doibase 10.1103/PhysRevLett.128.140501} {\bibfield  {journal} {\bibinfo  {journal} {Phys. Rev. Lett.}\ }\textbf {\bibinfo {volume} {128}},\ \bibinfo {pages} {140501} (\bibinfo {year} {2022})}\BibitemShut {NoStop}%
\bibitem [{\citenamefont {Gyhm}\ and\ \citenamefont {Fischer}(2024)}]{Gyhm-24}%
  \BibitemOpen
  \bibfield  {author} {\bibinfo {author} {\bibfnamefont {Ju-Yeon}\ \bibnamefont {Gyhm}}\ and\ \bibinfo {author} {\bibfnamefont {Uwe~R.}\ \bibnamefont {Fischer}},\ }\bibfield  {title} {\enquote {\bibinfo {title} {Beneficial and detrimental entanglement for quantum battery charging},}\ }\href {\doibase 10.1116/5.0184903} {\bibfield  {journal} {\bibinfo  {journal} {AVS Quantum Science}\ }\textbf {\bibinfo {volume} {6}} (\bibinfo {year} {2024}),\ 10.1116/5.0184903}\BibitemShut {NoStop}%
\bibitem [{\citenamefont {Gyhm}\ \emph {et~al.}(2024)\citenamefont {Gyhm}, \citenamefont {Rosa},\ and\ \citenamefont {\ifmmode~\check{S}\else \v{S}\fi{}afr\'anek}}]{Gyhm-24Min}%
  \BibitemOpen
  \bibfield  {author} {\bibinfo {author} {\bibfnamefont {Ju-Yeon}\ \bibnamefont {Gyhm}}, \bibinfo {author} {\bibfnamefont {Dario}\ \bibnamefont {Rosa}}, \ and\ \bibinfo {author} {\bibfnamefont {Dominik}\ \bibnamefont {\ifmmode~\check{S}\else \v{S}\fi{}afr\'anek}},\ }\bibfield  {title} {\enquote {\bibinfo {title} {Minimal time required to charge a quantum system},}\ }\href {\doibase 10.1103/PhysRevA.109.022607} {\bibfield  {journal} {\bibinfo  {journal} {Phys. Rev. A}\ }\textbf {\bibinfo {volume} {109}},\ \bibinfo {pages} {022607} (\bibinfo {year} {2024})}\BibitemShut {NoStop}%
\bibitem [{\citenamefont {Juli\`a-Farr\'e}\ \emph {et~al.}(2020)\citenamefont {Juli\`a-Farr\'e}, \citenamefont {Salamon}, \citenamefont {Riera}, \citenamefont {Bera},\ and\ \citenamefont {Lewenstein}}]{Lewenstein-20}%
  \BibitemOpen
  \bibfield  {author} {\bibinfo {author} {\bibfnamefont {Sergi}\ \bibnamefont {Juli\`a-Farr\'e}}, \bibinfo {author} {\bibfnamefont {Tymoteusz}\ \bibnamefont {Salamon}}, \bibinfo {author} {\bibfnamefont {Arnau}\ \bibnamefont {Riera}}, \bibinfo {author} {\bibfnamefont {Manabendra~N.}\ \bibnamefont {Bera}}, \ and\ \bibinfo {author} {\bibfnamefont {Maciej}\ \bibnamefont {Lewenstein}},\ }\bibfield  {title} {\enquote {\bibinfo {title} {Bounds on the capacity and power of quantum batteries},}\ }\href {\doibase 10.1103/PhysRevResearch.2.023113} {\bibfield  {journal} {\bibinfo  {journal} {Phys. Rev. Res.}\ }\textbf {\bibinfo {volume} {2}},\ \bibinfo {pages} {023113} (\bibinfo {year} {2020})}\BibitemShut {NoStop}%
\bibitem [{\citenamefont {Campaioli}\ \emph {et~al.}(2023)\citenamefont {Campaioli}, \citenamefont {Gherardini}, \citenamefont {Quach}, \citenamefont {Polini},\ and\ \citenamefont {Andolina}}]{Campaioli-23}%
  \BibitemOpen
  \bibfield  {author} {\bibinfo {author} {\bibfnamefont {Francesco}\ \bibnamefont {Campaioli}}, \bibinfo {author} {\bibfnamefont {Stefano}\ \bibnamefont {Gherardini}}, \bibinfo {author} {\bibfnamefont {James~Q.}\ \bibnamefont {Quach}}, \bibinfo {author} {\bibfnamefont {Marco}\ \bibnamefont {Polini}}, \ and\ \bibinfo {author} {\bibfnamefont {Gian~Marcello}\ \bibnamefont {Andolina}},\ }\href@noop {} {\enquote {\bibinfo {title} {Colloquium: Quantum batteries},}\ } (\bibinfo {year} {2023}),\ \Eprint {http://arxiv.org/abs/2308.02277} {arXiv:2308.02277 [quant-ph]} \BibitemShut {NoStop}%
\bibitem [{\citenamefont {Andolina}\ \emph {et~al.}(2018)\citenamefont {Andolina}, \citenamefont {Farina}, \citenamefont {Mari}, \citenamefont {Pellegrini}, \citenamefont {Giovannetti},\ and\ \citenamefont {Polini}}]{Andolina-18}%
  \BibitemOpen
  \bibfield  {author} {\bibinfo {author} {\bibfnamefont {Gian~Marcello}\ \bibnamefont {Andolina}}, \bibinfo {author} {\bibfnamefont {Donato}\ \bibnamefont {Farina}}, \bibinfo {author} {\bibfnamefont {Andrea}\ \bibnamefont {Mari}}, \bibinfo {author} {\bibfnamefont {Vittorio}\ \bibnamefont {Pellegrini}}, \bibinfo {author} {\bibfnamefont {Vittorio}\ \bibnamefont {Giovannetti}}, \ and\ \bibinfo {author} {\bibfnamefont {Marco}\ \bibnamefont {Polini}},\ }\bibfield  {title} {\enquote {\bibinfo {title} {Charger-mediated energy transfer in exactly solvable models for quantum batteries},}\ }\href {\doibase 10.1103/PhysRevB.98.205423} {\bibfield  {journal} {\bibinfo  {journal} {Phys. Rev. B}\ }\textbf {\bibinfo {volume} {98}},\ \bibinfo {pages} {205423} (\bibinfo {year} {2018})}\BibitemShut {NoStop}%
\bibitem [{\citenamefont {Le}\ \emph {et~al.}(2018)\citenamefont {Le}, \citenamefont {Levinsen}, \citenamefont {Modi}, \citenamefont {Parish},\ and\ \citenamefont {Pollock}}]{Thao-18}%
  \BibitemOpen
  \bibfield  {author} {\bibinfo {author} {\bibfnamefont {Thao~P.}\ \bibnamefont {Le}}, \bibinfo {author} {\bibfnamefont {Jesper}\ \bibnamefont {Levinsen}}, \bibinfo {author} {\bibfnamefont {Kavan}\ \bibnamefont {Modi}}, \bibinfo {author} {\bibfnamefont {Meera~M.}\ \bibnamefont {Parish}}, \ and\ \bibinfo {author} {\bibfnamefont {Felix~A.}\ \bibnamefont {Pollock}},\ }\bibfield  {title} {\enquote {\bibinfo {title} {Spin-chain model of a many-body quantum battery},}\ }\href {\doibase 10.1103/PhysRevA.97.022106} {\bibfield  {journal} {\bibinfo  {journal} {Phys. Rev. A}\ }\textbf {\bibinfo {volume} {97}},\ \bibinfo {pages} {022106} (\bibinfo {year} {2018})}\BibitemShut {NoStop}%
\bibitem [{\citenamefont {Zhang}\ \emph {et~al.}(2019)\citenamefont {Zhang}, \citenamefont {Yang}, \citenamefont {Fu},\ and\ \citenamefont {Wang}}]{Zhang-19}%
  \BibitemOpen
  \bibfield  {author} {\bibinfo {author} {\bibfnamefont {Yu-Yu}\ \bibnamefont {Zhang}}, \bibinfo {author} {\bibfnamefont {Tian-Ran}\ \bibnamefont {Yang}}, \bibinfo {author} {\bibfnamefont {Libin}\ \bibnamefont {Fu}}, \ and\ \bibinfo {author} {\bibfnamefont {Xiaoguang}\ \bibnamefont {Wang}},\ }\bibfield  {title} {\enquote {\bibinfo {title} {Powerful harmonic charging in a quantum battery},}\ }\href {\doibase 10.1103/PhysRevE.99.052106} {\bibfield  {journal} {\bibinfo  {journal} {Phys. Rev. E}\ }\textbf {\bibinfo {volume} {99}},\ \bibinfo {pages} {052106} (\bibinfo {year} {2019})}\BibitemShut {NoStop}%
\bibitem [{\citenamefont {Barra}(2019)}]{Barra-19}%
  \BibitemOpen
  \bibfield  {author} {\bibinfo {author} {\bibfnamefont {Felipe}\ \bibnamefont {Barra}},\ }\bibfield  {title} {\enquote {\bibinfo {title} {Dissipative charging of a quantum battery},}\ }\href {\doibase 10.1103/PhysRevLett.122.210601} {\bibfield  {journal} {\bibinfo  {journal} {Phys. Rev. Lett.}\ }\textbf {\bibinfo {volume} {122}},\ \bibinfo {pages} {210601} (\bibinfo {year} {2019})}\BibitemShut {NoStop}%
\bibitem [{\citenamefont {Santos}\ \emph {et~al.}(2019)\citenamefont {Santos}, \citenamefont {\ifmmode~\mbox{\c{C}}\else \c{C}\fi{}akmak}, \citenamefont {Campbell},\ and\ \citenamefont {Zinner}}]{Santos-19}%
  \BibitemOpen
  \bibfield  {author} {\bibinfo {author} {\bibfnamefont {Alan~C.}\ \bibnamefont {Santos}}, \bibinfo {author} {\bibfnamefont {Bar\ifmmode \imath \else \i \fi{}\ifmmode \mbox{\c{s}}\else~\c{s}\fi{}}\ \bibnamefont {\ifmmode~\mbox{\c{C}}\else \c{C}\fi{}akmak}}, \bibinfo {author} {\bibfnamefont {Steve}\ \bibnamefont {Campbell}}, \ and\ \bibinfo {author} {\bibfnamefont {Nikolaj~T.}\ \bibnamefont {Zinner}},\ }\bibfield  {title} {\enquote {\bibinfo {title} {Stable adiabatic quantum batteries},}\ }\href {\doibase 10.1103/PhysRevE.100.032107} {\bibfield  {journal} {\bibinfo  {journal} {Phys. Rev. E}\ }\textbf {\bibinfo {volume} {100}},\ \bibinfo {pages} {032107} (\bibinfo {year} {2019})}\BibitemShut {NoStop}%
\bibitem [{\citenamefont {Andolina}\ \emph {et~al.}(2019)\citenamefont {Andolina}, \citenamefont {Keck}, \citenamefont {Mari}, \citenamefont {Campisi}, \citenamefont {Giovannetti},\ and\ \citenamefont {Polini}}]{Andolina-19}%
  \BibitemOpen
  \bibfield  {author} {\bibinfo {author} {\bibfnamefont {Gian~Marcello}\ \bibnamefont {Andolina}}, \bibinfo {author} {\bibfnamefont {Maximilian}\ \bibnamefont {Keck}}, \bibinfo {author} {\bibfnamefont {Andrea}\ \bibnamefont {Mari}}, \bibinfo {author} {\bibfnamefont {Michele}\ \bibnamefont {Campisi}}, \bibinfo {author} {\bibfnamefont {Vittorio}\ \bibnamefont {Giovannetti}}, \ and\ \bibinfo {author} {\bibfnamefont {Marco}\ \bibnamefont {Polini}},\ }\bibfield  {title} {\enquote {\bibinfo {title} {Extractable work, the role of correlations, and asymptotic freedom in quantum batteries},}\ }\href {\doibase 10.1103/PhysRevLett.122.047702} {\bibfield  {journal} {\bibinfo  {journal} {Phys. Rev. Lett.}\ }\textbf {\bibinfo {volume} {122}},\ \bibinfo {pages} {047702} (\bibinfo {year} {2019})}\BibitemShut {NoStop}%
\bibitem [{\citenamefont {Crescente}\ \emph {et~al.}(2020{\natexlab{a}})\citenamefont {Crescente}, \citenamefont {Carrega}, \citenamefont {Sassetti},\ and\ \citenamefont {Ferraro}}]{Alba-20}%
  \BibitemOpen
  \bibfield  {author} {\bibinfo {author} {\bibfnamefont {Alba}\ \bibnamefont {Crescente}}, \bibinfo {author} {\bibfnamefont {Matteo}\ \bibnamefont {Carrega}}, \bibinfo {author} {\bibfnamefont {Maura}\ \bibnamefont {Sassetti}}, \ and\ \bibinfo {author} {\bibfnamefont {Dario}\ \bibnamefont {Ferraro}},\ }\bibfield  {title} {\enquote {\bibinfo {title} {Ultrafast charging in a two-photon dicke quantum battery},}\ }\href {\doibase 10.1103/PhysRevB.102.245407} {\bibfield  {journal} {\bibinfo  {journal} {Phys. Rev. B}\ }\textbf {\bibinfo {volume} {102}},\ \bibinfo {pages} {245407} (\bibinfo {year} {2020}{\natexlab{a}})}\BibitemShut {NoStop}%
\bibitem [{\citenamefont {Santos}\ \emph {et~al.}(2020)\citenamefont {Santos}, \citenamefont {Saguia},\ and\ \citenamefont {Sarandy}}]{Santos-20}%
  \BibitemOpen
  \bibfield  {author} {\bibinfo {author} {\bibfnamefont {Alan~C.}\ \bibnamefont {Santos}}, \bibinfo {author} {\bibfnamefont {Andreia}\ \bibnamefont {Saguia}}, \ and\ \bibinfo {author} {\bibfnamefont {Marcelo~S.}\ \bibnamefont {Sarandy}},\ }\bibfield  {title} {\enquote {\bibinfo {title} {Stable and charge-switchable quantum batteries},}\ }\href {\doibase 10.1103/PhysRevE.101.062114} {\bibfield  {journal} {\bibinfo  {journal} {Phys. Rev. E}\ }\textbf {\bibinfo {volume} {101}},\ \bibinfo {pages} {062114} (\bibinfo {year} {2020})}\BibitemShut {NoStop}%
\bibitem [{\citenamefont {Santos}(2021)}]{Santos-21}%
  \BibitemOpen
  \bibfield  {author} {\bibinfo {author} {\bibfnamefont {Alan~C.}\ \bibnamefont {Santos}},\ }\bibfield  {title} {\enquote {\bibinfo {title} {Quantum advantage of two-level batteries in the self-discharging process},}\ }\href {\doibase 10.1103/PhysRevE.103.042118} {\bibfield  {journal} {\bibinfo  {journal} {Phys. Rev. E}\ }\textbf {\bibinfo {volume} {103}},\ \bibinfo {pages} {042118} (\bibinfo {year} {2021})}\BibitemShut {NoStop}%
\bibitem [{\citenamefont {Dou}\ \emph {et~al.}(2022)\citenamefont {Dou}, \citenamefont {Lu}, \citenamefont {Wang},\ and\ \citenamefont {Sun}}]{Dou-22}%
  \BibitemOpen
  \bibfield  {author} {\bibinfo {author} {\bibfnamefont {Fu-Quan}\ \bibnamefont {Dou}}, \bibinfo {author} {\bibfnamefont {You-Qi}\ \bibnamefont {Lu}}, \bibinfo {author} {\bibfnamefont {Yuan-Jin}\ \bibnamefont {Wang}}, \ and\ \bibinfo {author} {\bibfnamefont {Jian-An}\ \bibnamefont {Sun}},\ }\bibfield  {title} {\enquote {\bibinfo {title} {Extended dicke quantum battery with interatomic interactions and driving field},}\ }\href {\doibase 10.1103/PhysRevB.105.115405} {\bibfield  {journal} {\bibinfo  {journal} {Phys. Rev. B}\ }\textbf {\bibinfo {volume} {105}},\ \bibinfo {pages} {115405} (\bibinfo {year} {2022})}\BibitemShut {NoStop}%
\bibitem [{\citenamefont {Barra}\ \emph {et~al.}(2022)\citenamefont {Barra}, \citenamefont {Hovhannisyan},\ and\ \citenamefont {Imparato}}]{Barra-22}%
  \BibitemOpen
  \bibfield  {author} {\bibinfo {author} {\bibfnamefont {Felipe}\ \bibnamefont {Barra}}, \bibinfo {author} {\bibfnamefont {Karen~V}\ \bibnamefont {Hovhannisyan}}, \ and\ \bibinfo {author} {\bibfnamefont {Alberto}\ \bibnamefont {Imparato}},\ }\bibfield  {title} {\enquote {\bibinfo {title} {Quantum batteries at the verge of a phase transition},}\ }\href {\doibase 10.1088/1367-2630/ac43ed} {\bibfield  {journal} {\bibinfo  {journal} {New Journal of Physics}\ }\textbf {\bibinfo {volume} {24}},\ \bibinfo {pages} {015003} (\bibinfo {year} {2022})}\BibitemShut {NoStop}%
\bibitem [{\citenamefont {Carrasco}\ \emph {et~al.}(2022)\citenamefont {Carrasco}, \citenamefont {Maze}, \citenamefont {Hermann-Avigliano},\ and\ \citenamefont {Barra}}]{Carrasco-22}%
  \BibitemOpen
  \bibfield  {author} {\bibinfo {author} {\bibfnamefont {Javier}\ \bibnamefont {Carrasco}}, \bibinfo {author} {\bibfnamefont {Jer\'onimo~R.}\ \bibnamefont {Maze}}, \bibinfo {author} {\bibfnamefont {Carla}\ \bibnamefont {Hermann-Avigliano}}, \ and\ \bibinfo {author} {\bibfnamefont {Felipe}\ \bibnamefont {Barra}},\ }\bibfield  {title} {\enquote {\bibinfo {title} {Collective enhancement in dissipative quantum batteries},}\ }\href {\doibase 10.1103/PhysRevE.105.064119} {\bibfield  {journal} {\bibinfo  {journal} {Phys. Rev. E}\ }\textbf {\bibinfo {volume} {105}},\ \bibinfo {pages} {064119} (\bibinfo {year} {2022})}\BibitemShut {NoStop}%
\bibitem [{\citenamefont {Shaghaghi}\ \emph {et~al.}(2022)\citenamefont {Shaghaghi}, \citenamefont {Singh}, \citenamefont {Benenti},\ and\ \citenamefont {Rosa}}]{Shaghaghi-22}%
  \BibitemOpen
  \bibfield  {author} {\bibinfo {author} {\bibfnamefont {Vahid}\ \bibnamefont {Shaghaghi}}, \bibinfo {author} {\bibfnamefont {Varinder}\ \bibnamefont {Singh}}, \bibinfo {author} {\bibfnamefont {Giuliano}\ \bibnamefont {Benenti}}, \ and\ \bibinfo {author} {\bibfnamefont {Dario}\ \bibnamefont {Rosa}},\ }\bibfield  {title} {\enquote {\bibinfo {title} {Micromasers as quantum batteries},}\ }\href {\doibase 10.1088/2058-9565/ac8829} {\bibfield  {journal} {\bibinfo  {journal} {Quantum Science and Technology}\ }\textbf {\bibinfo {volume} {7}},\ \bibinfo {pages} {04LT01} (\bibinfo {year} {2022})}\BibitemShut {NoStop}%
\bibitem [{\citenamefont {Rodr\'{\i}guez}\ \emph {et~al.}(2023)\citenamefont {Rodr\'{\i}guez}, \citenamefont {Rosa},\ and\ \citenamefont {Olle}}]{Carla-23}%
  \BibitemOpen
  \bibfield  {author} {\bibinfo {author} {\bibfnamefont {Carla}\ \bibnamefont {Rodr\'{\i}guez}}, \bibinfo {author} {\bibfnamefont {Dario}\ \bibnamefont {Rosa}}, \ and\ \bibinfo {author} {\bibfnamefont {Jan}\ \bibnamefont {Olle}},\ }\bibfield  {title} {\enquote {\bibinfo {title} {Artificial intelligence discovery of a charging protocol in a micromaser quantum battery},}\ }\href {\doibase 10.1103/PhysRevA.108.042618} {\bibfield  {journal} {\bibinfo  {journal} {Phys. Rev. A}\ }\textbf {\bibinfo {volume} {108}},\ \bibinfo {pages} {042618} (\bibinfo {year} {2023})}\BibitemShut {NoStop}%
\bibitem [{\citenamefont {Santos}\ \emph {et~al.}(2023)\citenamefont {Santos}, \citenamefont {de~Almeida},\ and\ \citenamefont {Santos}}]{Santos-23}%
  \BibitemOpen
  \bibfield  {author} {\bibinfo {author} {\bibfnamefont {Tiago F.~F.}\ \bibnamefont {Santos}}, \bibinfo {author} {\bibfnamefont {Yohan~Vianna}\ \bibnamefont {de~Almeida}}, \ and\ \bibinfo {author} {\bibfnamefont {Marcelo~F.}\ \bibnamefont {Santos}},\ }\bibfield  {title} {\enquote {\bibinfo {title} {Vacuum-enhanced charging of a quantum battery},}\ }\href {\doibase 10.1103/PhysRevA.107.032203} {\bibfield  {journal} {\bibinfo  {journal} {Phys. Rev. A}\ }\textbf {\bibinfo {volume} {107}},\ \bibinfo {pages} {032203} (\bibinfo {year} {2023})}\BibitemShut {NoStop}%
\bibitem [{\citenamefont {Kamin}\ \emph {et~al.}(2023)\citenamefont {Kamin}, \citenamefont {Abuali}, \citenamefont {Ness},\ and\ \citenamefont {Salimi}}]{Kamin-23}%
  \BibitemOpen
  \bibfield  {author} {\bibinfo {author} {\bibfnamefont {F~H}\ \bibnamefont {Kamin}}, \bibinfo {author} {\bibfnamefont {Z}~\bibnamefont {Abuali}}, \bibinfo {author} {\bibfnamefont {H}~\bibnamefont {Ness}}, \ and\ \bibinfo {author} {\bibfnamefont {S}~\bibnamefont {Salimi}},\ }\bibfield  {title} {\enquote {\bibinfo {title} {Quantum battery charging by non-equilibrium steady-state currents},}\ }\href {\doibase 10.1088/1751-8121/acdb11} {\bibfield  {journal} {\bibinfo  {journal} {Journal of Physics A: Mathematical and Theoretical}\ }\textbf {\bibinfo {volume} {56}},\ \bibinfo {pages} {275302} (\bibinfo {year} {2023})}\BibitemShut {NoStop}%
\bibitem [{\citenamefont {Downing}\ and\ \citenamefont {Ukhtary}(2023)}]{Downing-23}%
  \BibitemOpen
  \bibfield  {author} {\bibinfo {author} {\bibfnamefont {Charles~Andrew}\ \bibnamefont {Downing}}\ and\ \bibinfo {author} {\bibfnamefont {Muhammad~Shoufie}\ \bibnamefont {Ukhtary}},\ }\bibfield  {title} {\enquote {\bibinfo {title} {A quantum battery with quadratic driving},}\ }\href {\doibase 10.1038/s42005-023-01439-y} {\bibfield  {journal} {\bibinfo  {journal} {Communications Physics}\ }\textbf {\bibinfo {volume} {6}} (\bibinfo {year} {2023}),\ 10.1038/s42005-023-01439-y}\BibitemShut {NoStop}%
\bibitem [{\citenamefont {Hadipour}\ \emph {et~al.}(2023)\citenamefont {Hadipour}, \citenamefont {Haseli}, \citenamefont {Wang},\ and\ \citenamefont {Haddadi}}]{Hadipour-23a}%
  \BibitemOpen
  \bibfield  {author} {\bibinfo {author} {\bibfnamefont {Maryam}\ \bibnamefont {Hadipour}}, \bibinfo {author} {\bibfnamefont {Soroush}\ \bibnamefont {Haseli}}, \bibinfo {author} {\bibfnamefont {Dong}\ \bibnamefont {Wang}}, \ and\ \bibinfo {author} {\bibfnamefont {Saeed}\ \bibnamefont {Haddadi}},\ }\href@noop {} {\enquote {\bibinfo {title} {Practical scheme for realization of a quantum battery},}\ } (\bibinfo {year} {2023}),\ \Eprint {http://arxiv.org/abs/2312.06389} {arXiv:2312.06389 [quant-ph]} \BibitemShut {NoStop}%
\bibitem [{\citenamefont {Tabesh}\ \emph {et~al.}(2020)\citenamefont {Tabesh}, \citenamefont {Kamin},\ and\ \citenamefont {Salimi}}]{Kamin-20A}%
  \BibitemOpen
  \bibfield  {author} {\bibinfo {author} {\bibfnamefont {F.~T.}\ \bibnamefont {Tabesh}}, \bibinfo {author} {\bibfnamefont {F.~H.}\ \bibnamefont {Kamin}}, \ and\ \bibinfo {author} {\bibfnamefont {S.}~\bibnamefont {Salimi}},\ }\bibfield  {title} {\enquote {\bibinfo {title} {Environment-mediated charging process of quantum batteries},}\ }\href {\doibase 10.1103/PhysRevA.102.052223} {\bibfield  {journal} {\bibinfo  {journal} {Phys. Rev. A}\ }\textbf {\bibinfo {volume} {102}},\ \bibinfo {pages} {052223} (\bibinfo {year} {2020})}\BibitemShut {NoStop}%
\bibitem [{\citenamefont {Kamin}\ \emph {et~al.}(2020)\citenamefont {Kamin}, \citenamefont {Tabesh}, \citenamefont {Salimi},\ and\ \citenamefont {Santos}}]{Iran-20}%
  \BibitemOpen
  \bibfield  {author} {\bibinfo {author} {\bibfnamefont {F.~H.}\ \bibnamefont {Kamin}}, \bibinfo {author} {\bibfnamefont {F.~T.}\ \bibnamefont {Tabesh}}, \bibinfo {author} {\bibfnamefont {S.}~\bibnamefont {Salimi}}, \ and\ \bibinfo {author} {\bibfnamefont {Alan~C.}\ \bibnamefont {Santos}},\ }\bibfield  {title} {\enquote {\bibinfo {title} {Entanglement, coherence, and charging process of quantum batteries},}\ }\href {\doibase 10.1103/PhysRevE.102.052109} {\bibfield  {journal} {\bibinfo  {journal} {Phys. Rev. E}\ }\textbf {\bibinfo {volume} {102}},\ \bibinfo {pages} {052109} (\bibinfo {year} {2020})}\BibitemShut {NoStop}%
\bibitem [{\citenamefont {Pirmoradian}\ and\ \citenamefont {M\o{}lmer}(2019)}]{Pirmoradian-19}%
  \BibitemOpen
  \bibfield  {author} {\bibinfo {author} {\bibfnamefont {Faezeh}\ \bibnamefont {Pirmoradian}}\ and\ \bibinfo {author} {\bibfnamefont {Klaus}\ \bibnamefont {M\o{}lmer}},\ }\bibfield  {title} {\enquote {\bibinfo {title} {Aging of a quantum battery},}\ }\href {\doibase 10.1103/PhysRevA.100.043833} {\bibfield  {journal} {\bibinfo  {journal} {Phys. Rev. A}\ }\textbf {\bibinfo {volume} {100}},\ \bibinfo {pages} {043833} (\bibinfo {year} {2019})}\BibitemShut {NoStop}%
\bibitem [{\citenamefont {Zhang}\ and\ \citenamefont {Blaauboer}(2023)}]{Zhang-23}%
  \BibitemOpen
  \bibfield  {author} {\bibinfo {author} {\bibfnamefont {Xiang}\ \bibnamefont {Zhang}}\ and\ \bibinfo {author} {\bibfnamefont {Miriam}\ \bibnamefont {Blaauboer}},\ }\bibfield  {title} {\enquote {\bibinfo {title} {Enhanced energy transfer in a dicke quantum battery},}\ }\href {\doibase 10.3389/fphy.2022.1097564} {\bibfield  {journal} {\bibinfo  {journal} {Frontiers in Physics}\ }\textbf {\bibinfo {volume} {10}} (\bibinfo {year} {2023}),\ 10.3389/fphy.2022.1097564}\BibitemShut {NoStop}%
\bibitem [{\citenamefont {Crescente}\ \emph {et~al.}(2020{\natexlab{b}})\citenamefont {Crescente}, \citenamefont {Carrega}, \citenamefont {Sassetti},\ and\ \citenamefont {Ferraro}}]{Crescente-20}%
  \BibitemOpen
  \bibfield  {author} {\bibinfo {author} {\bibfnamefont {A}~\bibnamefont {Crescente}}, \bibinfo {author} {\bibfnamefont {M}~\bibnamefont {Carrega}}, \bibinfo {author} {\bibfnamefont {M}~\bibnamefont {Sassetti}}, \ and\ \bibinfo {author} {\bibfnamefont {D}~\bibnamefont {Ferraro}},\ }\bibfield  {title} {\enquote {\bibinfo {title} {Charging and energy fluctuations of a driven quantum battery},}\ }\href {\doibase 10.1088/1367-2630/ab91fc} {\bibfield  {journal} {\bibinfo  {journal} {New Journal of Physics}\ }\textbf {\bibinfo {volume} {22}},\ \bibinfo {pages} {063057} (\bibinfo {year} {2020}{\natexlab{b}})}\BibitemShut {NoStop}%
\bibitem [{\citenamefont {Joshi}\ and\ \citenamefont {Mahesh}(2022)}]{Joshi-22}%
  \BibitemOpen
  \bibfield  {author} {\bibinfo {author} {\bibfnamefont {Jitendra}\ \bibnamefont {Joshi}}\ and\ \bibinfo {author} {\bibfnamefont {T.~S.}\ \bibnamefont {Mahesh}},\ }\bibfield  {title} {\enquote {\bibinfo {title} {Experimental investigation of a quantum battery using star-topology nmr spin systems},}\ }\href {\doibase 10.1103/PhysRevA.106.042601} {\bibfield  {journal} {\bibinfo  {journal} {Phys. Rev. A}\ }\textbf {\bibinfo {volume} {106}},\ \bibinfo {pages} {042601} (\bibinfo {year} {2022})}\BibitemShut {NoStop}%
\bibitem [{\citenamefont {Mohan}\ and\ \citenamefont {Pati}(2021)}]{BrijM-21}%
  \BibitemOpen
  \bibfield  {author} {\bibinfo {author} {\bibfnamefont {Brij}\ \bibnamefont {Mohan}}\ and\ \bibinfo {author} {\bibfnamefont {Arun~K.}\ \bibnamefont {Pati}},\ }\bibfield  {title} {\enquote {\bibinfo {title} {Reverse quantum speed limit: How slowly a quantum battery can discharge},}\ }\href {\doibase 10.1103/PhysRevA.104.042209} {\bibfield  {journal} {\bibinfo  {journal} {Phys. Rev. A}\ }\textbf {\bibinfo {volume} {104}},\ \bibinfo {pages} {042209} (\bibinfo {year} {2021})}\BibitemShut {NoStop}%
\bibitem [{\citenamefont {Niedenzu}\ \emph {et~al.}(2018)\citenamefont {Niedenzu}, \citenamefont {Mukherjee}, \citenamefont {Ghosh}, \citenamefont {Kofman},\ and\ \citenamefont {Kurizki}}]{Niedenzu-18}%
  \BibitemOpen
  \bibfield  {author} {\bibinfo {author} {\bibfnamefont {Wolfgang}\ \bibnamefont {Niedenzu}}, \bibinfo {author} {\bibfnamefont {Victor}\ \bibnamefont {Mukherjee}}, \bibinfo {author} {\bibfnamefont {Arnab}\ \bibnamefont {Ghosh}}, \bibinfo {author} {\bibfnamefont {Abraham~G.}\ \bibnamefont {Kofman}}, \ and\ \bibinfo {author} {\bibfnamefont {Gershon}\ \bibnamefont {Kurizki}},\ }\bibfield  {title} {\enquote {\bibinfo {title} {Quantum engine efficiency bound beyond the second law of thermodynamics},}\ }\href {\doibase 10.1038/s41467-017-01991-6} {\bibfield  {journal} {\bibinfo  {journal} {Nature Communications}\ }\textbf {\bibinfo {volume} {9}} (\bibinfo {year} {2018}),\ 10.1038/s41467-017-01991-6}\BibitemShut {NoStop}%
\bibitem [{\citenamefont {Ferraro}\ \emph {et~al.}(2018)\citenamefont {Ferraro}, \citenamefont {Campisi}, \citenamefont {Andolina}, \citenamefont {Pellegrini},\ and\ \citenamefont {Polini}}]{Ferraro-18}%
  \BibitemOpen
  \bibfield  {author} {\bibinfo {author} {\bibfnamefont {Dario}\ \bibnamefont {Ferraro}}, \bibinfo {author} {\bibfnamefont {Michele}\ \bibnamefont {Campisi}}, \bibinfo {author} {\bibfnamefont {Gian~Marcello}\ \bibnamefont {Andolina}}, \bibinfo {author} {\bibfnamefont {Vittorio}\ \bibnamefont {Pellegrini}}, \ and\ \bibinfo {author} {\bibfnamefont {Marco}\ \bibnamefont {Polini}},\ }\bibfield  {title} {\enquote {\bibinfo {title} {High-power collective charging of a solid-state quantum battery},}\ }\href {\doibase 10.1103/PhysRevLett.120.117702} {\bibfield  {journal} {\bibinfo  {journal} {Phys. Rev. Lett.}\ }\textbf {\bibinfo {volume} {120}},\ \bibinfo {pages} {117702} (\bibinfo {year} {2018})}\BibitemShut {NoStop}%
\bibitem [{\citenamefont {Zhao}\ \emph {et~al.}(2021)\citenamefont {Zhao}, \citenamefont {Dou},\ and\ \citenamefont {Zhao}}]{Zhao-21}%
  \BibitemOpen
  \bibfield  {author} {\bibinfo {author} {\bibfnamefont {Fang}\ \bibnamefont {Zhao}}, \bibinfo {author} {\bibfnamefont {Fu-Quan}\ \bibnamefont {Dou}}, \ and\ \bibinfo {author} {\bibfnamefont {Qing}\ \bibnamefont {Zhao}},\ }\bibfield  {title} {\enquote {\bibinfo {title} {Quantum battery of interacting spins with environmental noise},}\ }\href {\doibase 10.1103/PhysRevA.103.033715} {\bibfield  {journal} {\bibinfo  {journal} {Phys. Rev. A}\ }\textbf {\bibinfo {volume} {103}},\ \bibinfo {pages} {033715} (\bibinfo {year} {2021})}\BibitemShut {NoStop}%
\bibitem [{\citenamefont {Rossini}\ \emph {et~al.}(2019)\citenamefont {Rossini}, \citenamefont {Andolina},\ and\ \citenamefont {Polini}}]{Rossini-19}%
  \BibitemOpen
  \bibfield  {author} {\bibinfo {author} {\bibfnamefont {Davide}\ \bibnamefont {Rossini}}, \bibinfo {author} {\bibfnamefont {Gian~Marcello}\ \bibnamefont {Andolina}}, \ and\ \bibinfo {author} {\bibfnamefont {Marco}\ \bibnamefont {Polini}},\ }\bibfield  {title} {\enquote {\bibinfo {title} {Many-body localized quantum batteries},}\ }\href {\doibase 10.1103/PhysRevB.100.115142} {\bibfield  {journal} {\bibinfo  {journal} {Phys. Rev. B}\ }\textbf {\bibinfo {volume} {100}},\ \bibinfo {pages} {115142} (\bibinfo {year} {2019})}\BibitemShut {NoStop}%
\bibitem [{\citenamefont {Zakavati}\ \emph {et~al.}(2021)\citenamefont {Zakavati}, \citenamefont {Tabesh},\ and\ \citenamefont {Salimi}}]{Zakavati-21}%
  \BibitemOpen
  \bibfield  {author} {\bibinfo {author} {\bibfnamefont {Shadab}\ \bibnamefont {Zakavati}}, \bibinfo {author} {\bibfnamefont {Fatemeh~T.}\ \bibnamefont {Tabesh}}, \ and\ \bibinfo {author} {\bibfnamefont {Shahriar}\ \bibnamefont {Salimi}},\ }\bibfield  {title} {\enquote {\bibinfo {title} {Bounds on charging power of open quantum batteries},}\ }\href {\doibase 10.1103/PhysRevE.104.054117} {\bibfield  {journal} {\bibinfo  {journal} {Phys. Rev. E}\ }\textbf {\bibinfo {volume} {104}},\ \bibinfo {pages} {054117} (\bibinfo {year} {2021})}\BibitemShut {NoStop}%
\bibitem [{\citenamefont {Zhao}\ \emph {et~al.}(2022)\citenamefont {Zhao}, \citenamefont {Dou},\ and\ \citenamefont {Zhao}}]{Zhao-22}%
  \BibitemOpen
  \bibfield  {author} {\bibinfo {author} {\bibfnamefont {Fang}\ \bibnamefont {Zhao}}, \bibinfo {author} {\bibfnamefont {Fu-Quan}\ \bibnamefont {Dou}}, \ and\ \bibinfo {author} {\bibfnamefont {Qing}\ \bibnamefont {Zhao}},\ }\bibfield  {title} {\enquote {\bibinfo {title} {Charging performance of the su-schrieffer-heeger quantum battery},}\ }\href {\doibase 10.1103/PhysRevResearch.4.013172} {\bibfield  {journal} {\bibinfo  {journal} {Phys. Rev. Res.}\ }\textbf {\bibinfo {volume} {4}},\ \bibinfo {pages} {013172} (\bibinfo {year} {2022})}\BibitemShut {NoStop}%
\bibitem [{\citenamefont {Quach}\ \emph {et~al.}(2022)\citenamefont {Quach}, \citenamefont {McGhee}, \citenamefont {Ganzer}, \citenamefont {Rouse}, \citenamefont {Lovett}, \citenamefont {Gauger}, \citenamefont {Keeling}, \citenamefont {Cerullo}, \citenamefont {Lidzey},\ and\ \citenamefont {Virgili}}]{Quach-22}%
  \BibitemOpen
  \bibfield  {author} {\bibinfo {author} {\bibfnamefont {James~Q.}\ \bibnamefont {Quach}}, \bibinfo {author} {\bibfnamefont {Kirsty~E.}\ \bibnamefont {McGhee}}, \bibinfo {author} {\bibfnamefont {Lucia}\ \bibnamefont {Ganzer}}, \bibinfo {author} {\bibfnamefont {Dominic~M.}\ \bibnamefont {Rouse}}, \bibinfo {author} {\bibfnamefont {Brendon~W.}\ \bibnamefont {Lovett}}, \bibinfo {author} {\bibfnamefont {Erik~M.}\ \bibnamefont {Gauger}}, \bibinfo {author} {\bibfnamefont {Jonathan}\ \bibnamefont {Keeling}}, \bibinfo {author} {\bibfnamefont {Giulio}\ \bibnamefont {Cerullo}}, \bibinfo {author} {\bibfnamefont {David~G.}\ \bibnamefont {Lidzey}}, \ and\ \bibinfo {author} {\bibfnamefont {Tersilla}\ \bibnamefont {Virgili}},\ }\bibfield  {title} {\enquote {\bibinfo {title} {Superabsorption in an organic microcavity: Toward a quantum battery},}\ }\href {\doibase 10.1126/sciadv.abk3160} {\bibfield  {journal} {\bibinfo  {journal} {Science Advances}\ }\textbf {\bibinfo {volume} {8}} (\bibinfo {year} {2022}),\
  10.1126/sciadv.abk3160}\BibitemShut {NoStop}%
\bibitem [{\citenamefont {Carabba}\ \emph {et~al.}(2022)\citenamefont {Carabba}, \citenamefont {Hörnedal},\ and\ \citenamefont {Campo}}]{Carabba-22}%
  \BibitemOpen
  \bibfield  {author} {\bibinfo {author} {\bibfnamefont {Nicoletta}\ \bibnamefont {Carabba}}, \bibinfo {author} {\bibfnamefont {Niklas}\ \bibnamefont {Hörnedal}}, \ and\ \bibinfo {author} {\bibfnamefont {Adolfo~del}\ \bibnamefont {Campo}},\ }\bibfield  {title} {\enquote {\bibinfo {title} {Quantum speed limits on operator flows and correlation functions},}\ }\href {\doibase 10.22331/q-2022-12-22-884} {\bibfield  {journal} {\bibinfo  {journal} {Quantum}\ }\textbf {\bibinfo {volume} {6}},\ \bibinfo {pages} {884} (\bibinfo {year} {2022})}\BibitemShut {NoStop}%
\bibitem [{\citenamefont {Caputa}\ \emph {et~al.}(2022)\citenamefont {Caputa}, \citenamefont {Magan},\ and\ \citenamefont {Patramanis}}]{Pawel-22}%
  \BibitemOpen
  \bibfield  {author} {\bibinfo {author} {\bibfnamefont {Pawel}\ \bibnamefont {Caputa}}, \bibinfo {author} {\bibfnamefont {Javier~M.}\ \bibnamefont {Magan}}, \ and\ \bibinfo {author} {\bibfnamefont {Dimitrios}\ \bibnamefont {Patramanis}},\ }\bibfield  {title} {\enquote {\bibinfo {title} {Geometry of krylov complexity},}\ }\href {\doibase 10.1103/PhysRevResearch.4.013041} {\bibfield  {journal} {\bibinfo  {journal} {Physical Review Research}\ }\textbf {\bibinfo {volume} {4}},\ \bibinfo {pages} {013041} (\bibinfo {year} {2022})}\BibitemShut {NoStop}%
\bibitem [{\citenamefont {Pandey}\ \emph {et~al.}(2023)\citenamefont {Pandey}, \citenamefont {Shrimali}, \citenamefont {Mohan}, \citenamefont {Das},\ and\ \citenamefont {Pati}}]{Divyansh-22}%
  \BibitemOpen
  \bibfield  {author} {\bibinfo {author} {\bibfnamefont {Vivek}\ \bibnamefont {Pandey}}, \bibinfo {author} {\bibfnamefont {Divyansh}\ \bibnamefont {Shrimali}}, \bibinfo {author} {\bibfnamefont {Brij}\ \bibnamefont {Mohan}}, \bibinfo {author} {\bibfnamefont {Siddhartha}\ \bibnamefont {Das}}, \ and\ \bibinfo {author} {\bibfnamefont {Arun~Kumar}\ \bibnamefont {Pati}},\ }\bibfield  {title} {\enquote {\bibinfo {title} {Speed limits on correlations in bipartite quantum systems},}\ }\href {\doibase 10.1103/PhysRevA.107.052419} {\bibfield  {journal} {\bibinfo  {journal} {Phys. Rev. A}\ }\textbf {\bibinfo {volume} {107}},\ \bibinfo {pages} {052419} (\bibinfo {year} {2023})}\BibitemShut {NoStop}%
\bibitem [{\citenamefont {Pati}\ \emph {et~al.}(2023)\citenamefont {Pati}, \citenamefont {Mohan}, \citenamefont {Sahil},\ and\ \citenamefont {Braunstein}}]{pati-23}%
  \BibitemOpen
  \bibfield  {author} {\bibinfo {author} {\bibfnamefont {Arun~K.}\ \bibnamefont {Pati}}, \bibinfo {author} {\bibfnamefont {Brij}\ \bibnamefont {Mohan}}, \bibinfo {author} {\bibnamefont {Sahil}}, \ and\ \bibinfo {author} {\bibfnamefont {Samuel~L.}\ \bibnamefont {Braunstein}},\ }\href@noop {} {\enquote {\bibinfo {title} {Exact quantum speed limits},}\ } (\bibinfo {year} {2023}),\ \Eprint {http://arxiv.org/abs/2305.03839} {arXiv:2305.03839 [quant-ph]} \BibitemShut {NoStop}%
\bibitem [{\citenamefont {D\"ur}\ \emph {et~al.}(2001)\citenamefont {D\"ur}, \citenamefont {Vidal}, \citenamefont {Cirac}, \citenamefont {Linden},\ and\ \citenamefont {Popescu}}]{rate_ent}%
  \BibitemOpen
  \bibfield  {author} {\bibinfo {author} {\bibfnamefont {Wolfgang}\ \bibnamefont {D\"ur}}, \bibinfo {author} {\bibfnamefont {Guifre}\ \bibnamefont {Vidal}}, \bibinfo {author} {\bibfnamefont {Juan~Ignacio}\ \bibnamefont {Cirac}}, \bibinfo {author} {\bibfnamefont {Noah}\ \bibnamefont {Linden}}, \ and\ \bibinfo {author} {\bibfnamefont {Sandu}\ \bibnamefont {Popescu}},\ }\bibfield  {title} {\enquote {\bibinfo {title} {Entanglement capabilities of nonlocal hamiltonians},}\ }\href {\doibase 10.1103/PhysRevLett.87.137901} {\bibfield  {journal} {\bibinfo  {journal} {Phys. Rev. Lett.}\ }\textbf {\bibinfo {volume} {87}},\ \bibinfo {pages} {137901} (\bibinfo {year} {2001})}\BibitemShut {NoStop}%
\bibitem [{\citenamefont {Bennett}\ \emph {et~al.}(2002)\citenamefont {Bennett}, \citenamefont {Cirac}, \citenamefont {Leifer}, \citenamefont {Leung}, \citenamefont {Linden}, \citenamefont {Popescu},\ and\ \citenamefont {Vidal}}]{bennett-02}%
  \BibitemOpen
  \bibfield  {author} {\bibinfo {author} {\bibfnamefont {C.~H.}\ \bibnamefont {Bennett}}, \bibinfo {author} {\bibfnamefont {J.~I.}\ \bibnamefont {Cirac}}, \bibinfo {author} {\bibfnamefont {M.~S.}\ \bibnamefont {Leifer}}, \bibinfo {author} {\bibfnamefont {D.~W.}\ \bibnamefont {Leung}}, \bibinfo {author} {\bibfnamefont {N.}~\bibnamefont {Linden}}, \bibinfo {author} {\bibfnamefont {S.}~\bibnamefont {Popescu}}, \ and\ \bibinfo {author} {\bibfnamefont {G.}~\bibnamefont {Vidal}},\ }\bibfield  {title} {\enquote {\bibinfo {title} {Optimal simulation of two-qubit hamiltonians using general local operations},}\ }\href {\doibase 10.1103/PhysRevA.66.012305} {\bibfield  {journal} {\bibinfo  {journal} {Physical Review A}\ }\textbf {\bibinfo {volume} {66}},\ \bibinfo {pages} {012305} (\bibinfo {year} {2002})}\BibitemShut {NoStop}%
\bibitem [{\citenamefont {Pati}(1995)}]{pati-95}%
  \BibitemOpen
  \bibfield  {author} {\bibinfo {author} {\bibfnamefont {Arun~Kumar}\ \bibnamefont {Pati}},\ }\bibfield  {title} {\enquote {\bibinfo {title} {New derivation of the geometric phase},}\ }\href {\doibase https://doi.org/10.1016/0375-9601(95)00299-I} {\bibfield  {journal} {\bibinfo  {journal} {Physics Letters A}\ }\textbf {\bibinfo {volume} {202}},\ \bibinfo {pages} {40} (\bibinfo {year} {1995})}\BibitemShut {NoStop}%
\bibitem [{\citenamefont {Deffner}\ and\ \citenamefont {Campbell}(2017{\natexlab{b}})}]{Deffner-17}%
  \BibitemOpen
  \bibfield  {author} {\bibinfo {author} {\bibfnamefont {Sebastian}\ \bibnamefont {Deffner}}\ and\ \bibinfo {author} {\bibfnamefont {Steve}\ \bibnamefont {Campbell}},\ }\bibfield  {title} {\enquote {\bibinfo {title} {Quantum speed limits: from heisenberg's uncertainty principle to optimal quantum control},}\ }\href {\doibase 10.1088/1751-8121/aa86c6} {\bibfield  {journal} {\bibinfo  {journal} {Journal of Physics A: Mathematical and Theoretical}\ }\textbf {\bibinfo {volume} {50}},\ \bibinfo {pages} {453001} (\bibinfo {year} {2017}{\natexlab{b}})}\BibitemShut {NoStop}%
\bibitem [{\citenamefont {Bravyi}(2007)}]{Bravyi-07}%
  \BibitemOpen
  \bibfield  {author} {\bibinfo {author} {\bibfnamefont {Sergey}\ \bibnamefont {Bravyi}},\ }\bibfield  {title} {\enquote {\bibinfo {title} {Upper bounds on entangling rates of bipartite hamiltonians},}\ }\href {\doibase 10.1103/PhysRevA.76.052319} {\bibfield  {journal} {\bibinfo  {journal} {Physical Review A}\ }\textbf {\bibinfo {volume} {76}},\ \bibinfo {pages} {052319} (\bibinfo {year} {2007})}\BibitemShut {NoStop}%
\bibitem [{\citenamefont {Pfeifer}(1993)}]{Pfeifer-93}%
  \BibitemOpen
  \bibfield  {author} {\bibinfo {author} {\bibfnamefont {Peter}\ \bibnamefont {Pfeifer}},\ }\bibfield  {title} {\enquote {\bibinfo {title} {How fast can a quantum state change with time?}}\ }\href {\doibase 10.1103/PhysRevLett.70.3365} {\bibfield  {journal} {\bibinfo  {journal} {Physical Review Letter}\ }\textbf {\bibinfo {volume} {70}},\ \bibinfo {pages} {3365} (\bibinfo {year} {1993})}\BibitemShut {NoStop}%
\bibitem [{\citenamefont {xiong Wu}\ and\ \citenamefont {shui Yu}(2018)}]{Wu-18}%
  \BibitemOpen
  \bibfield  {author} {\bibinfo {author} {\bibfnamefont {Shao}\ \bibnamefont {xiong Wu}}\ and\ \bibinfo {author} {\bibfnamefont {Chang}\ \bibnamefont {shui Yu}},\ }\bibfield  {title} {\enquote {\bibinfo {title} {Quantum speed limit for a mixed initial state},}\ }\href {https://doi.org/10.1103%2Fphysreva.98.042132} {\bibfield  {journal} {\bibinfo  {journal} {Physical Review A}\ }\textbf {\bibinfo {volume} {98}} (\bibinfo {year} {2018})}\BibitemShut {NoStop}%
\bibitem [{\citenamefont {Campaioli}\ \emph {et~al.}(2022)\citenamefont {Campaioli}, \citenamefont {shui Yu}, \citenamefont {Pollock},\ and\ \citenamefont {Modi}}]{Modi-20}%
  \BibitemOpen
  \bibfield  {author} {\bibinfo {author} {\bibfnamefont {Francesco}\ \bibnamefont {Campaioli}}, \bibinfo {author} {\bibfnamefont {Chang}\ \bibnamefont {shui Yu}}, \bibinfo {author} {\bibfnamefont {Felix~A}\ \bibnamefont {Pollock}}, \ and\ \bibinfo {author} {\bibfnamefont {Kavan}\ \bibnamefont {Modi}},\ }\bibfield  {title} {\enquote {\bibinfo {title} {Resource speed limits: maximal rate of resource variation},}\ }\href {\doibase 10.1088/1367-2630/ac7346} {\bibfield  {journal} {\bibinfo  {journal} {New Journal of Physics}\ }\textbf {\bibinfo {volume} {24}},\ \bibinfo {pages} {065001} (\bibinfo {year} {2022})}\BibitemShut {NoStop}%
\bibitem [{\citenamefont {Thakuria}\ \emph {et~al.}(2023)\citenamefont {Thakuria}, \citenamefont {Srivastav}, \citenamefont {Mohan}, \citenamefont {Kumari},\ and\ \citenamefont {Pati}}]{dimpi-22JPA}%
  \BibitemOpen
  \bibfield  {author} {\bibinfo {author} {\bibfnamefont {Dimpi}\ \bibnamefont {Thakuria}}, \bibinfo {author} {\bibfnamefont {Abhay}\ \bibnamefont {Srivastav}}, \bibinfo {author} {\bibfnamefont {Brij}\ \bibnamefont {Mohan}}, \bibinfo {author} {\bibfnamefont {Asmita}\ \bibnamefont {Kumari}}, \ and\ \bibinfo {author} {\bibfnamefont {Arun~Kumar}\ \bibnamefont {Pati}},\ }\bibfield  {title} {\enquote {\bibinfo {title} {Generalised quantum speed limit for arbitrary time-continuous evolution},}\ }\href {\doibase 10.1088/1751-8121/ad15ad} {\bibfield  {journal} {\bibinfo  {journal} {Journal of Physics A: Mathematical and Theoretical}\ }\textbf {\bibinfo {volume} {57}},\ \bibinfo {pages} {025302} (\bibinfo {year} {2023})}\BibitemShut {NoStop}%
\bibitem [{\citenamefont {Mohan}\ \emph {et~al.}(2022)\citenamefont {Mohan}, \citenamefont {Das},\ and\ \citenamefont {Pati}}]{mohan-22}%
  \BibitemOpen
  \bibfield  {author} {\bibinfo {author} {\bibfnamefont {Brij}\ \bibnamefont {Mohan}}, \bibinfo {author} {\bibfnamefont {Siddhartha}\ \bibnamefont {Das}}, \ and\ \bibinfo {author} {\bibfnamefont {Arun~Kumar}\ \bibnamefont {Pati}},\ }\bibfield  {title} {\enquote {\bibinfo {title} {Quantum speed limits for information and coherence},}\ }\href {\doibase 10.1088/1367-2630/ac753c} {\bibfield  {journal} {\bibinfo  {journal} {New Journal of Physics}\ }\textbf {\bibinfo {volume} {24}},\ \bibinfo {pages} {065003} (\bibinfo {year} {2022})}\BibitemShut {NoStop}%
\bibitem [{\citenamefont {Alicki}(1979)}]{Alicki-79}%
  \BibitemOpen
  \bibfield  {author} {\bibinfo {author} {\bibfnamefont {R}~\bibnamefont {Alicki}},\ }\bibfield  {title} {\enquote {\bibinfo {title} {The quantum open system as a model of the heat engine},}\ }\href {\doibase 10.1088/0305-4470/12/5/007} {\bibfield  {journal} {\bibinfo  {journal} {Journal of Physics A: Mathematical and General}\ }\textbf {\bibinfo {volume} {12}},\ \bibinfo {pages} {L103} (\bibinfo {year} {1979})}\BibitemShut {NoStop}%
\bibitem [{\citenamefont {Campaioli}\ \emph {et~al.}(2018{\natexlab{b}})\citenamefont {Campaioli}, \citenamefont {Pollock},\ and\ \citenamefont {Vinjanampathy}}]{campaioli-18}%
  \BibitemOpen
  \bibfield  {author} {\bibinfo {author} {\bibfnamefont {Francesco}\ \bibnamefont {Campaioli}}, \bibinfo {author} {\bibfnamefont {Felix~A.}\ \bibnamefont {Pollock}}, \ and\ \bibinfo {author} {\bibfnamefont {Sai}\ \bibnamefont {Vinjanampathy}},\ }\href@noop {} {\enquote {\bibinfo {title} {Quantum batteries - review chapter},}\ } (\bibinfo {year} {2018}{\natexlab{b}}),\ \Eprint {http://arxiv.org/abs/1805.05507} {arXiv:1805.05507 [quant-ph]} \BibitemShut {NoStop}%
\end{thebibliography}%

\begin{widetext}

\appendix

\section{SQSL bound for Entanglement Capacity}\label{ec}

SQSL bound for entanglement generation using the capacity of entanglement under Schr\"{o}dinger picture requires the functional form of $R(t)$ and $\ket{\Psi^{\perp}}$ as defined in Eq.\eqref{eqn:perp_prescription}

\begin{equation}
    R(t) = \frac{\left|\sqrt{-\arctanh(\alpha)^{2} \beta}/\sqrt{2} + \arctanh(\alpha) \hspace{0.1cm}{\rm Sign}((1-2p)\theta)\hspace{0.1cm}(-2 i \sqrt{p(1-p)}\hspace{0.1cm}\cos(2\theta t)-\sin(2\theta t))\right|^{2}}{\left|\arctanh(\alpha)^{2}\hspace{0.1cm}\beta\right|} ,
\end{equation}
\noindent where $\alpha=(2p-1)\cos(2\theta t)$, $\beta = -1 - 4p(1-p) + (1-2p)^{2}\cos(4\theta t)$. Also,
\begin{equation}
    |\Psi^{\perp}\rangle = 
\left(
\begin{array}{c}
\frac{{\sqrt{2} e^{-i t \mu_3} \operatorname{arctan}\left[(-1 + 2 p) \cos\left(2 t \theta\right)\right] \left(-1 + (-1 + 2 p) \cos\left(2 t \theta\right)\right) \left(\sqrt{p} \cos\left(t \theta\right) - i \sqrt{1 - p} \sin\left(t \theta\right)\right)}}{{\sqrt{-\operatorname{arctanh}\left[(-1 + 2 p) \cos\left(2 t \theta\right)\right]^2 \left(-1 + 4 (-1 + p) p + (1 - 2 p)^2 \cos\left(4 t \theta\right)\right)}}} \\
0 \\
0 \\
\frac{{\sqrt{2} e^{-i t \mu_3} \operatorname{arctanh}\left[(-1 + 2 p) \cos\left(2 t \theta\right)\right] \left(1 + (-1 + 2 p) \cos\left(2 t \theta\right)\right) \left(\sqrt{1 - p} \cos\left(t \theta\right) - i \sqrt{p} \sin\left(t \theta\right)\right)}}{{\sqrt{-\operatorname{arctanh}\left[(-1 + 2 p) \cos\left(2 t \theta\right)\right]^2 \left(-1 + 4 (-1 + p) p + (1 - 2 p)^2 \cos\left(4 t \theta\right)\right)}}}
\end{array}
\right) ,
\end{equation}

\noindent where \(\theta = (\mu_1 - \mu_2 )\).\\

Similarly, SQSLO bound for generation of modular energy using the composite modular Hamiltonian under the Heisenberg picture requires the functional form of $R(t)$ and $\ket{\Psi^{\perp}}$ as defined in Eq.\eqref{eqn:perp_prescription}.

\begin{equation}
   |\Psi^{\perp}\rangle = 
\left(
\begin{array}{c}
\frac{\log{\left(-1 + \frac{1}{p}\right)} \left(-2 \left(-1 + p\right) \sqrt{p} \cos{\left(2\hspace{0.05cm}\theta\hspace{0.05cm}t\right)} - i \sqrt{1 - p} \sin{\left(2\hspace{0.05cm}\theta\hspace{0.05cm}t\right)}\right)}{\sqrt{\log^2{\left(-1 + \frac{1}{p}\right)} \left(-4 \left(-1 + p\right) p \cos^{2}{\left(2\hspace{0.05cm}\theta\hspace{0.05cm}t\right)} + \sin^{2}{\left(2\hspace{0.05cm}\theta\hspace{0.05cm}t\right)}\right)}} \\
0 \\
0 \\
\frac{\log{\left(-1 + \frac{1}{p}\right)} \left(-2 \sqrt{\left(1 - p\right) p} \cos{\left(2\hspace{0.05cm}\theta\hspace{0.05cm}t\right)} + i \sin{\left(2\hspace{0.05cm}\theta\hspace{0.05cm}t\right)}\right)}{\sqrt{\frac{\log^2{\left(-1 + \frac{1}{p}\right)} \left(-4 \left(-1 + p\right) p \cos^{2}{\left(2\hspace{0.05cm}\theta\hspace{0.05cm}t\right)} + \sin^{2}{\left(2\hspace{0.05cm}\theta\hspace{0.05cm}t\right)}\right)}{p}}} \\
\end{array}
\right) ,
\label{eqn:psi_perp_H_cap}
\end{equation}

For $p$ = $0.1$ and $\Theta$ = $1$, $R$ is given as:

\begin{equation}
R(t) = \frac{0.63 \left|1.89 + \cos(2t) \left(1.67i \sqrt{0.68 - 0.32 \cos(4t)}\right) - 0.89 \cos(4t) + 2.78 \sqrt{0.68 - 0.32 \cos(4t)} \sin(2t) \right|^2}{(2.13 - \cos(4t))^2}  ,
\label{eqn:R01_H}
\end{equation}

Again, for $p$ = $0.1$ and $\Theta$ = $0.5$, $R$ is given as:

\begin{equation}
    R(t) = - \frac{{0.15 \left| \cos(t)^2 + \cos(t) \left(1.67i) \sqrt{0.68 - 0.32 \cos(2 t)}\right) + \sin(t) \left(2.78 \sqrt{0.68 - 0.32 \cos(2 t)} + 2.78 \sin(t)\right)\right|^2}}{-1.18 + \cos(2 t) - 0.12 \cos(4 t)} ,
\label{eqn:R015_H}
\end{equation}

Again, for $p$ = $0.4$ and $\Theta$ = $1$, $R$ is given as:

\begin{equation}
    R(t) = 0.5 \left| \frac{1}{(-49 + \cos(4t))} \left(-49 +  \cos(4t) - ( 48.99i) \sqrt{0.98 - 0.02 \cos(4t)} \left( \cos(2t) - (2.04i) \cos(t) \sin(t)\right) \right) \right|^2 ,
\label{eqn:R04_H}
\end{equation}

At last, for $p$ = $0.4$ and $\Theta$ = $0.5$, $R$ is given as:

\begin{equation}
    R(t) = - \frac{{11.76 \left| \cos(t)^2 + \cos(t) \left(-1.02i \sqrt{0.98 - 0.02 \cos(2t)}\right) + \sin(t) \left( 1.04 \sqrt{0.98 - 0.02 \cos(2t)} + 1.04 \sin(t)\right)  \right|^2}}{{-24.51 + \cos(2t) - 0.005 \cos(4t)}}
\label{eqn:R045_H} .
\end{equation}

\section{Related to Quantum Battery}\label{qba}

There are three cases in which we have calculated the SQSLO for QBs. For each case, we require the values of $|\Psi\rangle$, and $|\Psi^{\perp}\rangle$. We will define each of these components individually.

Here, for all cases the initial state is:

\begin{equation}
    |\Psi\rangle = (0, 0, 0, 1)^T ,
\end{equation}

% \begin{equation}
% |\Psi^{\perp}\rangle = \left(0, \frac{e^{-2i+\sqrt{5}t}(-2 + \sqrt{5} + 4e^{2i+\sqrt{5}t} - (2 + \sqrt{5}) \cdot e^{4i\sqrt{5}t} )}{4 \left|\sin(\sqrt{5} \cdot t) \cdot \sqrt{9 + \cos(2\sqrt{5} \cdot t)}\right|}, \frac{e^{-2i+\sqrt{5}t}(-2 + \sqrt{5} + 4e^{2i+\sqrt{5}t} - (2 + \sqrt{5}) \cdot e^{4+\sqrt{5}} \cdot t)}{4 \left|\sin(\sqrt{5} \cdot t) \cdot \sqrt{9 + \cos(2\sqrt{5} \cdot t)}\right|}, 0\right)
% \end{equation}

For $J = 1$, $\omega = 2$ and $\Omega = 1$, the function $R(t)$ in Eq.~\eqref{eqn:R_coup}, $|\Psi^{\perp}\rangle$ is provided below:

\begin{equation}
|\Psi^{\perp}\rangle = \left(
\begin{array}{c}
0 \\
\frac{2 - 2\cos(2\sqrt{5} t) - i  \sqrt{5} \sin(2\sqrt{5} t)}{2\left|\sin(\sqrt{5} t)\right|  \sqrt{9 + \cos(2\sqrt{5}  t)}} \\
\frac{2 - 2\cos(2\sqrt{5} t) - i  \sqrt{5} \sin(2\sqrt{5} t)}{2\left|\sin(\sqrt{5} t)\right| \sqrt{9 + \cos(2\sqrt{5} t)}} \\
0
\end{array}
\right) .
\end{equation}

% \begin{equation}
% |\Psi^{\perp}\rangle = \left(0, \frac{2 - 2\cos(2\sqrt{5} t) - i  \sqrt{5} \sin(2\sqrt{5} t)}{2\left|\sin(\sqrt{5} t)\right| \cdot \sqrt{9 + \cos(2\sqrt{5} \cdot t)}} ,\frac{2 - 2\cos(2\sqrt{5} t) - i  \sqrt{5} \sin(2\sqrt{5} t)}{2\left|\sin(\sqrt{5} t)\right| \cdot \sqrt{9 + \cos(2\sqrt{5} \cdot t)}} ,0\right).
% \end{equation}
 
For $J = 1$, $\omega = 2$ and $\Omega = 4$, the function $R(t)$ in Eq.~\eqref{eqn:R_nocoup}, $\Psi^{\perp}\rangle$ is provided below:

\begin{equation}
|\Psi^{\perp}\rangle = \left(
\begin{array}{c}
0 \\
\frac{1 - \cos(4\sqrt{5} t) - t  \sqrt{5} \sin(4\sqrt{5}  t)}{2\left|\sin(2\sqrt{5}  t)\right|  \sqrt{6 + 4\cos(4\sqrt{5} t)}} \\
\frac{1 - \cos(4\sqrt{5} t) - t = \sqrt{5}  \sin(4\sqrt{5} t)}{2\left|\sin(2\sqrt{5}  t)\right|  \sqrt{6 + 4\cos(4\sqrt{5}  t)}} \\
0
\end{array}
\right) .
\end{equation}

For $J = 0$, $\omega = 2$ and $\Omega = 1$, the functions $R(t)$ in Eq.~\eqref{R_J0}, $|\Psi^{\perp}\rangle$ is provided below:

\begin{equation}
|\Psi^{\perp}\rangle = \left(
\begin{array}{c}
0 \\
\frac{e^{-2i+\sqrt{5}t}(-2 + \sqrt{5} + 4e^{2i+\sqrt{5}t} - (2 + \sqrt{5}) e^{4i\sqrt{5}t} )}{4 \left|\sin(\sqrt{5}\hspace{0.1cm}t)\right|\sqrt{9 + \cos(2\sqrt{5} t)}} \\
\frac{e^{-2i+\sqrt{5}t}(-2 + \sqrt{5} + 4e^{2i+\sqrt{5}t} - (2 + \sqrt{5}) e^{4+\sqrt{5}} t)}{4 \left|\sin(\sqrt{5}\hspace{0.1cm}t) \right| \sqrt{9 + \cos(2\sqrt{5} t)}} \\
0
\end{array}
\right) .
\end{equation}

\end{widetext}

% \begin{appendices}

% \section{Related to Entanglement Capacity}\label{ec}
% SQSL bound for entanglement generation using capacity of entanglement under Schrodinger picture requires functional form of $R(t)$ and $\ket{\Psi^{\perp}}$ as defined in~\eqref{eqn:perp_prescription}

% \end{appendices}

\end{document}